\documentclass[ review]{elsarticle}
\usepackage{setspace}
\usepackage{enumerate}
\usepackage{graphicx}
\usepackage{amsmath, bm}
\usepackage{float}
\usepackage{mdframed}
\usepackage{multirow}
\usepackage{textcomp}
\usepackage{subcaption}
\usepackage{url}
\usepackage{makecell}
\usepackage{amssymb}
\usepackage{bm}
\usepackage{textcmds}
 \usepackage{hyperref}
\biboptions{numbers,sort&compress}

\usepackage{tikz}
\usetikzlibrary{shapes.geometric, arrows}
\usepackage{caption}  

\def\be{\begin{equation}}\def\ee{\end{equation}}
\def\ba{\begin{array}}\def\ea{\end{array}}
\def\bfg{\begin{figure}}\def\efg{\end{figure}}
\makeatletter
\def\fps@figure{htbp}
\makeatother

\usepackage{stackengine}
\stackMath
\newcommand\tenq[2][1]{%
 \def\useanchorwidth{T}%
  \ifnum#1>1%
    \stackunder[0pt]{\tenq[\numexpr#1-1\relax]{#2}}{\scriptscriptstyle\sim}%
  \else%
    \stackunder[1pt]{#2}{\scriptscriptstyle\sim}%
  \fi%
}

\RequirePackage[margin=20mm]{geometry}

\journal{International journal of solids and structures}
\begin{document}

\begin{frontmatter}

\title{Fractional-Order Thermo-Piezoelectric Modelling of qP-Wave Interaction and Energy Partition at Welded Interface}

\author[label1]{Hriticka Dhiman}
\author[label1]{Soniya Chaudhary*}
\cortext[cor1]{Corresponding author: soniyachaudhary18@gmail.com}
\address[label1]{Department of Mathematics and Scientific Computing, National Institute of Technology Hamirpur, Himachal Pradesh, 177005, India}

\begin{abstract}
An analytical model is developed to investigate the interaction of quasi-longitudinal (qP) waves with a perfectly bonded interface between a thermo-piezoelectric half-space and a functionally graded piezoelectric half-space. The formulation is based on the fractional-order Lord–Shulman generalized thermoelasticity theory, which provides an enhanced description of coupled thermo-electro-mechanical wave behaviour. Rotational effects are incorporated into the constitutive relations and equations of motion for both media, while the lower half-space is assumed to be subjected to initial stress. Closed-form solutions for reflection and transmission coefficients are obtained, together with associated energy partition factors, allowing a comprehensive assessment of interface-wave characteristics. Numerical simulations carried out using MATLAB demonstrate that the reflection and transmission responses are strongly influenced by initial stress, fractional-order parameter, and thermal relaxation time. The calculated energy ratios of scattered waves satisfy the energy conservation condition, confirming the mathematical consistency of the formulation. The findings of this study are relevant to the design and analysis of smart sensors, rotating and aerospace structures, vibration control systems, and energy-harvesting devices employing functionally graded thermo-piezoelectric materials under fractional-order effects.
\end{abstract}

\end{frontmatter}
\section{Introduction}
A wave is a physical mechanism through which energy is conveyed from one location to another without any net transport of the material medium itself. When a wave encounters an interface separating two distinct media, part of its energy may be redirected back into the original medium, a phenomenon known as reflection, which results in change in the direction of the wavefront. At the same interface, another portion of the wave may continue to propagate into the second medium; this process is referred to as transmission, during which the incident wave may be partially or completely absorbed. In practical situations, reflection and transmission generally occur simultaneously. The study of waves at interfaces between dissimilar medium is of considerable importance in fields such as seismology and geophysics, where it aids in understanding subsurface structures and material properties. In recent years, a wide range of seismic devices have been developed using smart materials  owing to their superior performance and enhanced responsiveness under elastic conditions. Surface wave have the low frequency than the body wave when they go through the crust. They can be sorted as a type of mechanical surface wave.
A new surface wave in piezoelectric materials has been studied by JL Bleustein \cite{bleustein1968new}.
\subsection{Wave propagation in piezoelectric media}
The concept of piezoelectricity was first identified in 1880 by the French physiscists Jacques and Pierrre Curie. Piezoelectricity is a marvel in which mechanical vitality is changed over into electrical vitality and the other way around. The most familiar piezoelectric material is quartz. 
 Abd-alla and Alsheikh \cite{abd2009reflection} studied a problem of reflection and refraction of quasi-longitudinal waves under initial stresses at an interface of two anisotropic piezoelectric media with different properties. Yuan and Zhu \cite{yuan2012reflection} examined plane-wave scattering at the interface between dissimilar piezoelectric media. Fang et al. \cite{fang2001surface} analyzed surface acoustic wave propagation in a rotating piezoelectric substrate. Abd-alla et al. \cite{abd2012reflection} investigated the propagation of plane vertical transverse waves at the interface of a semi-infinite piezoelectric elastic medium under initial stress conditions. Singh \cite{singh2011effect} has shown the effect of initial stresses on incident qSV-waves in prestressed elastic half-spaces.  Abo-el-nour N. Abd-alla \cite{abd2014mathematical} develops a mathematical model to analyze the scattering behavior of longitudinal elastic waves at the interface of thermo-piezoelectric materials, incorporating both thermal and piezoelectric coupling effects. Yang \cite{yang2006review} reviewed few topics in piezoelectricity.
 Pang et al. \cite{pang2008reflection}examined plane-wave interactions at the interface of piezoelectric and piezomagnetic materials, focusing on reflection and refraction phenomena.
 Piezoelectric materials are extensively employed in signal processing, sensing, and frequency regulation systems, where understanding wave reflection and transmission across interfaces between different media is essential. In particular, the analysis of surface wave characteristics, such as Rayleigh and leaky waves, is essential for the effective design and optimization of Seismic Acoustic Wave (SAW) devices.
\subsection{Significance of functionally graded piezoelectric media in wave propagation}
Functionally graded materials are composites with spatially varying properties that smoothly transition between different phases, resulting in reduced stress concentrations and enhanced thermal performance compared to laminated materials. They are commonly fabricated using isotropic constituents such as metals and ceramics, particularly due to their effectiveness as thermal barrier materials in applications involving extreme temperature gradients. Saroj et al.\cite{saroj2015love} analyzed Love-type wave propagation in FGPM structures subjected to initial stress and supported by an elastic substrate. Arani et al.\cite{arani2011effect} investigated the influence of material inhomogeneity on electro-thermo-mechanical response of a rotating FGP shaft. Sahu et al. \cite{sahu2019scattering} analyzed surface wave propagation in functionally graded piezoelectric media using an analytical approach.
\subsection{Fractional-order theories for wave propagation in coupled media} Fractional calculus concerns the study of derivatives and integrals of arbitrary real or complex orders, thereby unifying and extending the classical concepts of integer-order differentiation and repeated integration. Fractional calculus is a mathematical framework that extends the classical concepts of differentiation and integration to non-integer orders, in a manner similar to the extension of integer-valued exponents to fractional powers.  It is evident that numerous applications of fractional calculus have emerged, particularly during the 20th century. Although the physical interpretation of fractional-order operators is often not straightforward, the underlying mathematical formulations are no less rigorous than their integer-order counterparts. Rajneesh Kumar and Poonam Sharma 
\cite{kumar2017effect} studied fractional-order effects on energy ratios at the boundary of elastic–piezothermoelastic media. Within the context of two-phase-lag and three-phase-lag thermoelasticity, Kumar and Gupta \cite{kumar2013plane} analyzed plane wave propagation in anisotropic thermoelastic media with voids and fractional-order derivatives. Kaur \cite{kaur2019effect} investigated the influence of Hall current on plane wave propagation in a transversely isotropic thermoelastic medium considering a two-temperature theory and fractional-order heat conduction. Kaur further extended her research to analyze the reflection and refraction of plane waves in piezo-thermoelastic diffusive half-spaces incorporating three-phase-lag memory-dependent derivatives and two-temperature theory \cite{kaur2022reflection}. Reflection of plane waves in a fraction-order generalized magneto-thermoelasticity in a rotating triclinic solid half-space has been studied by Yadav \cite{yadav2022reflection}. Y Kang \cite{kang2021modeling} analyzed the reflection and transmission behavior of elastic waves in a configuration where a partially saturated porous layer is embedded between two elastic half-spaces. Bibi \cite{bibi2023propagation} examined thermoelastic wave propagation and reflection in a rotating nonlocal porous medium incorporating fractional-order effects and Hall current. The application of fractional calculus in mechanics has been discussed by \cite{atanackovic2014fractional}.  Behavior of higher-order MDD on energy ratios at the interface of thermoelastic and piezothermoelastic mediums has been studied by B.M.Barak \cite{barak2023behavior}. SM Said \cite{said2024reflection} focuses on the  analysis of the reflection of the waves through a fiber-reinforced thermoelastic medium under the effect of the magnetic field, gravity, and the initial stress. Effects of external magnetic field on the dispersion and attenuation feature as well as the reflection and transmission behavior of thermoelastic coupled waves are studied by Ying Li \cite{li2025effects}.
\subsection{Motivation, objectives and contributions of the present work}

A critical review of the existing literature reveals that, although wave propagation in piezoelectric and functionally graded piezoelectric media has been extensively studied, most investigations address the influencing factors independently. In particular, the combined effects of initial stress, rotation, material gradation, relaxation time, and fractional-order derivatives on wave reflection and refraction have not yet been examined within a unified theoretical framework. This shortcoming in the literature provides the primary motivation for the present study.

The aim of this study is to investigate the influence of initial stress, rotational parameters, material gradient, order of fractional derivatives, and relaxation time on the scattering characteristics of qP waves at the interface between FGPM and thermo-piezoelectric medium. The amplitude and energy ratios of scattered wave modes are evaluated numerically and their variation with the angle of incidence is illustrated graphically. The individual and combined effects of the governing parameters on the wave behavior of the considered structure are also systematically analyzed through graphical results.

To model the underlying physics more realistically, the formulation incorporates fractional-order derivatives, which are well-suited for describing memory-dependent and nonlocal effects in wave propagation, thermoelasticity, and piezoelectric media. Fractional calculus enables an accurate representation of anomalous dispersion, attenuation, and thermal relaxation phenomena, particularly in complex and functionally graded materials. The earliest application of fractional derivatives dates back to Abel, who employed them in solving an integral equation related to the tautochrone problem. In this context, the present study adopts fractional-order operators to develop a mathematically generalized and physically consistent framework.

The findings of this work are expected to be useful for the design and analysis of functionally graded thermo-piezoelectric structures operating under coupled mechanical, thermal, rotational, and fractional-order effects, with potential applications in smart sensing, structural health monitoring, aerospace and rotating systems, vibration control, thermal protection, geophysics, and energy harvesting.
\subsection{Organization of the manuscript}
The structure of the present paper is outlined as follows. Section \ref{01} is devoted to the mathematical formulation of the problem, including the governing equations of the model, kinematic and electric field relations, constitutive equations for both half-spaces, material gradation of the functionally graded piezoelectric material, and the governing equations of motion for the FGPM and thermo-piezoelectric media. Section \ref{02} is devoted to the solution of the problem based on appropriate wave-type assumptions. Section \ref{03} presents the formulation of the boundary conditions at the interface, where two distinct cases are considered. Section \ref{04} derives the amplitude ratios of the reflected and transmitted waves by enforcing the boundary conditions, resulting in a system of four coupled linear algebraic equations for each case. Section \ref{05} presents the expressions for the energy ratios of the scattered waves and verifies the conservation of energy through numerical results, with tables showing that the sum of the energy ratios is approximately one. Section \ref{06} discusses the numerical results and their graphical representation. Section \ref{07} concludes the paper by outlining the key findings of the investigation.

\section{Mathematical formulation of the problem}
\label{01}
In the present model, the Cartesian coordinate system is defined such that the $x$-axis aligns with the wave propagation direction, while the $z$-axis is directed downward. The model involves two half-spaces, with the functionally graded piezoelectric medium occupying the upper region and a transversely isotropic thermo-piezoelectric medium, accounting for rotation and initial stress, occupying lower region.

The displacement field is assumed to be of the form
\[
\vec{w} = (w_1, 0, w_3),
\]
and $\psi$, the electric potential is taken as a function of the spatial coordinates $x$, $z$, and time $t$.
In view of the above assumptions on the displacement field and electric potential, the governing equations of motion together with the relevant thermo-piezoelectric constitutive relations are established and discussed in the subsequent subsection.
\subsection{Governing equations for the model}
The governing dynamical equations of motion, incorporating the effects of rotation and initial stress, together with the electric field equation, are expressed as \cite{sahu2019scattering}
\begin{equation}
(\tau_{ij}+w_{j,k} \sigma_{ik}),_i +\rho_1f_i=\rho_{1} \ddot{\vec{w}} +\big( \vec{\Omega} \times  (\vec{\Omega} \times \vec{w}) \big) +\big( 2 \vec{\Omega} \times \dot{\vec{w}}\big),
\label{9}
\end{equation}
\begin{equation}
D_{i,i} = 0
\label{10}
\end{equation}
\begin{figure}[H]
    \centering
    \includegraphics[width=0.8\linewidth]{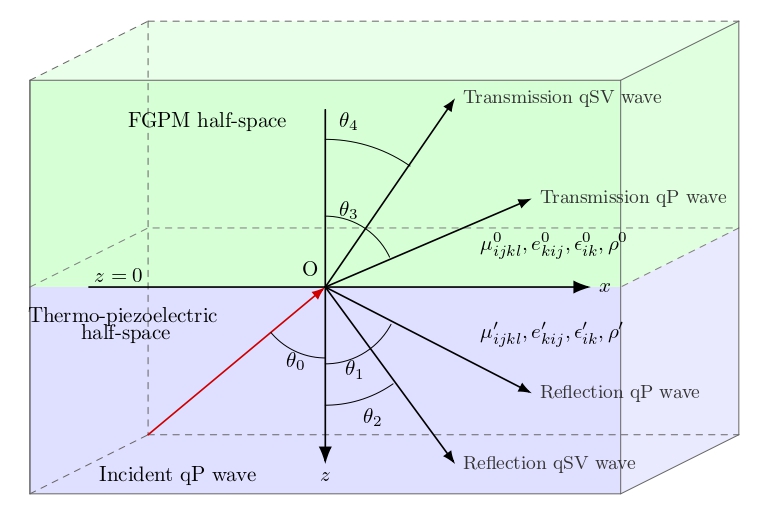}
    \caption{Geometry of the Problem}
    \label{diagram}
\end{figure}
The constitutive relations for the coupled thermo-piezoelectric medium are given by
\begin{equation}
\tau_{ij} = \mu_{ijkl} S_{kl} - e_{kij} E_{k} - \beta_{ij}T,
\label{11}
\end{equation}
\begin{equation}
D_{i} = e_{kij} \varepsilon_{kl} + \epsilon_{ik} E_{k}+p_iT.
\label{12}
\end{equation}
The heat conduction equation with fractional-order thermal relaxation is written as \cite{kumar2017effect}
\begin{equation}
K_{ij}T_{,ij}-(\frac{\partial}{\partial t}+\tau_0\frac{\partial^{\gamma+1}}{\partial t^{\gamma+1}})(\rho C_eT+\beta_{ij}w_{i,j}T_0-p_i\dot \phi_{,i}T_0)=0
\label{13}
\end{equation}
Here, $\tau_{ij}$, $\sigma_{ij}$, $\varepsilon_{ij}$, $D_i$, and $E_i$ denote the stress tensor, initial stress tensor, strain tensor, electric displacement differentiation vector, and electric field vector, respectively. The tensors $\mu_{ijkl}$, $\beta_{ij}$, $\epsilon_{ik}$, and $e_{kij}$ represent the elastic, thermal, dielectric, and piezoelectric coefficients. Furthermore, $\rho$ and $C_e$ denote the mass density and specific heat at constant strain, $T_0$ is the reference temperature, $\tau_0$ is the thermal relaxation time, and $\gamma$ $(0<\gamma \le 1)$ is the fractional-order parameter. The quantities $p_i$ represent the pyroelectric constants.
The overdot denotes differentiation with respect to time, whereas a comma indicates partial differentiation with respect to spatial coordinates. The terms $\vec{\Omega} \times (\vec{\Omega} \times \vec{u})$ and $2 \vec{\Omega} \times \dot{\vec{u}}$ correspond to the centripetal and Coriolis accelerations, respectively.
\subsection{Kinematic and electric field relations}

The strain-displacement relation is given by:
\begin{equation}
 \varepsilon_{ij}=\frac{1}{2} \left(\frac{\partial w_i}{\partial x_j}+\frac{\partial w_j}{\partial x_i} \right)
 \label{14}
 \end{equation}
 and, under the quasi-static approximation of Maxwell’s equations, the electric field intensity is related to the electric potential as
 \begin{equation}
E_k=-\frac{\partial\psi}{\partial x_k}
\label{15}
\end{equation}
\subsection{Constitutive relations for the two half-spaces}
For the transversely isotropic thermo-piezoelectric half-space, the constitutive relations in the $x-z$ plane take the form \cite{kumar2017effect}
\begin{align*}
\tau_{11} &= \mu_{11} w_{1,1} + \mu_{13} w_{3,3} + e_{31} \psi_3 - \beta_{11} T, \\
\tau_{13} &= \mu_{44} (w_{1,3} + w_{3,1}) + e_{15} \psi_1, \\
\tau_{33} &= \mu_{13} w_{1,1} + \mu_{33} w_{3,3} + e_{33} \psi_3 - \beta_{33} T, \\
D_1 &= e_{15} (w_{1,3} + w_{3,1}) - \epsilon_{11} \psi_{,1}, \\
D_3 &= e_{31} w_{1,1} + e_{33} w_{3,3} - \epsilon_{33} \psi_{,3} + p_3 T, \\
E_1 &= -\psi_{,1}, \\
E_3 &= -\psi_{,3}.
\end{align*}
For the functionally graded piezoelectric material, the constitutive relations are expressed as
\begin{align*}
\tau_{11} &= \mu_{11} w_{1,1} + \mu_{13} w_{3,3} + e_{31} \psi_3 , \\
\tau_{13} &= \mu_{44} (w_{1,3} + w_{3,1}) + e_{15} \psi_1, \\
\tau_{33} &= \mu_{13} w_{1,1} + \mu_{33} u_{3,3} + e_{33} \psi_3, \\
D_1 &= e_{15} (w_{1,3} + w_{3,1}) - \epsilon_{11} \psi_{,1}, \\
D_3 &= e_{31} w_{1,1} + e_{33} w_{3,3} - \epsilon_{33} \psi_{,3}, \\
E_1 &= -\psi_{,1}, \\
E_3 &= -\psi_{,3}.
\end{align*}
\subsection{Material gradation and wave assumptions}
To study the transmission of qP wave in the FGPM substrate, the wave is assumed to propagate along the positive $x$-direction. Accordingly, the displacement and electric potential fields are taken as
\begin{equation}
w_1 = w_1(x,z,t), \quad
w_2 = 0, \quad
w_3 = w_3(x,z,t), \quad
\psi = \psi(x,z,t).
\label{16}
\end{equation}
The material properties, including elastic, piezoelectric, dielectric, and density parameters, of FGPM half-space are modeled as exponentially varying functions of the $z$-coordinate.
\begin{equation}
\mu_{ij}(z) = \mu_{ij}^0 e^{\alpha z}, \quad
e_{ij}(z) = e_{ij}^0 e^{\alpha z}, \quad
\rho(z) = \rho^0 e^{\alpha z}, \quad
\epsilon(z) = \epsilon^0 e^{\alpha z},
\label{17}
\end{equation}
In the above expressions, $\alpha$ characterizes the material gradation, and the superscript $(0)$ indicates the material properties at the reference surface. Furthermore, $\mu_{ij}^0$, $e_{ij}^0$, $\rho^0$, and $\epsilon^0$ denote the elastic constants, piezoelectric coefficients, mass density, and dielectric permittivity of the upper half-space, respectively.
\begin{equation}
\mu_{11}^0\frac{\partial^2 w_1}{\partial x^2} + (\mu_{13}^0+\mu_{44}^0)\frac{\partial^2w_3}{\partial x\partial z} + (e_{31}^0+e_{15}^0)\frac{\partial^2\psi}{\partial x \partial z} + \mu_{44}^0\alpha\!\left(\frac{\partial w_1}{\partial z}+\frac{\partial w_3}{\partial x}\right) + \mu_{44}^0\frac{\partial^2w_1}{\partial z^2} + e_{15}^0\alpha\frac{\partial \psi}{\partial x} = \rho^0\!\left(\frac{\partial^2 w_1}{\partial t^2}-\Omega^2 w_1+2\Omega \frac{\partial w_3}{\partial t}\right)
\label{18}
\end{equation}

\begin{equation}
(\mu_{13}^0+\mu_{44}^0)\frac{\partial^2w_1}{\partial x \partial z} + \mu_{44}^0\frac{\partial^2w_3}{\partial x^2} + e_{15}^0 \frac{\partial^2\phi}{\partial x^2} + \mu_{13}^0\alpha \frac{\partial w_1}{\partial x} + \mu_{33}^0\frac{\partial^2w_3}{\partial z^2} + \mu_{33}^0\alpha \frac{\partial w_3}{\partial x} + e_{33}^0\frac{\partial^2\psi}{\partial z^2} + e_{33}^0 \alpha \frac{\partial \psi}{\partial z} = \rho^0\!\left(\frac{\partial^2 w_3}{\partial t^2}-\Omega^2w_3+2\Omega\frac{\partial w_1}{\partial t}\right)
\label{19}
\end{equation}

\begin{equation}
e_{15}^0\frac{\partial^2w_3}{\partial x^2} + (e_{15}^0+e_{13}^0)\frac{\partial^2w_1}{\partial x \partial z} + e_{33}^0\frac{\partial^2w_3}{\partial z^2} + e_{31}^0 \alpha \frac{\partial w_1}{\partial x} + e_{33}^0 \alpha \frac{\partial w_3}{\partial z} - \epsilon_{11}^0\frac{\partial^2\psi}{\partial x^2} - \epsilon_{33}^0\frac{\partial^2\psi}{\partial z^2} - \epsilon_{33}^0\alpha \frac{\partial \psi}{\partial t} = 0
\label{20}
\end{equation}

where $\mu_{ij}^0$ are the elastic constants; $w_1$ and $w_3$ denote the mechanical displacement components; and $\phi$ represents the electric potential. The quantities $\epsilon_{11}^0$ and $\epsilon_{33}^0$ are the dielectric constants, while $\rho^0$ and $\Omega$ denote the mass density and rotation parameter of the FGPM half-space, respectively.

For a transversely isotropic thermo-piezoelectric medium, neglecting body forces, the equations of motion take the following form:
\begin{equation}
(\mu_{11}^\prime + \sigma_{11}^\prime)\frac{\partial^2 w_1^\prime}{\partial x^2}
+ (\mu_{31}^\prime+\mu_{44}^\prime)\frac{\partial^2 w_3^\prime}{\partial x \partial z}
+ (\mu_{44}^\prime + \sigma_{33}^\prime)\frac{\partial^2 w_1^\prime}{\partial z^2}
+ (e_{31}^\prime + e_{15}^\prime)\frac{\partial^2 \psi^\prime}{\partial x \partial z}
- \beta_{11}\frac{\partial T}{\partial x}
= \rho^\prime\left(\frac{\partial^2 w_1^\prime}{\partial t^2} - {{\Omega}^\prime}^2 w_1^\prime + 2 {{\Omega}^\prime} \frac{\partial w_3^\prime}{\partial t} \right)
\label{21}
\end{equation}
\begin{equation}
(\mu_{44}^\prime+\sigma_{11}^\prime) \frac{\partial^2 w_1^\prime}{\partial x^2}
+ (\mu_{31}^\prime+\mu_{44}^\prime)\frac{\partial^2 w_3^\prime}{\partial x \partial z}
+ (\mu_{33}^\prime+\sigma_{33}^\prime)\frac{\partial^2 w_1^\prime}{\partial z^2}
+ e_{15}^\prime \frac{\partial^2 \psi^\prime}{\partial x^2}
+ e_{33}^\prime\frac{\partial^2 \psi^\prime}{\partial z^2}
- \beta_{33} \frac{\partial T}{\partial z}
= \rho^\prime\left(\frac{\partial^2w_3^\prime}{\partial t^2}-{{\Omega}^\prime}^2 w_3^\prime+2 {{\Omega}^\prime}\frac{\partial w_1^\prime}{\partial t}\right)
\label{22}
\end{equation}
\begin{equation}
e_{15}^\prime \frac{\partial^2u_3^\prime}{\partial x^2} + (e_{15}^\prime+e_{31}^\prime)\frac{\partial^2w_1^\prime}{\partial x \partial z} + e_{33}^\prime \frac{\partial^2 u_3^\prime}{\partial z^2} - \epsilon_{11}^\prime \frac{\partial^2 \psi^\prime}{\partial x^2} - \epsilon_{33}^\prime \frac{\partial ^2 \psi^\prime}{\partial z^2} + p_3 \frac{\partial T}{\partial z} = 0
\label{23}
\end{equation}
\begin{equation}
K_{11}\frac{\partial^2 T}{\partial x^2} + K_{33}\frac{\partial^2T}{\partial z^2} - \left(\frac{\partial}{\partial t}-\tau_0\frac{\partial^{\gamma+1}}{\partial t^{\gamma+1}}\right) \left(\rho C_eT + T_0\beta_{11} \frac{\partial w_1}{\partial x} + T_0\beta_{33}\frac{\partial w_3}{\partial z} - T_0p_3\frac{\partial \psi}{\partial z} \right) = 0
\label{24}
\end{equation}
where $\mu_{ij}^\prime$ are the elastic constants, $\rho^\prime$ denotes the mass density, $\Omega$ is the rotation parameter, $\epsilon_{11}^\prime$ and $\epsilon_{33}^\prime$ are the dielectric constants, $\sigma_{11}^1$ and $\sigma_{33}^1$ represent the initial stresses, and $w_1^\prime$, $w_3^\prime$, and $\psi^\prime$ denote the mechanical displacement components and the electric potential in the thermo-piezoelectric half-space, respectively.\\
To account for memory-dependent and nonlocal thermal effects in the present system, fractional calculus is employed to model the heat conduction behavior. In the next subsection, we discuss the fractional-order heat conduction model in detail and recall the essential definitions and properties of fractional differential operators required for the subsequent analysis \cite{balachandran2023introduction}.
\subsection{Fractional-order heat conduction model}
\textbf{Riemann-Liouville fractional integrals and derivatives
}\\
Consider a bounded interval $J=[a,b] \subset  \mathbb{R}$.
The corresponding Riemann–Liouville fractional integral operators of order $\gamma \in \mathbb{C}$ with $\Re(\gamma) > 0$ are defined as

\begin{equation}
I_{a+}^\gamma f(x) = \frac{1}{\Gamma(\gamma)} \int_{a}^{x} (x-t)^{\gamma-1} f(t) \, dt, \quad x > a, \quad \Re(\gamma) > 0
\label{1}
\end{equation}

\begin{equation}
I_{b-}^\gamma f(x) = \frac{1}{\Gamma(\gamma)} \int_{x}^{b} (t-x)^{\gamma-1} f(t) \, dt, \quad x < b, \quad \Re(\gamma) > 0
\label{2}
\end{equation}

These are the left-sided and right-sided fractional integrals, respectively, where the gamma function $\Gamma(z)$ is defined by
\begin{equation}
\Gamma(z) = \int_{0}^{\infty} t^{z-1} e^{-t} \, dt, \quad \Re(z) > 0.
\end{equation}
The Riemann--Liouville fractional derivatives are defined as

\begin{equation}
\begin{aligned}
D_{a+}^{\gamma} f(x) &= D^n I_{a+}^{\,n-\gamma} f(x) \\
&= \frac{1}{\Gamma(n-\gamma)} \frac{d^n}{dx^n} \int_a^x \frac{f(t)}{(x-t)^{\gamma-n+1}} \, dt, \quad n = [\Re(\gamma)] + 1, \quad x > a,
\label{3}
\end{aligned}
\end{equation}

\begin{equation}
\begin{aligned}
D_{b-}^{\gamma} f(x) &= \left(-\frac{d}{dx}\right)^n I_{b-}^{\,n-\gamma} f(x) \\
&= \frac{1}{\Gamma(n-\gamma)} \left(-\frac{d}{dx}\right)^n \int_x^b \frac{f(t)}{(t-x)^{\gamma-n+1}} \, dt, \quad n = [\Re(\gamma)] + 1, \quad x < b,
\label{4}
\end{aligned}
\end{equation}

where $[\Re(\gamma)]$ denotes the integer part of the real part of $\gamma$, $D = \frac{d}{dx}$, and when $\gamma = n \in \mathbb{N}_0 = \{0, 1, 2, \dots\}$, these definitions reduce to the classical integer-order derivatives.\\
\textbf{Caputo fractional derivatives} \\
The left-sided and right-sided Caputo fractional derivatives of order $\gamma>0$, with $1<\gamma \le n$, are defined as

\begin{equation}
\begin{aligned}
{}^C D_{a+}^{\gamma} f(x) &= I_{a+}^{\,n-\gamma} D^n f(x) \\
&= \frac{1}{\Gamma(n-\gamma)} \int_a^x (x-t)^{n-\gamma-1} f^{(n)}(t) \, dt,
\label{5}
\end{aligned}
\end{equation}

\begin{equation}
\begin{aligned}
{}^C D_{b-}^{\gamma} f(x) &= (-1)^n I_{b-}^{\,n-\gamma} D^n f(x) \\
&= \frac{(-1)^n}{\Gamma(n-\gamma)} \int_x^b (t-x)^{\,n-\gamma-1} f^{(n)}(t) \, dt,
\label{6}
\end{aligned}
\end{equation}

where $f(t)$ is absolutely continuous with derivatives up to order $(n-1)$ on $[a,b]$.  

In particular, when $0 < \gamma < 1$, these reduce to

\begin{equation}
{}^C D_{a+}^{\gamma} f(x) = \frac{1}{\Gamma(1-\gamma)} \int_a^x (x-t)^{-\gamma} f'(t) \, dt,
\label{7}
\end{equation}

\begin{equation}
{}^C D_{b-}^{\gamma} f(x) = - \frac{1}{\Gamma(1-\gamma)} \int_x^b (t-x)^{-\gamma} f'(t) \, dt.
\label{8}
\end{equation}

Having established the governing equations for the FGPM and thermo-piezoelectric half-spaces, along with the fractional-order heat conduction model incorporating memory-dependent thermal effects, the wave propagation characteristics of the system can now be determined. \textcolor{black}{In the following section, the analytical solution of the present problem is assumed for the displacement, electric potential and temperature fields .}
\section{Formulation of the solution}
\label{02}
The solution of equations \eqref{18}-\eqref{20} and \eqref{21}-\eqref{24} is expressed in the form \cite{abd2014mathematical}:
\begin{equation}
    (\vec w^{(n)},\psi^{(n)},T^{(n)}) = (A_n, B_n, C_n)\, \vec d_n \,  e^{i \xi_n (\vec x \cdot \vec p - c_n t)},
    \label{25}
\end{equation}
where $n$ indicates the type of wave, either reflected or transmitted. Here, $\vec d_n$ is the unit vector specifying the displacement direction, and $\vec p$ is normalized propagation vector. The position vector $\vec x$ satisfies $\vec x \cdot \vec p = \text{constant}$.  

The coefficients $A_n$, $B_n$, and $C_n$ correspond to the amplitudes of the quasi-longitudinal, electric potential ($\psi$), and thermal ($T$) modes, respectively. The constant $c_n$ represents the phase velocity of the $n$-th wave, while $\xi_n$ denotes its associated wave number.
As the wave propagation is confined to a two-dimensional plane, the vectors can be expressed as
\[
\vec{w}^{(n)} = (w_1^{(n)}, w_3^{(n)}), \quad
\vec{d}^{(n)} = (d_1^{(n)}, d_3^{(n)}), \quad
\vec{p}^{(n)} = (p_1^{(n)}, p_3^{(n)}),
\]
with the normalization condition
\[
(p_1^{(n)})^2 + (p_3^{(n)})^2 = 1.
\]
Accordingly, the displacement components, electric potential, and temperature field implied by Equation \eqref{25} can be written as
\begin{equation}
\begin{bmatrix}
u^{(n)}_{1} \\
u^{(n)}_{3} \\
\psi^{(n)} \\
T^{(n)}
\end{bmatrix}
=
\begin{bmatrix}
A_{n} d^{(n)}_{1} \\A_{n} d^{(n)}_{3} \\
B_{n} \\
C_{n}
\end{bmatrix}
 e^{i \xi_{n} (x_1 p^{(n)}_1 + x_3 p^{(n)}_3 - c_n t) }.
\label{26}
\end{equation}
where $n$ denotes the type of wave (incident, reflected, or transmitted), $A_n$, $B_n$, and $C_n$ are the vibrational amplitudes, $\vec{d}^{(n)}$ is the unit displacement direction, and $\vec{p}^{(n)}$ is the normalized propagation vector.
\subsection{Incident wave (n=0)}
The incident qP wave with displacement components $(w_1^{(0)}, w_3^{(0)})$, electric potential $\psi^{(0)}$, and thermal field $T^{(0)}$ can be expressed as
\begin{equation}
\begin{aligned}
(w_1^{(0)}, w_3^{(0)}, \psi^{(0)}, T^{(0)}) 
= (A_0 \sin\theta_0,\, A_0 \cos\theta_0,\, B_0,\, C_0) 
 e^{ i \xi_0 (x \sin\theta_0 + z \cos\theta_0 - c_{I_0} t) },
\label{27}
\end{aligned}
\end{equation}
Here $c_{I_0} = \omega / \xi_0$ denotes the phase velocity of incident wave, $\omega$ denotes the angular frequency, $\vec{d}^{(0)}$ is unit displacement vector, and $\vec{p}^{(0)}$ is normalized propagation vector.

The interface boundary conditions are satisfied by considering the reflected and transmitted waves.
\subsection{Reflected wave (n=1)}
\begin{equation}
    (w_1^{(1)},w_3^{(1)},\psi^{(1)},T^{(1)}) = 
    (A_1 \sin\theta_1, -A_1 \cos\theta_1, B_1, C_1)
     e^{ i \xi_1 (x \sin\theta_1 - z \cos\theta_1 - c_{I_1} t) }
    \label{28}
\end{equation}

The phase velocity corresponding to the reflected qP-wave can be written as
\begin{equation}
    c_{I_1} = \frac{\omega}{\xi_1} = 
    \frac{1}{\sqrt{2 \rho^\prime}} 
    \sqrt{\mu_{44}^0 + \mu_{11}^0 \sin^2 \beta_1 + \mu_{33}^0 \cos^2 \beta_1 + c_1},
\end{equation}
with
\begin{equation}
    c_1 = \sqrt{\big[(\mu_{11}^0-\mu_{44}^0)\sin^2\beta_1 + (\mu_{44}^0-\mu_{33}^0)\cos^2\beta_1\big]^2 + (\mu_{44}^0+\mu_{33}^0)^2 \sin^2 2\beta_1},
\end{equation}
where $\beta_1$ represents the angle formed between the symmetry axis and the propagation direction, while the displacement direction vector takes the form $d^1 = p^1 = (\sin\theta_1, -\cos\theta_1)$.
\subsection{Reflected wave (n=2)}
\begin{equation}
    (w_1^{(2)}, w_3^{(2)}, \psi^{(2)}, T^{(2)}) =
    (A_2 \cos\theta_2, A_2 \sin\theta_2, B_2, C_2)
    e^{i \xi_2 (x \sin\theta_2 - z \cos\theta_2 - c_{T_2} t)}
    \label{29}
\end{equation}

The phase velocity corresponding to the reflected quasi-transverse wave is
\begin{equation}
    c_{T_2} = \frac{\omega}{\xi_2} = 
    \frac{1}{2 \rho^\prime} 
    \sqrt{\mu_{44}^0 + \mu_{11}^0 \sin^2 \beta_2 + \mu_{33}^0 \cos^2 \beta_2 - c_1},
\end{equation}
where $c_1$ is defined as in the previous section, and the displacement and propagation vectors satisfy $d^2 \neq p^2$. These waves correspond to the quasi-transverse mode.
\subsection{Refracted wave (n=3)}
\begin{equation}
    (w_1^{(3)}, w_3^{(3)}, \psi^{(3)}, T^{(3)}) =
    (A_3 \sin\theta_3, A_3 \cos\theta_3, B_3, 0)
    e^{i \xi_3 (x \sin\theta_3 + z \cos\theta_3 - c_{I_3} t)}
    \label{30}
\end{equation}

The phase velocity of the refracted quasi-primary (qP) wave is
\begin{equation}
    c_{I_3} = \frac{\omega}{\xi_3} =
    \frac{1}{\sqrt{2 \rho^0}}
    \sqrt{\mu_{44}^\prime + \mu_{11}^\prime \sin^2 \beta_2 + \mu_{33}^\prime \cos^2 \beta_2 + c_2},
\end{equation}
with the displacement and propagation vectors given by $d^3 = p^3 = (\sin\theta_3, \cos\theta_3)$, and $\beta_2$ denoting the angle between the symmetry axis and direction of propagation of wave.
\subsection{Refracted wave (n=4)}
\begin{equation}
    (w_1^{(4)}, w_3^{(4)}, \psi^{(4)}, T^{(4)}) =
    (-A_4 \cos\theta_4, A_4 \sin\theta_4, B_4, 0)
   e^{i \xi_4 (x \sin\theta_4 + z \cos\theta_4 - c_{T_4} t)}
    \label{31}
\end{equation}

The phase velocity of the refracted quasi-shear vertical (qSV) wave is
\begin{equation}
    c_{T_4} = \frac{\omega}{\xi_4} =
    \frac{1}{\sqrt{2 \rho^0}}
    \sqrt{\mu_{44}^\prime + \mu_{11}^\prime \sin^2 \beta_2 + \mu_{33}^\prime \cos^2 \beta_2 - c_2},
\end{equation}
where
\begin{equation}
    c_2 = \sqrt{\big[(\mu_{11}^\prime - \mu_{44}^\prime) \sin^2 \beta_2 + (\mu_{44}^\prime - \mu_{33}^\prime) \cos^2 \beta_2\big]^2 + (\mu_{44}^\prime + \mu_{33}^\prime)^2 \sin^2 2\beta_2}.
\end{equation}

The displacement and propagation vectors satisfy $d^4 \neq p^4$.  
As thermal effects are considered only in the thermo-piezoelectric material, $T^{(3)} = 0$ and $T^{(4)} = 0$.
\section{Boundary conditions}
\label{03}
The fields must satisfy certain continuity conditions between two different half-spaces ( the FGPM and thermo-piezoelectric half-spaces) at the interface, to ensure the physical behavior is realistic. These boundary conditions govern how waves reflect and transmit at the interface.
\begin{enumerate}
\item Displacement continuity 
\begin{align}
    w_3^{(0)}+w_3^{(1)}+w_3^{(2)}=w_3^{(3)}+w_3^{(4)}
    \label{32}
\end{align}
This condition ensures that the normal displacement along the interface is continuous. Physically, it means the two materials remain in contact without separation or interpenetration, and the interface moves smoothly under the influence of incident, reflected, and transmitted waves.
 \item Stress continuity 

      \begin{align}
      \tau_{33}^{(0)} + \tau_{33}^{(1)} + \tau_{33}^{(2)} &= \tau_{33}^{(3)} + \tau_{33}^{(4)}
      \label{33}
  \end{align}
The normal stress acting perpendicular to the interface must be equal on both sides. This guarantees that the total force exerted by all waves on the interface is balanced, preventing unphysical jumps in stress and maintaining mechanical equilibrium at the boundary.
   \begin{align}
      \tau_{31}^{(0)} + \tau_{31}^{(1)} + \tau_{31}^{(2)} &= \tau_{31}^{(3)} + \tau_{31}^{(4)}
      \label{34}
      \end{align}
Similarly, the shear stress along the interface must be continuous. This condition ensures that tangential forces are transmitted consistently, preserving the rotational equilibrium of the interface and enabling proper propagation of shear waves.
  \item Electric potential continuity
  \begin{equation}
      \psi^{(0)} + \psi^{(1)} + \psi^{(2)} = \psi^{(3)} + \psi^{(4)}
      \label{35}
  \end{equation}
In piezoelectric and thermo-piezoelectric materials, the electric potential must not experience sudden jumps at the interface. This condition allows the proper transfer of electromechanical coupling effects, ensuring the electric field and mechanical response remain consistent across the boundary.
  \item Thermal insulated boundary
  \begin{equation}
      T^{(0)} +T^{(1)}+T^{(2)}=T^{(3)}+T^{(4)}
      \label{36}
  \end{equation}
This condition imposes thermal insulation at the interface, meaning that the net heat flux across the boundary is zero. It ensures that the incident and reflected thermal fields in the FGPM medium match the transmitted thermal fields in the thermo-piezoelectric medium, accurately modeling the thermal interaction between the two materials.
 \end{enumerate}
 Now, we consider two distinct cases for applying the boundary conditions:\\
 \textbf{Case 1:} In the first case, we impose the normal stress and shear stress continuity, along with the continuity of electric potential and thermal insulated boundary. That is, the boundary conditions used are Equations \eqref{33}, \eqref{34}, \eqref{35}, and \eqref{36}. These conditions ensure that both mechanical stresses and coupled electro-thermal effects are properly matched across the interface.\label{case1}\\
 \textbf{Case 2:} In the second case, we consider the continuity of displacement instead of normal stress, along with shear stress continuity, continuity of electric potential, and thermal insulated boundary. Accordingly, the boundary conditions are Equations \eqref{32}, \eqref{34}, \eqref{35}, and \eqref{36}. This scenario is useful for analyzing cases where displacement continuity is enforced rather than stress continuity, affecting how waves are transmitted and reflected at the interface.\label{case2}
 \section{Amplitude ratios corresponding to reflected and transmitted waves}
 \label{04}
Imposing the continuity of normal displacement at the interface, i.e., Equation \ref{32}, results in the following equation:
 \begin{align}
     A_0\cos\theta_0 e^{\varpi_0} - A_1\cos\theta_1  e^{\varpi_1} + A_2\sin\theta_2  e^{\varpi_2} 
   = A_3\cos\theta_3  e^{\varpi_3} + A_4\sin\theta_4  e^{\varpi_4} 
   \label{37}
 \end{align}
 Using the normal stress boundary condition, Equation \ref{33}, leads to the following expression:
\begin{equation}
\begin{aligned}
& \xi_0\Big[A_0(\mu_{31}^\prime \sin^2\theta_0 + \mu_{33}^\prime \cos^2\theta_0)
+ e_{33}^\prime B_0 \cos\theta_0 - \beta_{33} C_0\Big]  e^{\varpi_0} \\
& + \xi_1\Big[A_1(\mu_{31}^\prime \sin^2\theta_1 + \mu_{33}^\prime \cos^2\theta_1)
+ e_{33}^\prime B_1 \cos\theta_1 - \beta_{33} C_1\Big]  e^{\varpi_1} \\
& + \xi_2\Big[A_2(\mu_{31}^\prime - \mu_{33}^\prime)\sin\theta_2\cos\theta_2
- e_{33}^\prime B_2 \cos\theta_2 - \beta_{33} C_2\Big]  e^{\varpi_2} \\
& = \xi_3\Big[A_3(\mu_{31}^0 \sin^2\theta_3 + \mu_{33}^0 \cos^2\theta_3)
+ e_{33}^0 B_3 \cos\theta_3\Big]  e^{\varpi_3} \\
& \quad + \xi_4\Big[A_4(\mu_{31}^0 - \mu_{33}^0)\sin\theta_4\cos\theta_4
+ e_{33}^0 B_4 \cos\theta_4\Big]  e^{\varpi_4}
\end{aligned}
\label{38}
\end{equation}

Enforcing the continuity of shear stress across the interface, as given by Equation \ref{34}, yields the following expression:
\begin{equation}
\begin{aligned}
& \xi_0\Big[ A_0 \mu_{44}^\prime \sin 2\theta_0
+ B_0 e_{15}^\prime \sin\theta_0 \Big]  e^{\varpi_0} \\
& - \xi_1\Big[ A_1 \mu_{44}^\prime \sin 2\theta_1
- B_1 e_{15}^\prime \sin\theta_1 \Big]  e^{\varpi_1} \\
& - \xi_2\Big[ A_2 \mu_{44}^\prime \sin 2\theta_2
+ B_2 e_{15}^\prime \sin\theta_2 \Big]  e^{\varpi_2} \\
& = \xi_3\Big[ A_3 \mu_{44}^0 \sin 2\theta_3
+ B_3 e_{15}^0 \sin\theta_3 \Big] \ e^{\varpi_3} \\
& \quad + \xi_4\Big[ A_4 \mu_{44}^0 \sin 2\theta_4
+ B_4 e_{15}^0 \sin\theta_4 \Big]  e^{\varpi_4}
\end{aligned}
\label{39}
\end{equation}

Using the continuity condition for the electric potential at the interface (Equation \ref{35}), the following relation is derived: 
\begin{align}
    B_0 e^{\varpi_0}+B_1 e^{\varpi_1}+B_2 e^{\varpi_2}=B_3 e^{\varpi_3}+B_4 e^{\varpi_4}
    \label{40}
\end{align}
Applying the thermally insulated boundary condition at the interface, as expressed by Equation \ref{36}, leads to the following expression:
\begin{align}
    C_0 e^{\varpi_0}+C_1 e^{\varpi_1}+C_2 e^{\varpi_2}=0
    \label{41}
\end{align}
The boundary conditions are satisfied identically iff the phase factors of all incident, reflected, and refracted waves at the interface are equal, i.e.,
\begin{equation}
\varpi_0 = \varpi_1 = \varpi_2 = \varpi_3 = \varpi_4.
\end{equation}
Here,
\begin{align}
\varpi_0 &= i \xi_0 \big( x \sin\theta_0 - c_{I_0} t \big), \quad
\varpi_1 = i \xi_1 \big( x \sin\theta_1 - c_{I_1} t \big), \quad
\varpi_2 = i \xi_2 \big( x \sin\theta_2 - c_{T_2} t \big), \nonumber\\
\varpi_3 &= i \xi_3 \big( x \sin\theta_3 - c_{I_3} t \big), \quad
\varpi_4 = i \xi_4 \big( x \sin\theta_4 - c_{T_4} t \big).
\end{align}
Since these relations are required to be valid for all spatial coordinates $x$ and time $t$, the horizontal components of the wave vectors must be identical. Consequently, the following relation is obtained:
\begin{equation}
\xi_0 \sin\theta_0
= \xi_1 \sin\theta_1
= \xi_2 \sin\theta_2,\:  \xi_0c_{I_0}=\xi_1c_{I_1}=\xi_2c_{T_2}=\omega
\label{42}
\end{equation}
From the above relations i.e. \ref{42} it follows that
\begin{align}
    \xi_0=\xi_1, \theta_0=\theta_1,\: c_{I_0}=c_{I_1},\: \zeta_1=\frac{\xi_2}{\xi_0}, \zeta_2=\frac{\xi_3}{\xi_0}, \:\zeta_3=\frac{\xi_4}{\xi_0}, \\ \sin\theta_2=\frac{1}{\zeta_1}\sin\theta_0,\: \sin\theta_3=\frac{1}{\zeta_2}\sin\theta_0,\:\sin\theta_4=\frac{1}{\zeta_3}\sin\theta_0
    \label{44}
\end{align}
By substituting equations \ref{27} to \ref{31} into equations \ref{18},\ref{19}, \ref{21},\ref{22}, the expressions for the incident, reflected, and refracted qP-waves are thus obtained.
\begin{equation}
\left.
\begin{array}{l}
\chi_0 A_0 + M_0 B_0 + \nu_0 C_0 = 0,\\[4pt]
\chi_1 A_1 + M_1 B_1 + \nu_1 C_1 = 0,\\[4pt]
\chi_2 A_2 + M_2 B_2 + \nu_2 C_2 = 0,\\[4pt]
\chi_3 A_3 + M_3 B_3 + \nu_3 C_3 = 0,\\[4pt]
\chi_4 A_4 + M_4 B_4 + \nu_4 C_4 = 0
\end{array}
\right\}
\label{45}
\end{equation}
The parameters $\chi_i$, $M_i$, and $\nu_i$ are given in \ref{appendixa}. Substituting equations~(\ref{27})–(\ref{31}) into equation~(\ref{23}) yields:
\begin{equation}
\left.
\begin{array}{l}
L_0 A_0 + G_0 B_0 + S_0 C_0 = 0,\\[4pt]
L_1 A_1 + G_1 B_1 + S_1 C_1 = 0,\\[4pt]
L_2 A_2 + G_2 B_2 + S_2 C_2 = 0,\\[4pt]
L_3 A_3 + G_3 B_3 + S_3  C_3 = 0,\\[4pt]
L_4 A_4 + G_4 B_4 + S_4  C_4 = 0
\end{array}
\right\}
\label{46}
\end{equation}
The quantities $L_i$, $G_i$, and $S_i$ are listed in \ref{B}. By substituting equations~(\ref{27})–(\ref{31}) into equation~(\ref{24}), the following expression is derived:
\begin{equation}
\left.
\begin{array}{l}
E_0 A_0 + D_0 B_0 + F_0  C_0 = 0,\\[4pt]
E_1 A_1 + D_1 B_1 + F_1  C_1 = 0,\\[4pt]
E_2 A_2 + D_2 B_2 + F_2  C_2 = 0,
\end{array}
\right\}
\label{47}
\end{equation}
The values of $E_i$, $D_i$, and $F_i$ are provided in \ref{C}.\\
\textbf{Case 1.} For Case 1, using equations~(\ref{38})–(\ref{41}), we obtain
\begin{equation}
\left.
\begin{array}{l}
  (a_{11}A_1+a_{12}A_2+a_{13}A_3+a_{14}A_4)/A_0 = m_1\\
  (a_{21}A_1+a_{22}A_2+a_{23}A_3+a_{24}A_4)/A_0 = m_2\\
  (a_{31}A_1+a_{32}A_2+a_{33}A_3+a_{34}A_4)/A_0 = m_3\\
  (a_{41}A_1+a_{42}A_2+a_{43}A_3+a_{44}A_4)/A_0 = m_4\\
  \end{array}
\right\}
\label{48}
\end{equation}
The coefficients $a_{ij}$ are listed in \ref{D}. From equation~(\ref{49}), the reflection and refraction coefficients are obtained as
    \begin{equation}
    \begin{aligned}
        \frac{A_1}{A_0}=\frac{X_1}{X},\frac{A_2}{A_0}=\frac{X_2}{X},\frac{A_3}{A_0}=\frac{X_3}{X},\frac{A_4}{A_0}=\frac{X_4}{X}
        \label{49}
    \end{aligned}
\end{equation}
The quantities $X_1$, $X_2$, $X_3$, $X_4$, and $X$ are given in \ref{E}.
From equations~(\ref{36})–(\ref{38}), the following expression is obtained:
\begin{align}
\frac{B_1}{B_0} = \frac{A_1}{A_0}, 
\frac{B_2}{B_0} = \frac{N_2A_2}{N_0A_0}, 
\frac{B_3}{B_0} = 0, 
\frac{B_4}{B_0} = 0, \notag\\[6pt]
\frac{ C_1}{ C_0} = \frac{A_1}{A_0}, 
\frac{ C_2}{ C_0} = \frac{R_2A_2}{R_0A_0}, 
\frac{ C_3}{ C_0} = 0, 
\frac{ C_4}{ C_0} = 0.
\label{50}
\end{align}\\
\textbf{Case 2.} From equations~(\ref{37})–(\ref{41}) for Case 2, the following expressions are obtained:

\begin{equation}
\left.
\begin{array}{l}
  (b_{11}A_1+b_{12}A_2+b_{13}A_3+b_{14}A_4)/A_0 = s_1\\
  (b_{21}A_1+b_{22}A_2+b_{23}A_3+b_{24}A_4)/A_0 = s_2\\
  (b_{31}A_1+b_{32}A_2+b_{33}A_3+b_{34}A_4)/A_0 = s_3\\
  (b_{41}A_1+b_{42}A_2+b_{43}A_3+b_{44}A_4)/A_0 = s_4\\
  \end{array}
\right\}
\label{51}
\end{equation}
The values of $b_{ij}$ and $s_i$ are provided in\ref{D}.
Equation~(\ref{49}) yields the reflection and refraction coefficients as:
    \begin{equation}
    \begin{aligned}
        \frac{A_1}{A_0}=\frac{Y_1}{Y},\frac{A_2}{A_0}=\frac{Y_2}{Y},\frac{A_3}{A_0}=\frac{Y_3}{Y},\frac{A_4}{A_0}=\frac{Y_4}{Y}
        \label{52}
    \end{aligned}
\end{equation}
The quantities $Y_1$, $Y_2$, $Y_3$, $Y_4$, and $Y$ are given in \ref{E}. From equations~(\ref{36})–(\ref{38}), the following expressions are obtained:
\begin{align}
\frac{B_1}{B_0} = \frac{A_1}{A_0}, 
\frac{B_2}{B_0} = \frac{N_2A_2}{N_0A_0}, 
\frac{B_3}{B_0} = 0, 
\frac{B_4}{B_0} = 0, \notag\\[6pt]
\frac{C_1}{C_0} = \frac{A_1}{A_0}, 
\frac{C_2}{C_0} = \frac{R_2A_2}{R_0A_0}, 
\frac{C_3}{C_0} = 0, 
\frac{C_4}{C_0} = 0.
\label{53}
\end{align}\\
In this section, the amplitude ratios corresponding to the reflected and transmitted waves have been obtained by applying the appropriate boundary conditions. 
These amplitude ratios form the basis for analyzing the distribution of wave energy across the interface. 
Accordingly, the following section is devoted to the study of energy transmission and partition among the reflected and transmitted modes.

\section{Energy transmission and distribution}
\label{05}
The time-averaged energy flux associated with a propagating wave is defined as
\begin{equation}
    W=\frac{\omega}{2\pi} \int_{-\frac{\pi}{\omega}}^{\frac{\pi}{\omega}} P(t)\, dt,
\end{equation}
where $P(t)$ denotes the instantaneous energy flux density, which varies with time, and $\omega$ represents the angular frequency. 
For the thermo-piezoelectric medium under consideration, the corresponding energy flux density is evaluated following the formulation presented in \cite{kumar2021response}.

\begin{equation}
    P(t)=-\tau_{31}\dot{w_1}-\tau_{33}\dot{w_3}+\dot{D_3}\psi-K_{33}T_{,3}\frac{T}{T_0}
\end{equation}
The energy flux corresponding to the incident quasi-longitudinal wave is expressed as
\begin{equation}
    P_0 = \xi_0 \, \omega \, A_0^2 \, \phi_0 \, 
    e^{i \xi_0 \big(x \sin\theta_0 + z \cos\theta_0 - c_{I_0} t \big)},
\end{equation}
Here, $A_0$ is the amplitude of the incident wave, $\omega = \xi_0 c_{I_0}$, and the factor $\phi_0$ is given by
\begin{align}
\phi_0 ={}& 2 \mu_{44}^\prime \sin^2\theta_0 \cos\theta_0
+ e_{15}^\prime N_0 \sin^2\theta_0
+ \mu_{13}^\prime \sin^2\theta_0 \cos\theta_0
+ \mu_{33}^\prime \cos^3\theta_0 \nonumber\\
&+ e_{33}^\prime N_0 \cos^2\theta_0
+ e_{31}^\prime \sin^2\theta_0
+ e_{33}^\prime \cos^2\theta_0
+ \frac{K_{33}}{T_0} R_0^2 \cos\theta_0.
\end{align}
Energy partition refers to the ratio between the energy transported by each scattered wave and the energy of the incident wave.
\begin{align*}
\phi_1 ={}& 2\mu_{44}^\prime \sin^2\theta_1 \cos\theta_1
- e_{15}^\prime N_1 \sin^2\theta_1
- \mu_{13}^\prime \sin^2\theta_1 \cos\theta_1
+ \mu_{33}^\prime \cos^2\theta_1 \sin^2\theta_1 \\
&\quad
+ e_{33}^\prime N_1 \cos^2\theta_1
+ e_{31}^\prime \sin^3\theta_1
+ e_{33}^\prime \cos^3\theta_1
- \frac{K_{33}}{T_0} R_1^2 \cos\theta_1
\end{align*}
\begin{align*}
\phi_2 ={}& \mu_{44}^\prime \bigl(\cos\theta_2 \sin^2\theta_2 - \cos^3\theta_2 \bigr)
+ e_{15}^\prime N_2 \sin\theta_2 \cos\theta_2
+ \mu_{13}^\prime \sin^2\theta_2 \cos\theta_2
- \mu_{33}^\prime \cos\theta_2 \sin^2\theta_2 \\
&\quad
- e_{33}^\prime N_2 \cos\theta_2 \sin\theta_2
+ e_{33}^\prime \cos\theta_2 \sin^2\theta_2
- \frac{K_{33}}{T_0} R_2^2 \cos\theta_2
\end{align*}
\begin{align*}
\phi_3 ={}& -(\mu_{44}^0 + \alpha \mu_{44}^0)\sin^2\theta_3 \cos\theta_3
+ e_{15}^0 \alpha N_3 \sin^2\theta_3
+ \mu_{13}^0 \sin^2\theta_3 \cos\theta_3
+ \mu_{33}^0 \cos^2\theta_3 \\
&\quad
+ e_{33}^0 \alpha N_3 \cos^2\theta_3
+ e_{31}^0 \sin^2\theta_3
+ e_{33}^0 \alpha \cos^2\theta_3
\end{align*}
\begin{align*}
\phi_4 ={}& \mu_{44}^0 \bigl(\cos\theta_4 \sin^2\theta_4 - \cos^3\theta_4 \bigr)
+ e_{15}^0 \alpha N_4 \sin\theta_4 \cos\theta_4
+ \mu_{13}^0 \sin^2\theta_4 \cos\theta_4
+ \mu_{33}^0 \cos\theta_4 \sin^2\theta_4 \\
&\quad
+ e_{33}^0 N_4 \cos\theta_4 \sin\theta_4
+ e_{33}^0 \cos\theta_4 \sin^2\theta_4
- e_{31}^0 \alpha \cos^2\theta_4 \sin\theta_4
\end{align*}
The magnitudes of the energy partition coefficients for the various reflected and transmitted waves are expressed as follows:
\begin{equation}
    \left|\frac{P_1}{P_0}\right|=\left|\frac{\phi_1}{\phi_0}\right||R_{qP}|^2, \left|\frac{P_2}{P_0}\right|=\left|\frac{\phi_2}{\phi_0}\right||R_{qSV}|^2,\left|\frac{P_3}{P_0}\right|=\left|\frac{\phi_3}{\phi_0}\right||T_{qP}|^2,\left|\frac{P_4}{P_0}\right|=\left|\frac{P_4}{P_0}\right|=\left|\frac{\phi_4}{\phi_0}\right||T_{qSV}|^2
\end{equation}
Here, $R_{qP}$ and $R_{qSV}$ denote the reflection coefficients of the quasi-longitudinal (qP) and quasi-shear vertical (qSV) waves at the lower interface, while $T_{qP}$ and $T_{qSV}$ represent the transmission coefficients of the qP and qSV waves at the upper interface.\\
Energy conservation is satisfied when
\begin{equation}
    \sum_{n=1}^{4} \frac{P_n}{P_0} = 1,
\end{equation}
which can serve as a verification check for the numerical computations.Consequently, the validity of the energy conservation principle is established by numerical computation of the energy ratios for different incident angles.
\begin{table}[!htbp]
\centering
\caption{Energy ratio variation with incident angle for case 1}
\label{tab1}
\begin{tabular*}{\textwidth}{@{\extracolsep{\fill}} c c c c c c}
\hline
Incident angle 
& $\dfrac{P_1}{P_0}$ 
& $\dfrac{P_2}{P_0}$ 
& $\dfrac{P_3}{P_0}$ 
& $\dfrac{P_4}{P_0}$ 
& $\displaystyle\sum_{i=1}^{4}\dfrac{P_i}{P_0}$ \\
\hline
0.11 & 0.99 & 0.00 & $-0.01$  & $-0.10$ & 0.88$\approx 1$ \\
1.19 & 0.64 & 0.00 & $-0.04$  & 0.64    & 1.24 $\approx 1$\\
1.50 & 0.94 & 0.00 & $-0.009$ & 0.14    & 1.07$\approx 1$ \\
2.95 & 0.80 & 0.00 & 0.01     & 0.17    & 0.98 $\approx 1$\\
\hline
\end{tabular*}

\end{table}
\begin{table}[!htbp]
\centering
\caption{Energy ratio variation with incident angle for case 2}
\label{tab2}
\begin{tabular*}{\textwidth}{@{\extracolsep{\fill}} c c c c c c}
\hline
Incident angle 
& $\dfrac{P_1}{P_0}$ 
& $\dfrac{P_2}{P_0}$ 
& $\dfrac{P_3}{P_0}$ 
& $\dfrac{P_4}{P_0}$ 
& $\displaystyle\sum_{i=1}^{4}\dfrac{P_i}{P_0}$ \\
\hline
0.12 & 0.69 & 0.00 & 0.69 & $-0.25$ & 1.13$\approx 1$ \\
1.54 & 0.90 & 0.00 & 0.00 & 0.00    & 0.90$\approx 1$ \\
3.27 & 0.72 & 0.00 & 0.72 & $-0.26$ & 1.18$\approx 1$ \\
4.68 & 0.90 & 0.00 & 0.00 & 0.00    & 0.90$\approx 1$ \\
\hline
\end{tabular*}

\end{table}
The total energy ratios of all scattered waves are approximately equal to unity, satisfying the principle of energy conservation, as illustrated in Tables~\ref{tab1} and~\ref{tab2}.
\section{Numerical examples and discussion}
\label{06}
\begin{table}[h!]
\centering
\caption{Elastic constants for upper and lower FGPM half-space.}
\label{tab:elastic_constants}
\begin{tabular}{|c|c|c|}
\hline
\textbf{Parameters} & \textbf{FGPM Half- space} & \textbf{thermo-piezoelectric Half-space} \\ \hline
$\rho$ (kg/m$^{3}$) & $7.5\times10^3 $ & $7750$\\ \hline
$\mu_{11}$ (N/m$^{2}$) & $13.5 \times 10^{10}$ & $13.9 \times 10^{10}$ \\ \hline
$\mu_{13}$ (N/m$^{2}$) & $6.79 \times 10^{10}$ & $7.54 \times 10^{10}$ \\ \hline
$\mu_{44}$ (N/m$^{2}$) & $2.22\times 10^{10}$ & $2.56 \times 10^{10}$ \\ \hline
$\mu_{33}$ (N/m$^{2}$) & $11.3 \times 10^{10}$ & $11.3 \times 10^{10}$ \\ \hline
$e_{33}$ (C/m$^{2}$) & $9$ & $13.8$ \\ \hline
$e_{13}$ (C/m$^{2}$) & $-1.9$ & $-6.98$ \\ \hline
$e_{15}$ (C/m$^{2}$) & $9.8$ & $13.4$ \\ \hline
$\epsilon_{31}$ (C$^{2}$/N·m$^{2}$) & $9.9\times 10^{-11}$ & $60 \times 10^{-11}$ \\ \hline
$\epsilon_{33}$ (C$^{2}$/N·m$^{2}$) & $4.5 \times 10^{-11}$ & $54.7 \times 10^{-11}$ \\ \hline
$K_{33}=K_{11}$ (W$m^{-1}K^{-1}$)&$-$&    1.5 \\ \hline
$p_3$ (C$K^{-1}m^{-2}$) & $-$& $-452 \times 10^{-6}
$  \\ \hline
$\beta_{11}$ (N/Km$^{2}$) & $-$ & $1.52\times 10^{6}$ \\ \hline
$\beta_{33}$ (N/Km$^{2}$) & $-$ & $0.551 \times 10^{6}$ \\ \hline
\end{tabular}
\label{table 2}
\end{table}
The derived expressions are provided in closed form. Nevertheless, to carry out numerical evaluation and to present explicit results, it is necessary to assign particular material parameters, despite the general applicability of the preceding analysis. Accordingly, the material parameters for the piezoelectric substrate (PZT-5A) and the functionally graded piezoelectric material (FGPM) substrate (Weiss and Gaylord \cite{weis1985lithium}, Cao et al. \cite{cao2008dispersion}) are adopted, as summarized in Table~\ref{table 2}.
\subsection{Analysis of reflected and transmitted waves for stress and thermally-driven interface (Case 1)}
\begin{figure}[htbp]
\centering

\begin{subfigure}[b]{0.47\textwidth}
\includegraphics[width=\textwidth]{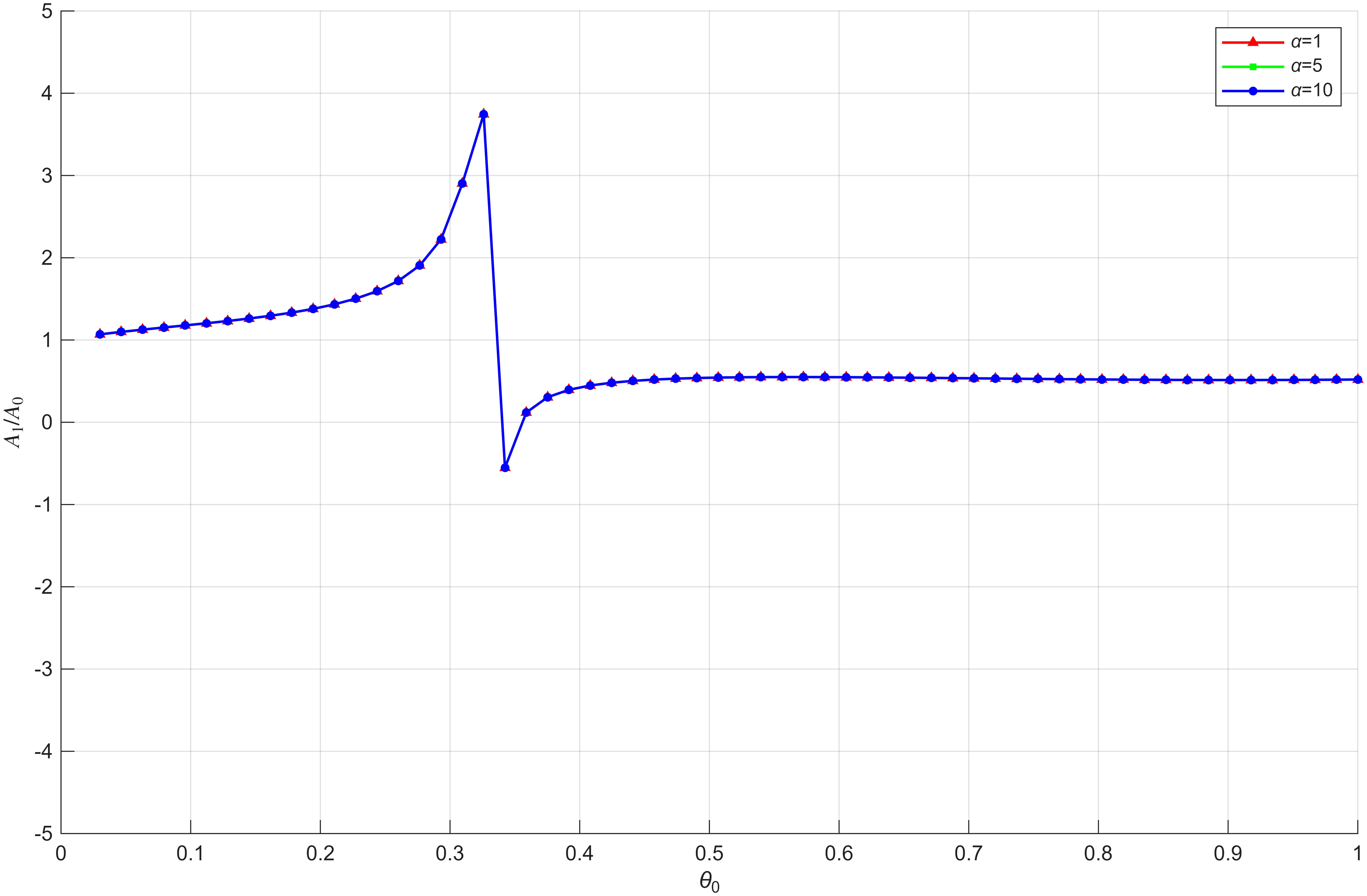}
\caption{}
\label{2a}
\end{subfigure}
\hfill
\begin{subfigure}[b]{0.47\textwidth}
\includegraphics[width=\textwidth]{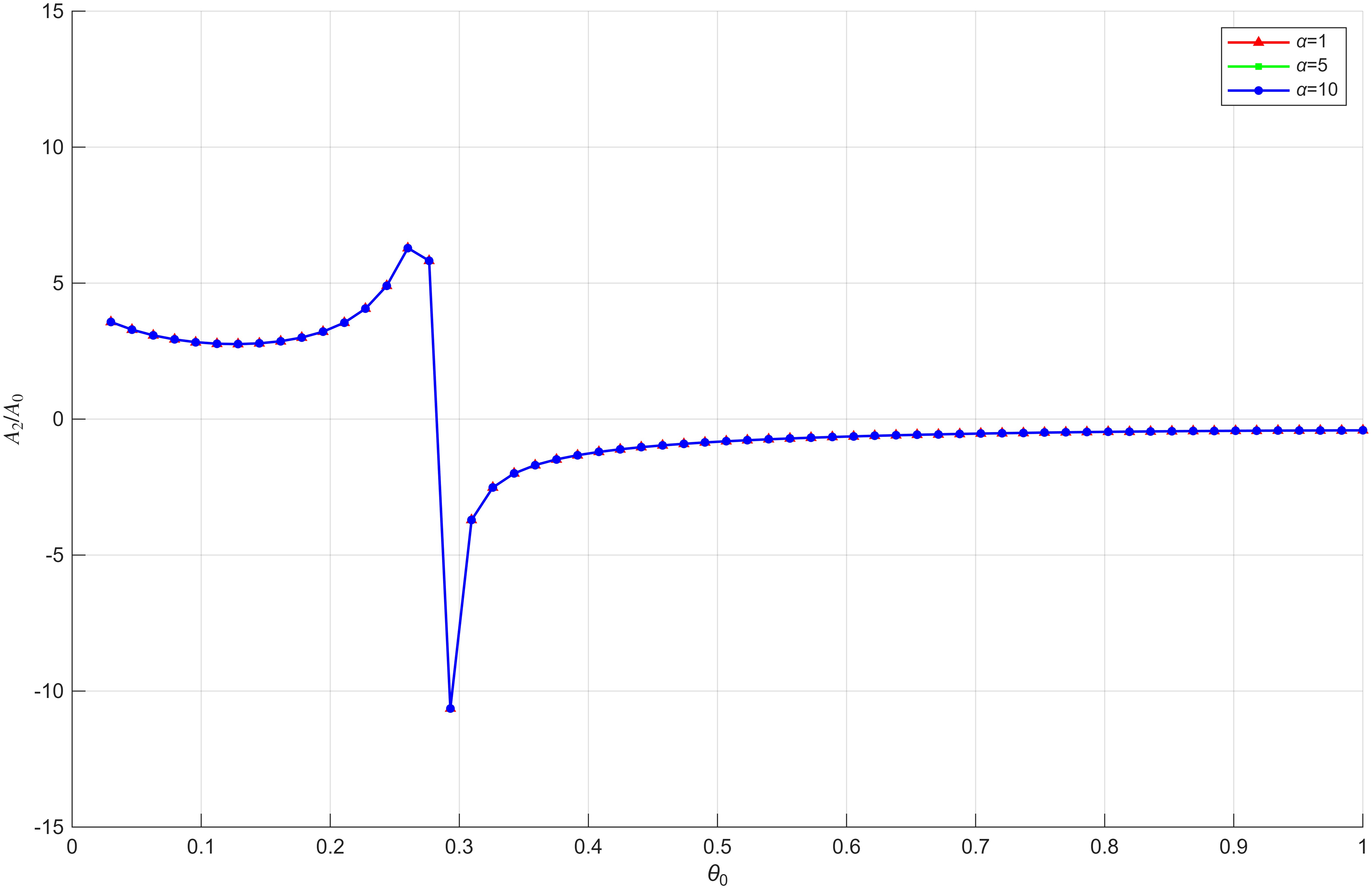}
\caption{}
\label{2b}
\end{subfigure}

\medskip

\begin{subfigure}[b]{0.47\textwidth}
\includegraphics[width=\textwidth]{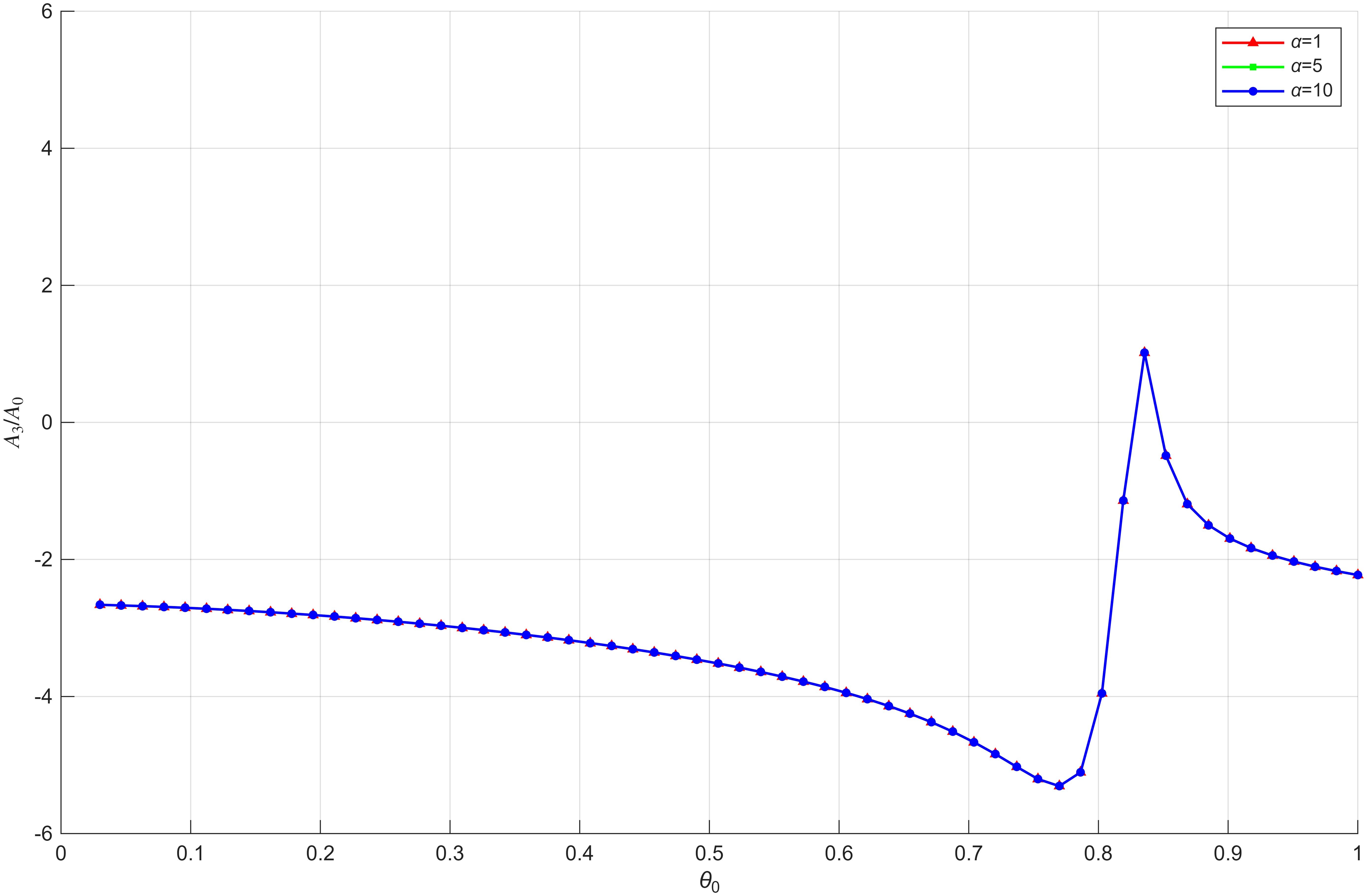}
\caption{}
\label{2c}
\end{subfigure}
\hfill
\begin{subfigure}[b]{0.47\textwidth}
\includegraphics[width=\textwidth]{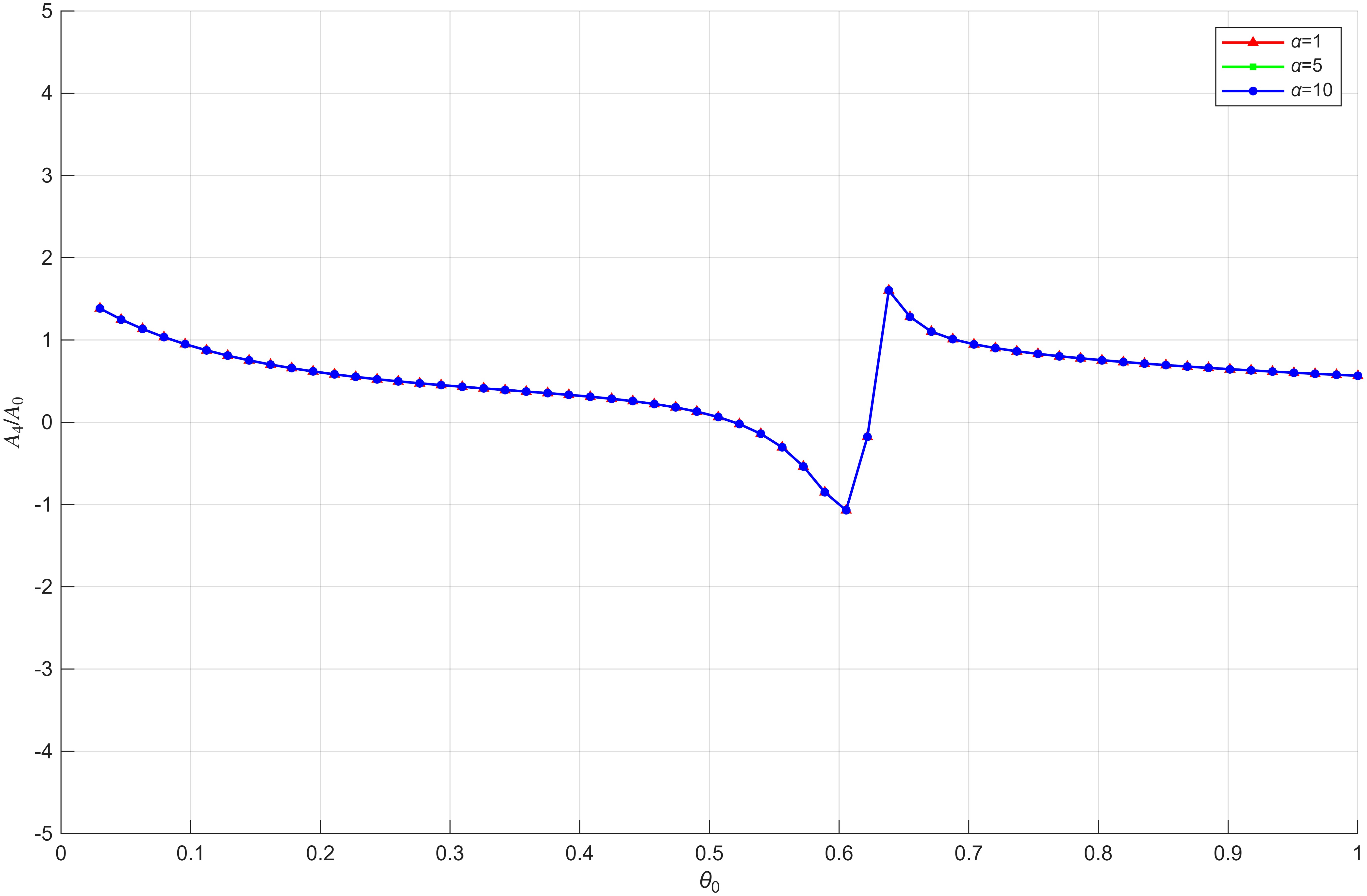}
\caption{}
\label{2d}
\end{subfigure}

\caption{Amplitude ratio versus incidence angle $\theta_0$ for varying material gradient parameters in FGPM media}
\label{Figure 2}
\end{figure}

\begin{figure}[htbp]
\centering

\begin{subfigure}[b]{0.47\textwidth}
\includegraphics[width=\textwidth]{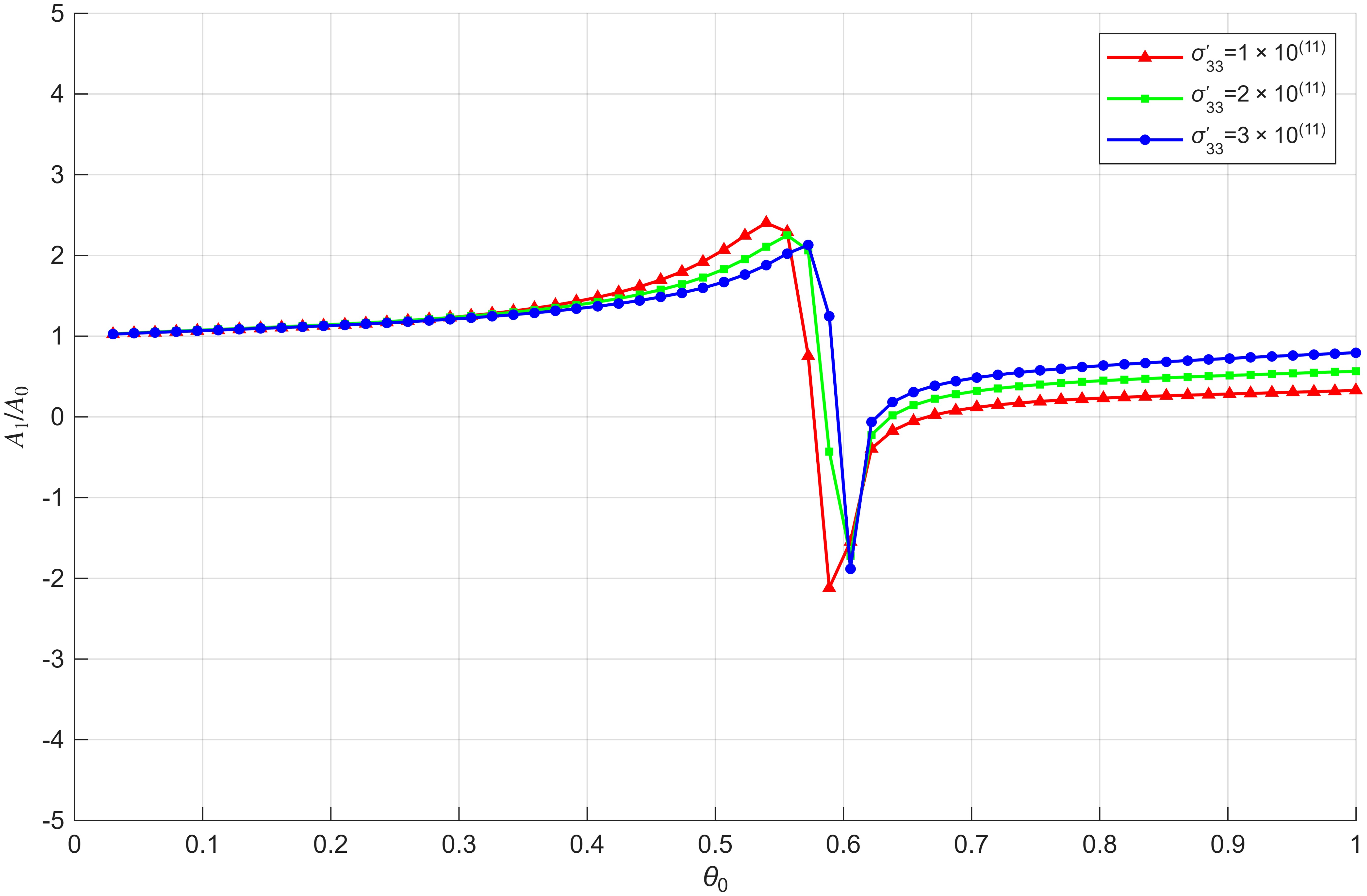}
\caption{}
\label{3a}
\end{subfigure}
\hfill
\begin{subfigure}[b]{0.47\textwidth}
\includegraphics[width=\textwidth]{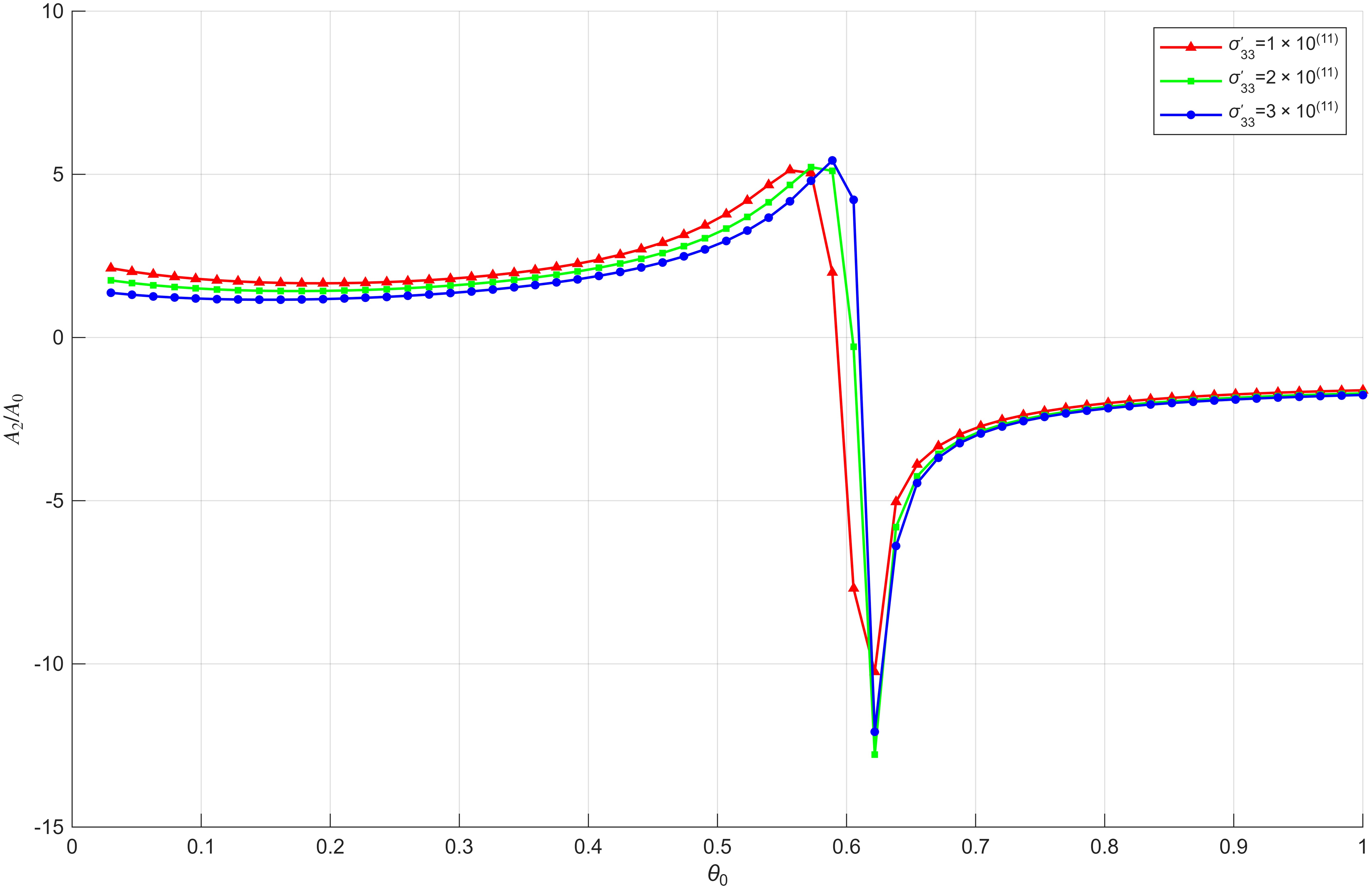}
\caption{}
\label{3b}
\end{subfigure}

\medskip

\begin{subfigure}[b]{0.47\textwidth}
\includegraphics[width=\textwidth]{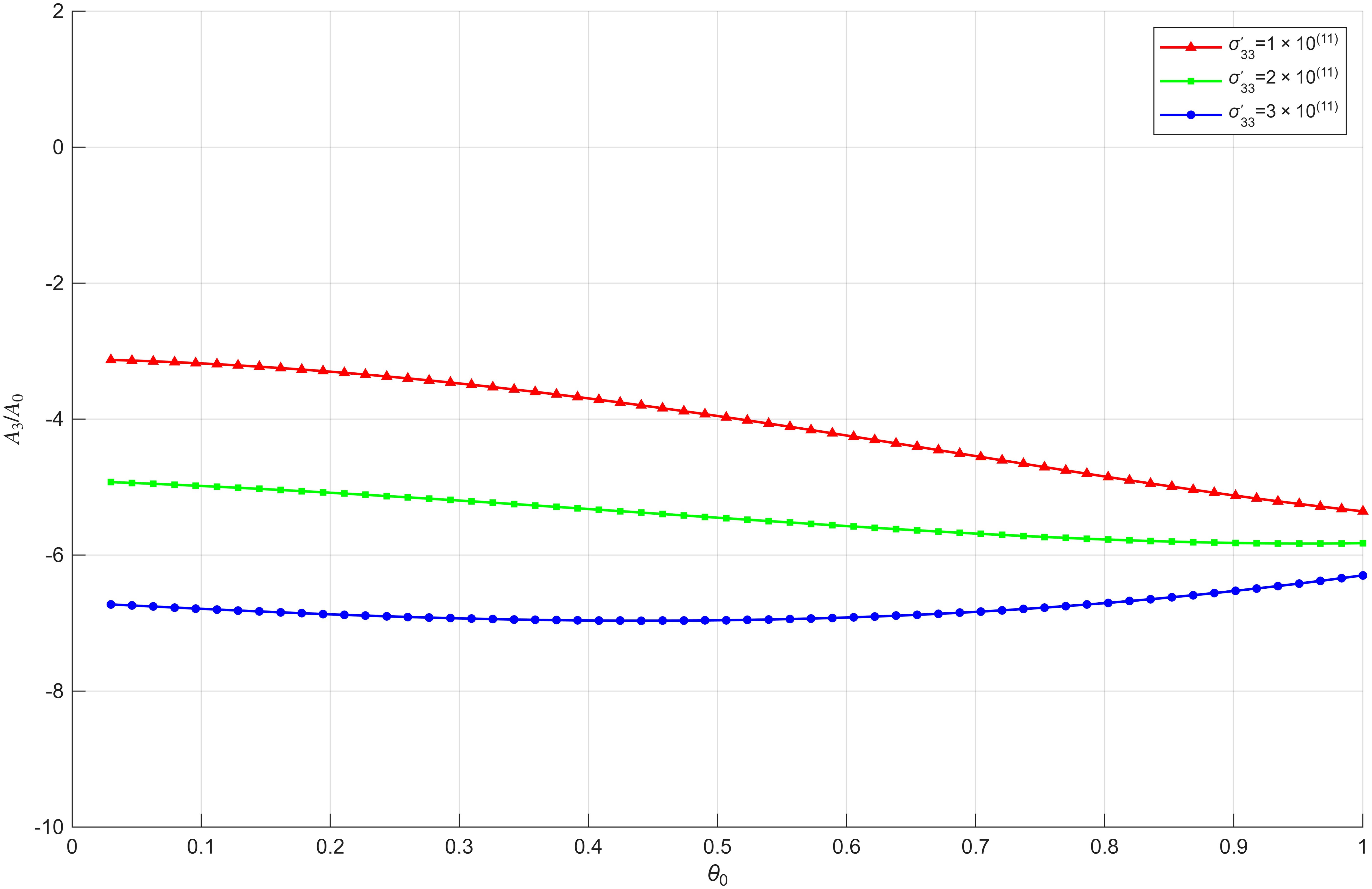}
\caption{}
\label{3c}
\end{subfigure}
\hfill
\begin{subfigure}[b]{0.47\textwidth}
\includegraphics[width=\textwidth]{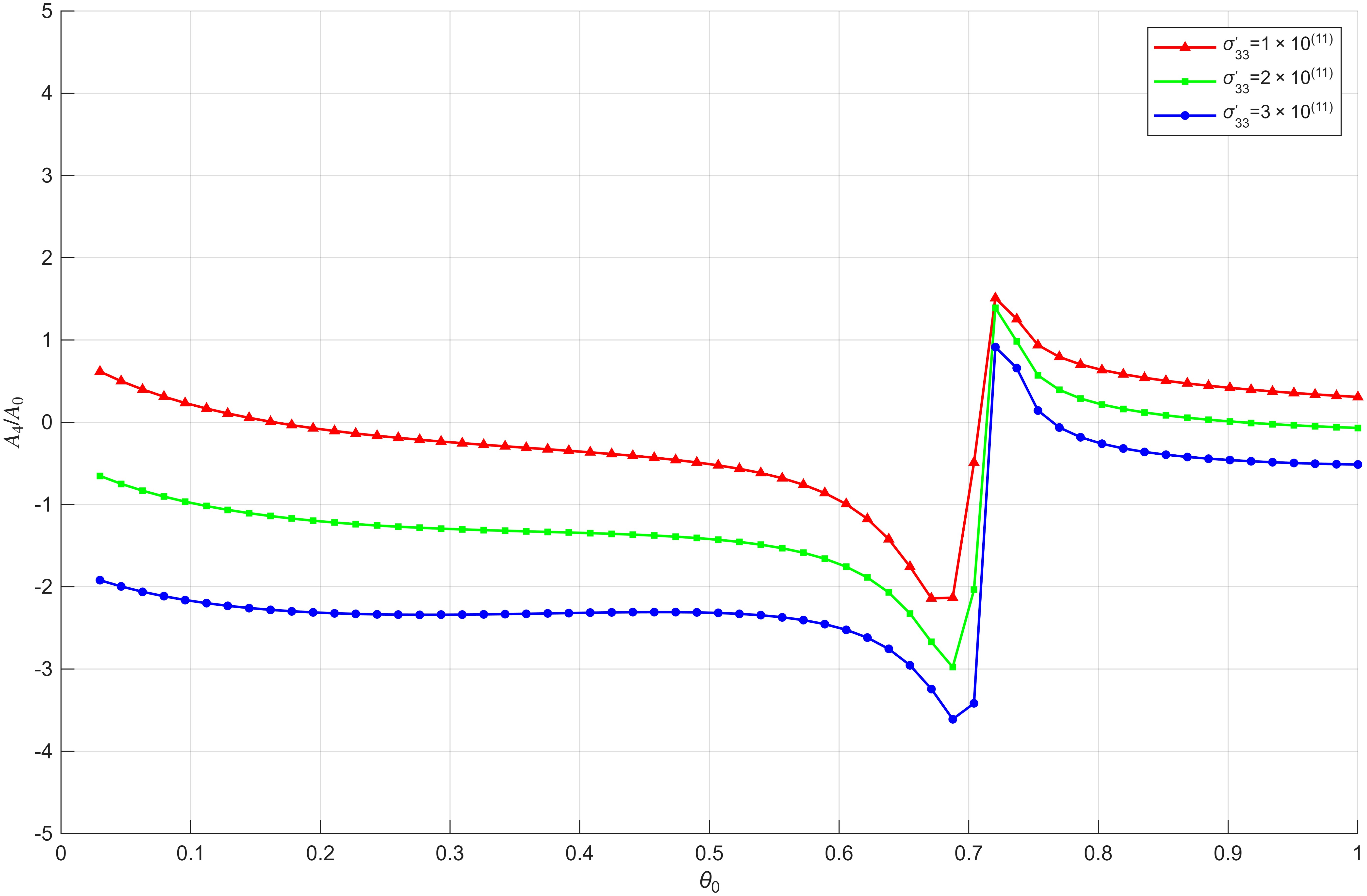}
\caption{}
\label{3d}
\end{subfigure}

\caption{Amplitude ratio versus incidence angle $\theta_0$ for varying initial stress in the thermo-piezoelectric media $(M_1)$.}
\label{Figure 3}
\end{figure}

\begin{figure}[htbp]
\centering

\begin{subfigure}[b]{0.47\textwidth}
\includegraphics[width=\textwidth]{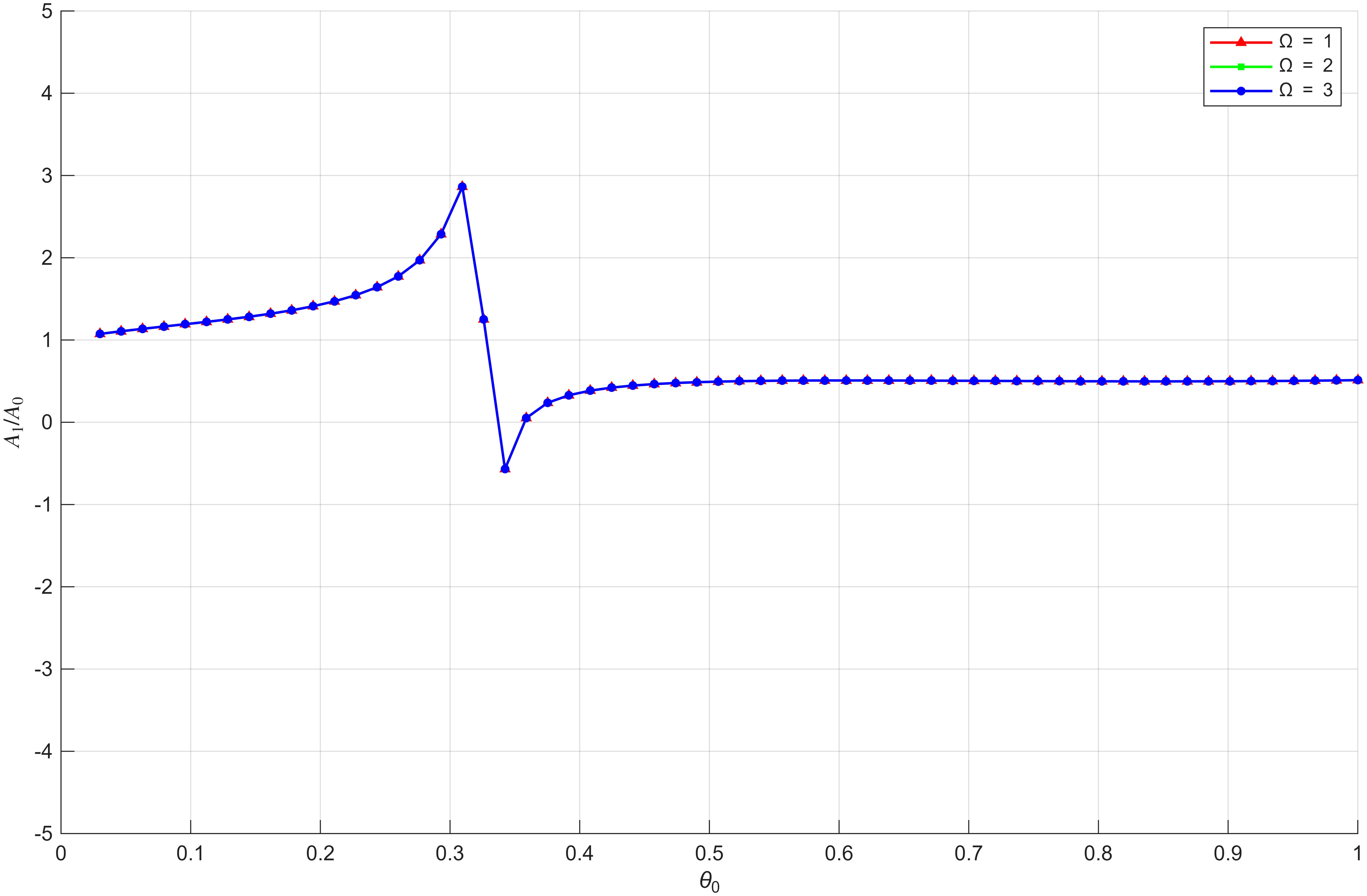}
\caption{}
\label{4a}
\end{subfigure}
\hfill
\begin{subfigure}[b]{0.47\textwidth}
\includegraphics[width=\textwidth]{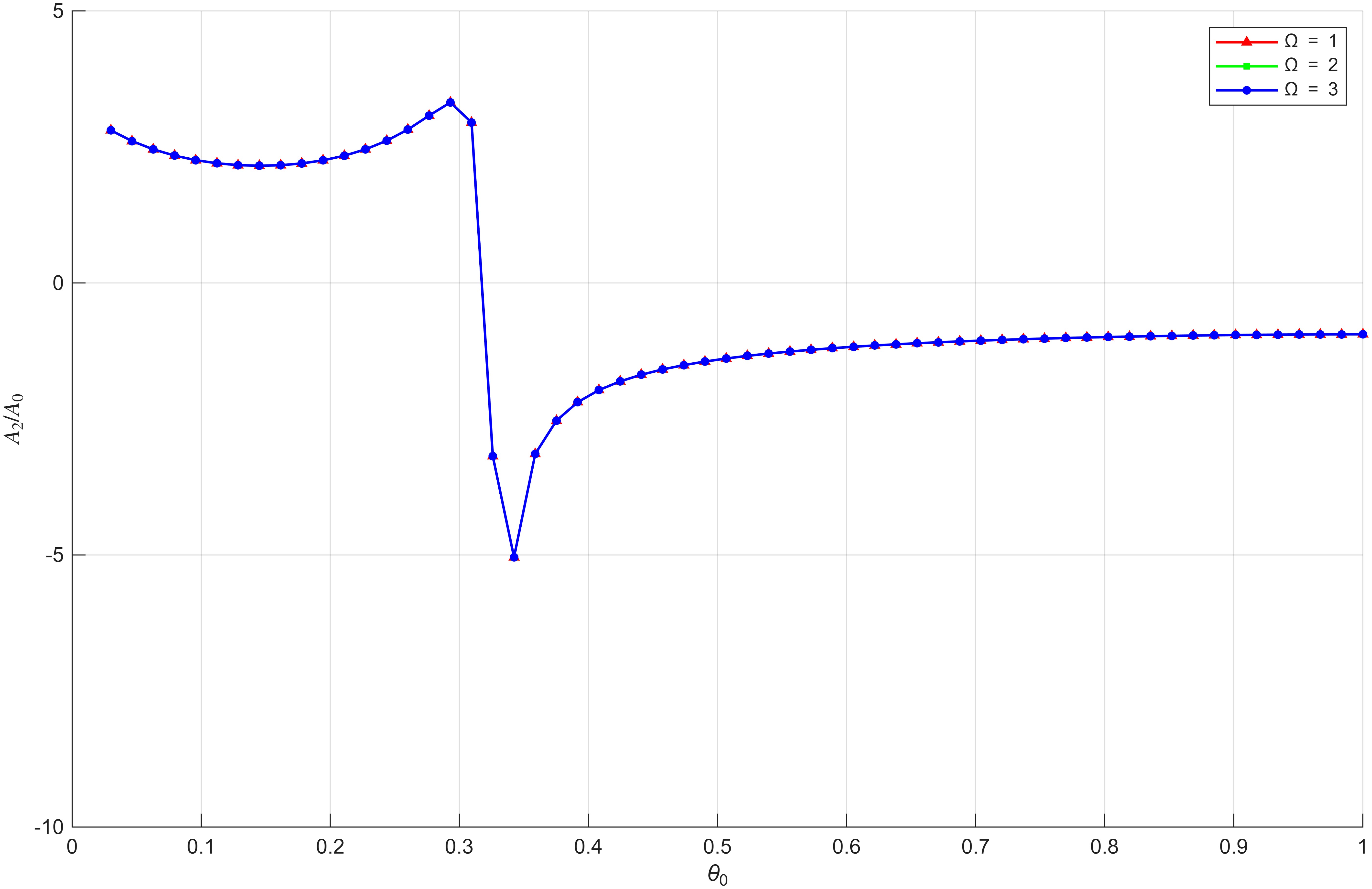}
\caption{}
\label{4b}
\end{subfigure}

\medskip

\begin{subfigure}[b]{0.47\textwidth}
\includegraphics[width=\textwidth]{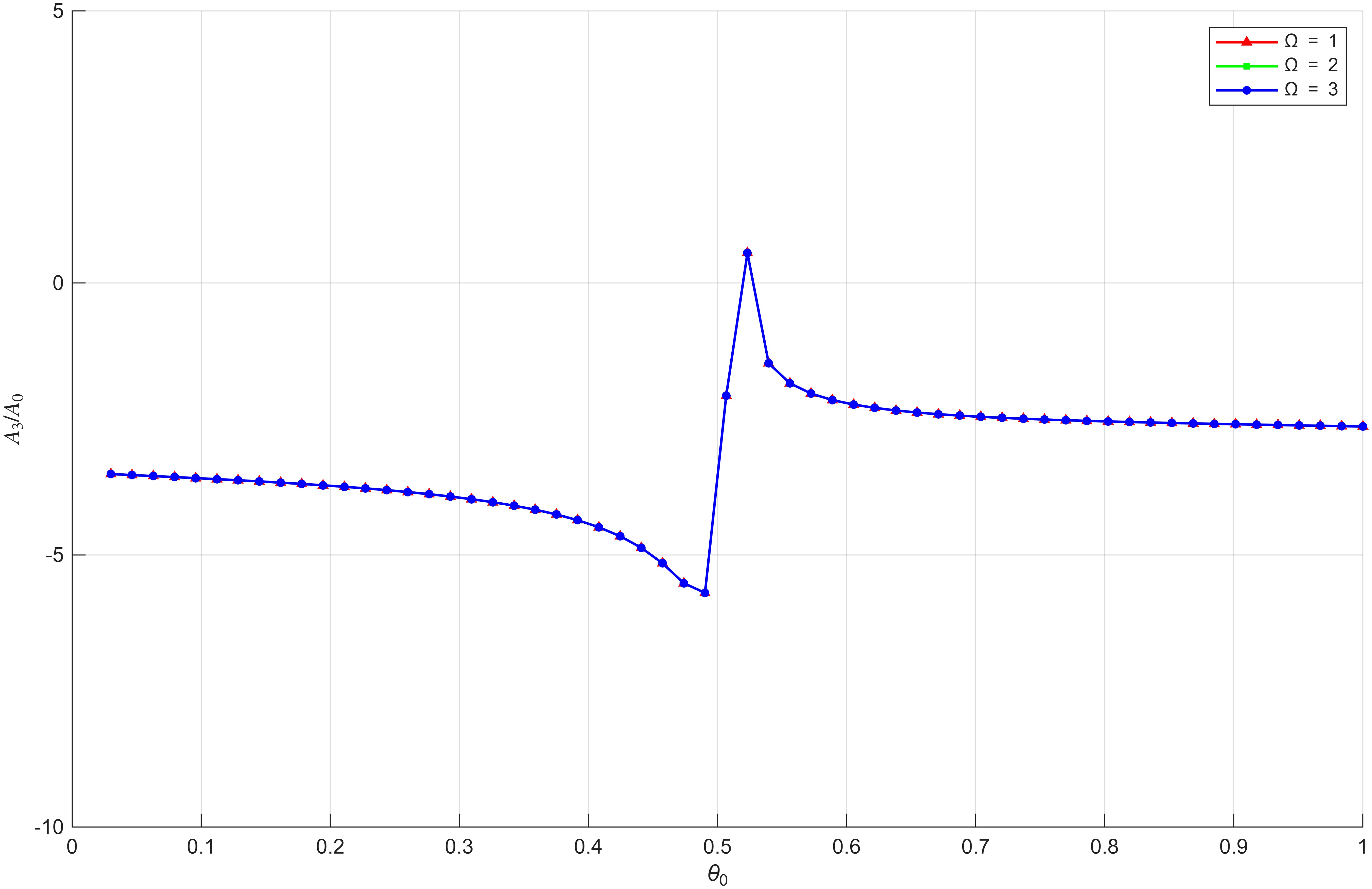}
\caption{}
\label{4c}
\end{subfigure}
\hfill
\begin{subfigure}[b]{0.47\textwidth}
\includegraphics[width=\textwidth]{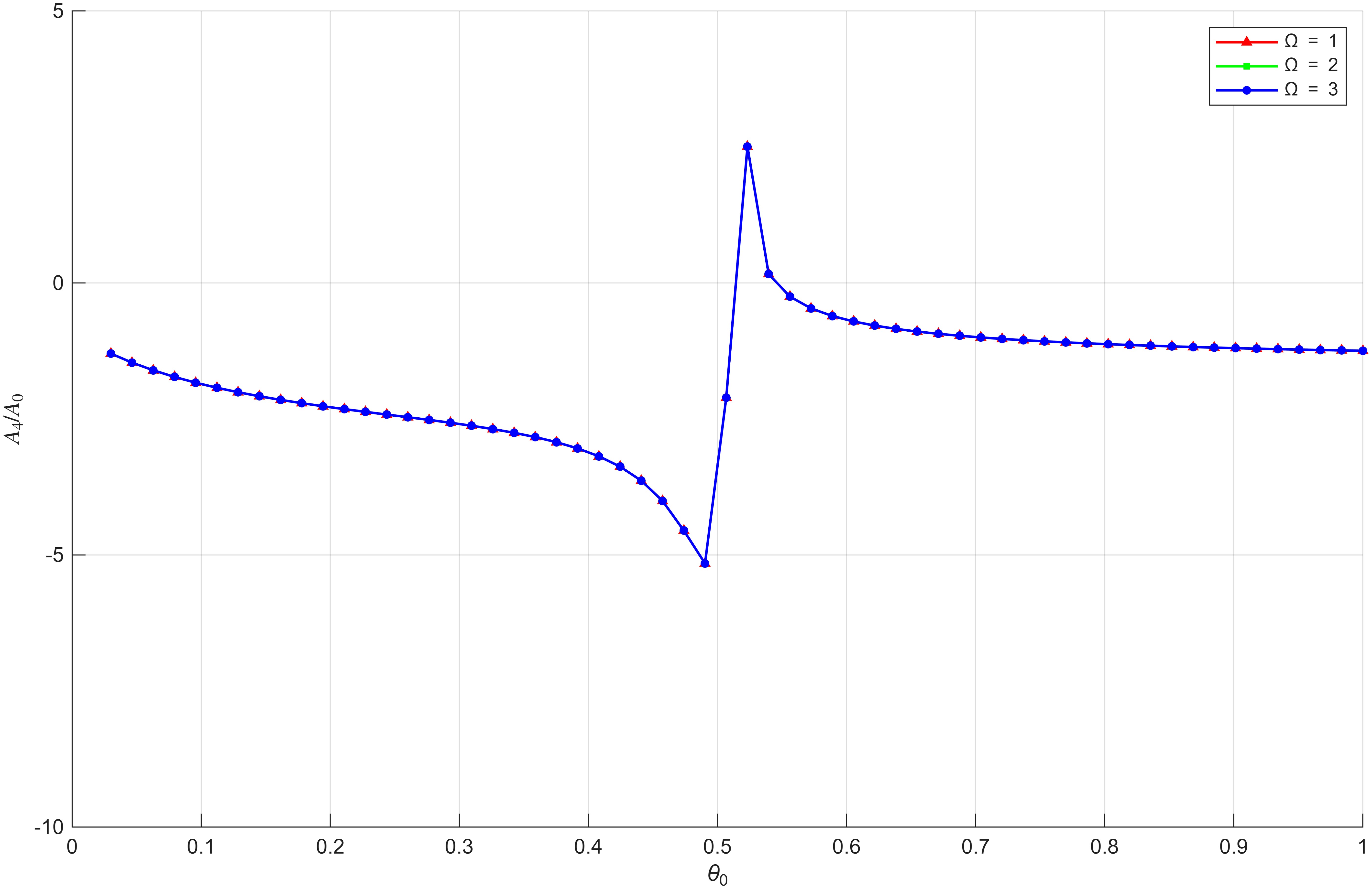}
\caption{}
\label{4d}
\end{subfigure}

\caption{Amplitude ratio versus incidence angle $\theta_0$ for varying rotation parameters in the thermo-piezoelectric media}
\label{Figure 4}
\end{figure}

\begin{figure}[htbp]
\centering

\begin{subfigure}[b]{0.47\textwidth}
\includegraphics[width=\textwidth]{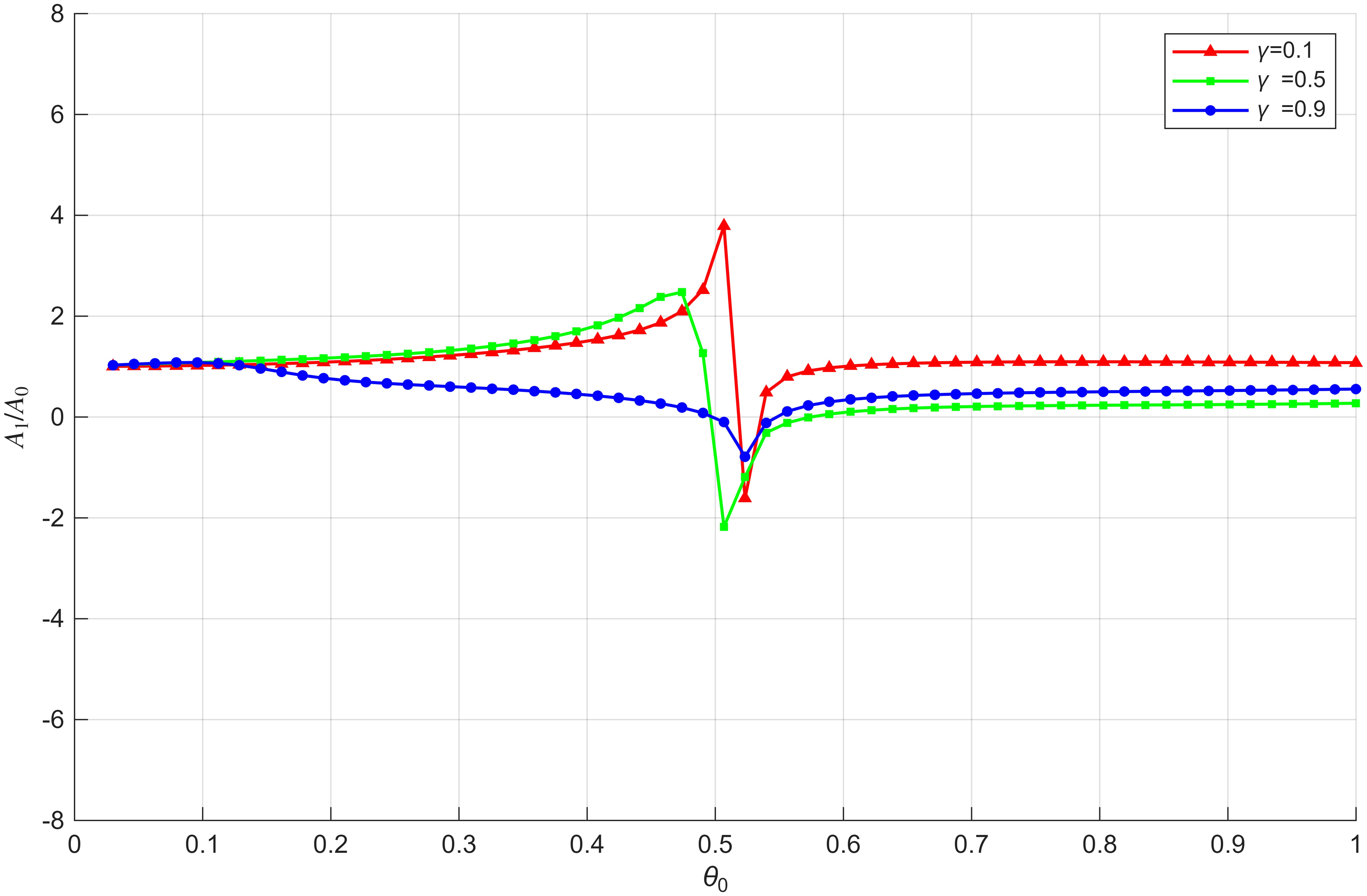}
\caption{}
\label{5a}
\end{subfigure}
\hfill
\begin{subfigure}[b]{0.47\textwidth}
\includegraphics[width=\textwidth]{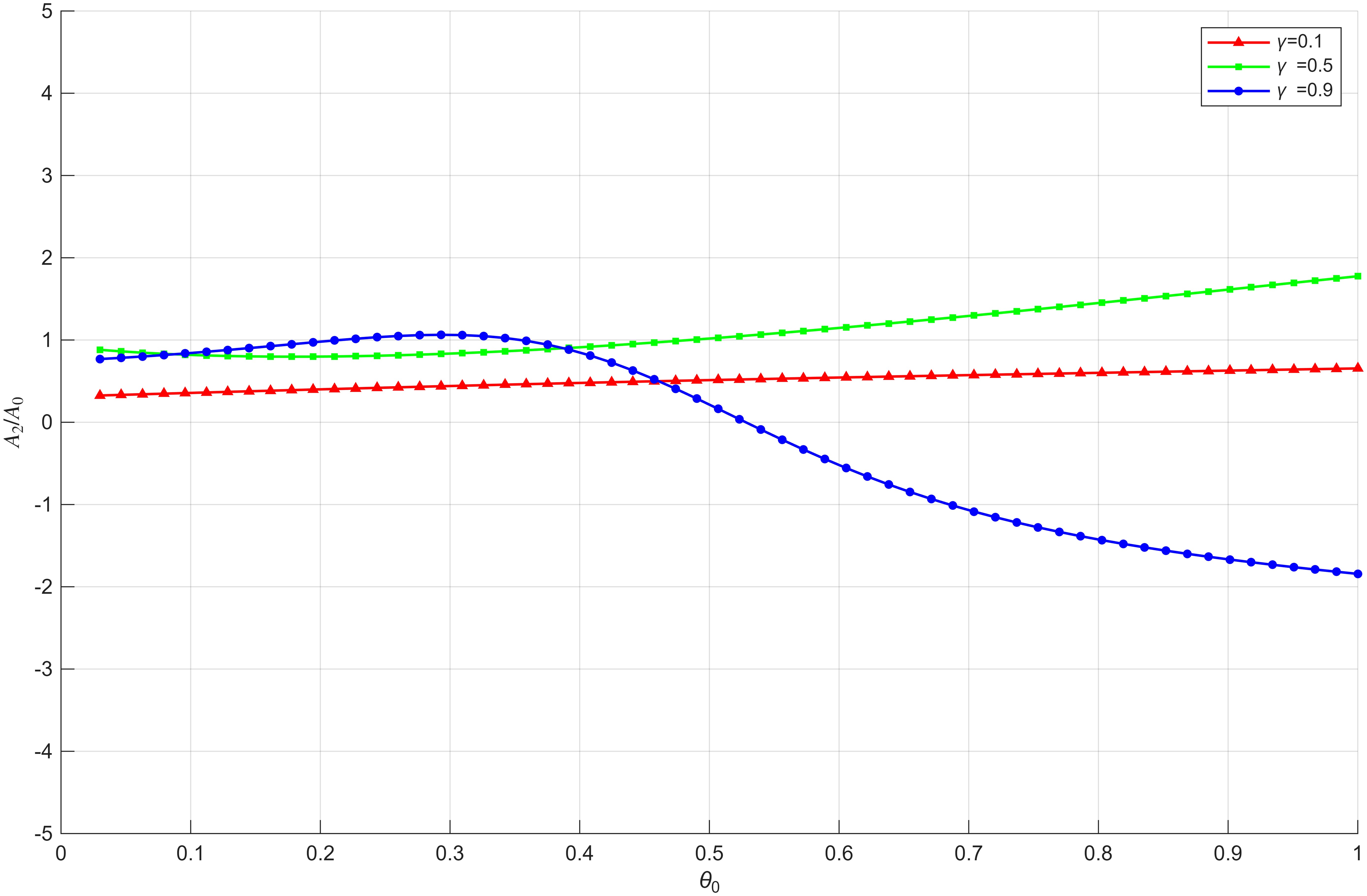}
\caption{}
\label{5b}
\end{subfigure}

\medskip

\begin{subfigure}[b]{0.47\textwidth}
\includegraphics[width=\textwidth]{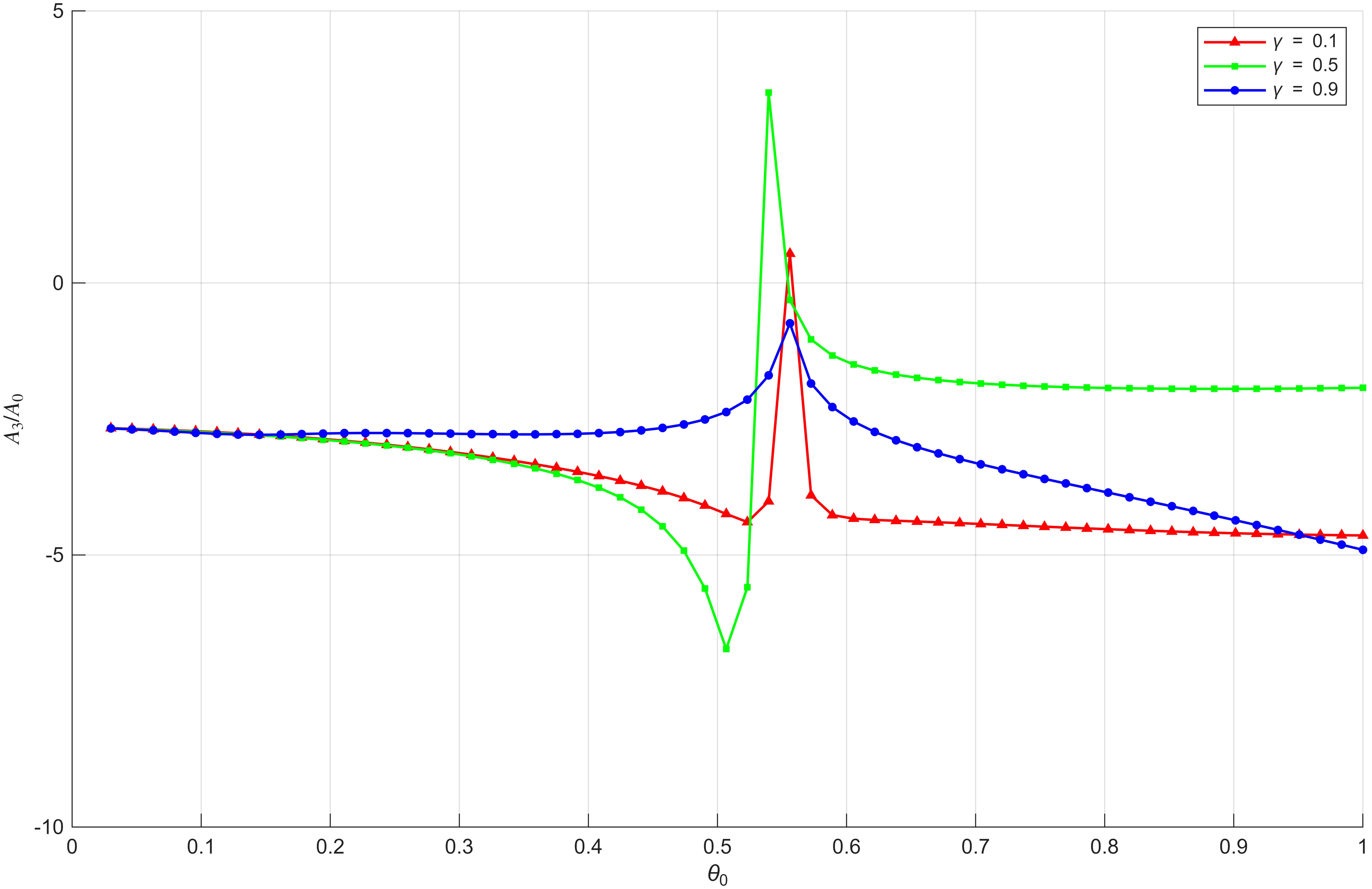}
\caption{}
\label{5c}
\end{subfigure}
\hfill
\begin{subfigure}[b]{0.47\textwidth}
\includegraphics[width=\textwidth]{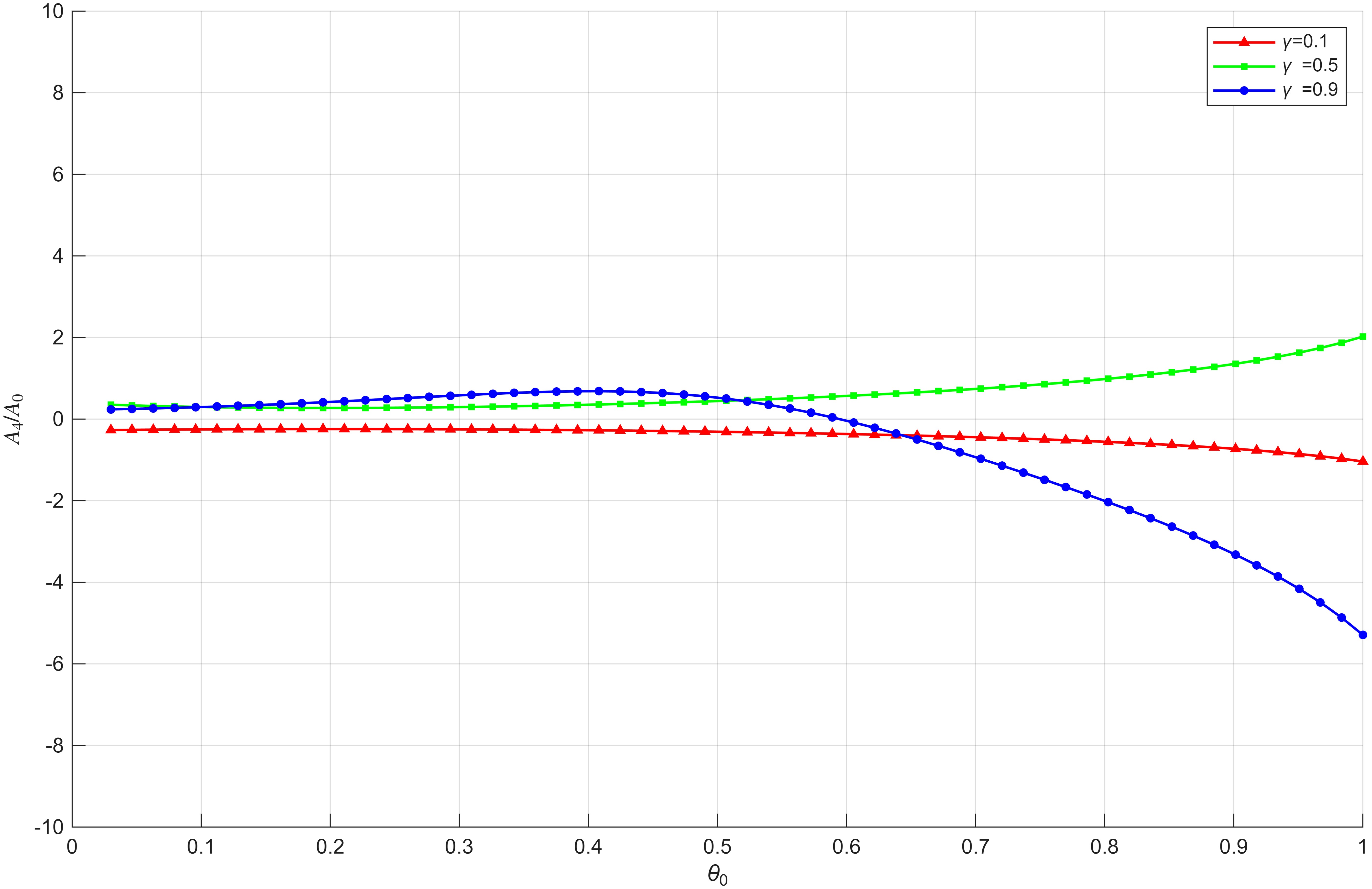}
\caption{}
\label{5d}
\end{subfigure}

\caption{Amplitude ratio versus incidence angle $\theta_0$ for varying order of fractional derivative in the thermo- piezoelectric media}
\label{Figure 5}
\end{figure}

Figure \ref{Figure 2} illustrate how the amplitude ratio varies with the angle of incidence $\theta_0$ for different values of material gradient in FGPM half space with $\sigma_{33}=2\times10^{11}$ and $\Omega^\prime=1$. The curves essentially overlap for different values of material gradient $\alpha$, indicating that $\alpha$ has negligible influence in this configuration. In figure \ref{2a}, for small values of $\theta_0$, the amplitude ratio remains positive and increases slightly, demonstrating a weak dependence on the incidence angle. After attaining its maximum value at $\theta_0=0.34$, the amplitude ratio $\frac{A_1}{A_0}$ begins to decrease and reaches its minimum value at $\theta_0=0.35$. Beyond this point, at $\theta_0=0.35$, the amplitude ratio exhibits a small rise before attaining a constant value. Figure \ref{2b} shows the amplitude ratio $\frac{A_2}{A_0}$ is positive for small values of $\theta_0$, increasing slowly, reaching its maximum value between $\theta_0=0.25$ to $\theta_0=0.3$, rapidly decreases and reaches its minimum value at $\theta_0=0.29$. After reaching its minimum value, the amplitude ratio $\frac{A_2}{A_0}$  increases and then becomes constant. From the figure \ref{2c}, it is observed that the amplitude ratio is negative for small value of $\theta_0$. As $\theta_0$ increases, the value of $\frac{A_3}{A_0}$ decreasing slowly and reaching its minimum value between $\theta_0=0.7$ to $\theta_0=0.8$ and then increases suddenly and reaches its maximum value at $\theta_0=0.83$ and started decreasing slowly. In figure \ref{2d}, when $\theta_0<0.5$, the amplitude ratio $\frac{A_4}{A_0}$ is positive and started decreasing slowly and reaching its minimum value at $\theta_0=0.61$. After attaining its minimum value, the amplitude ratio $\frac{A_4}{A_0}$ increases suddenly and reaches its maximum value at $\theta_0=0.63$ and then started decreasing slowly and becomes almost constant.
Figure \ref{Figure 3} presents the variation of amplitude ratio with incident angle for distinct initial stress when $\alpha=0.5$ and $\Omega^\prime=1$. In figure \ref{3a}, the value of amplitude ratio $\frac{A_1}{A_0}$ is approximately 1 when $\theta_0<0.1$. The value of the amplitude ratio started to gradually increase with the increasing value of $\theta_0$. After reaching its maximum value between $\theta_0=0.5$ to $\theta_0=0.6$, the curve starts to decrease and reaches its minimum value between $\theta_0=0.55$ to $\theta_0=0.65$ and eventually becomes constant. In figure \ref{3b} , for small value of $\theta_0$ the amplitude ratio remains positive and for $\theta_0>0.6$, the amplitude ratio $\frac{A_2}{A_0}$ starts decreasing suddenly. After reaching its minimum value at $\theta_0=0.62$, it increases a little bit and becomes constant. In figure \ref{3c}, for small values of $\theta_0$, the value of the amplitude ratio $\frac{A_3}{A_0}$ is negative and only decreases with the increasing value of $\theta_0$ for small values of initial stress, but for $\sigma_{33}^\prime=3\times 10^{11}$ the curve is increasing monotonically. In figure \ref{3d}, for some values of initial stress, the amplitude ratio is negative and for $\sigma_{33}^\prime=1\times10^{(11)}$ the value of amplitude ratio is positive. The value of $\frac{A_4}{A_0}$ decreases with increasing value of $\theta_0$, reaching its minimum value between $\theta_0=0.65$ to $\theta_0=0.7$, suddenly starts increasing and reached its maximum value at $\theta_0=0.72$. After reaching its maximum value, the value of amplitude ratio decreases slowly and eventually becomes constant.
Figure \ref{Figure 4} demonstrate how amplitude ratio vary for different values of rotation parameter when $\sigma_{33}=2\times10^{11}$ and $\alpha=1$. The curves essentially overlap for different values of the material gradient $\Omega^\prime$, indicating that $\Omega^\prime$ has negligible influence on the behavior of $A_1$ in this configuration. In figure \ref{4a}, for small values of $\theta_0$ the value of amplitude ratio remains positive. The value of amplitude ratio increases slowly and reaching its maximum value at $\theta_0=0.32$ and then started decreasing suddenly. After attaining its minimum value between $\theta_0=0.3$ to $\theta_0=0.35$, the amplitude ratio starts increasing slowly and eventually becoming constant. In figure \ref{4b}, the amplitude ratio increases with the increasing value of $\theta_0$. The curve attains its maximum value between $\theta_0=0.3$ to $\theta_0=0.35$ and then suddenly decreases, reached its minimum value at $\theta_0=0.3$. After attaining its minimum value, the value of reflection coefficient $\frac{A_2}{A_0}$ increases and eventually becoming constant. In figure \ref{4c}, the value of refraction coefficient $\frac{A_3}{A_0}$ is negative for small value of $\theta_0$, suddenly increases after $\theta_0=0.5$ to its maximum value at $\theta_0=0.54$. After reaching its maximum value the curve decreases slowly and becomes constant. In figure \ref{4d}, the curve starts from the negative value of refraction coefficient $\frac{A_4}{A_0}$, slowly decreasing and reached to its minimum value at $\theta_0=0.5$. After reaching its minimum value, the curve suddenly increases and reached its maximum value at $\theta_0=0.53$, decreasing slowly and then becomes constant. Figure \ref{Figure 5} shows the variation of amplitude ratio with the angle of incidence $\theta_0$ for distinct values of order of fractional derivative in the thermo-piezoelectric media when $\Omega^\prime=1$ and $\sigma_{33}=1\times10^{11}$. The results show a sharp resonance like behavior near a critical angle, followed by a rapid sign change in the amplitude ratio.

All curves begin from the same initial amplitude, but as $\theta_0$ increases, the influence of the parameter becomes evident and the curves spread apart, each showing its characteristic behavior. In \ref{5a}, for $\gamma=0.1$, the amplitude ratio remains close to unity for small and moderate angles, showing only a mild increase up to $\theta_0=0.45$. As the incident angle approaches the critical region around $\theta_0=0.51$, the amplitude ratio grows steeply and reaches a sharp positive peak of nearly at $\theta_0=0.52$. Immediately after this point, the curve dips dramatically to a large negative value and then gradually stabilizes to positive values for larger angles. After attaining its minimum value between $\theta_0=0.5$ to $\theta_0=0.55$, the curve becomes constant after $\theta_0=0.6$.  For $\gamma=0.5$, the curve shows the similar behavior. The curve reached its maximum value at $\theta_0=0.48$ and then suddenly decreases. After attaining its minimum value at $\theta_0=0.51$, the curve becomes constant after $\theta_0=0.6$. For $\gamma=0.9$, the amplitude ratio shows a distinctly different response. Instead of producing a large positive spike, the amplitude decreases smoothly as the incident angle increases, reaching a minimum just before the critical region. Although a small dip still occurs, the behavior is significantly less abrupt compared to smaller values of $\gamma$. Figure \ref{5b} shows the variation of reflection coefficient $\frac{A_2}{A_0}$ with respect to the angle of incidence $\theta_0$ for different values of $\gamma$. For $\gamma=0.1$, the amplitude ratio remains almost constant throughout the entire angular range. The curve shows only a slight upward trend for larger $\theta_0$, indicating that weak coupling produces minimal change in the amplitude. For $\gamma=0.5$, the amplitude ratio initially decreases slightly before gradually increasing for larger angles. Relative to the $\gamma=0.1$, the curve exhibits a more noticeable angular dependence, showing that medium coupling has a measurable but controlled coupling has a measurable but controlled effect on wave conversion. For $\gamma=0.9$, the amplitude ratio initially increases up to $\theta_0=0.32$ reaching its peak. Beyond this point, the amplitude ratio decreases sharply, crosses zero near $\theta_0=0.58$

\begin{figure}[htbp]
\centering

\begin{subfigure}[b]{0.47\textwidth}
\includegraphics[width=\textwidth]{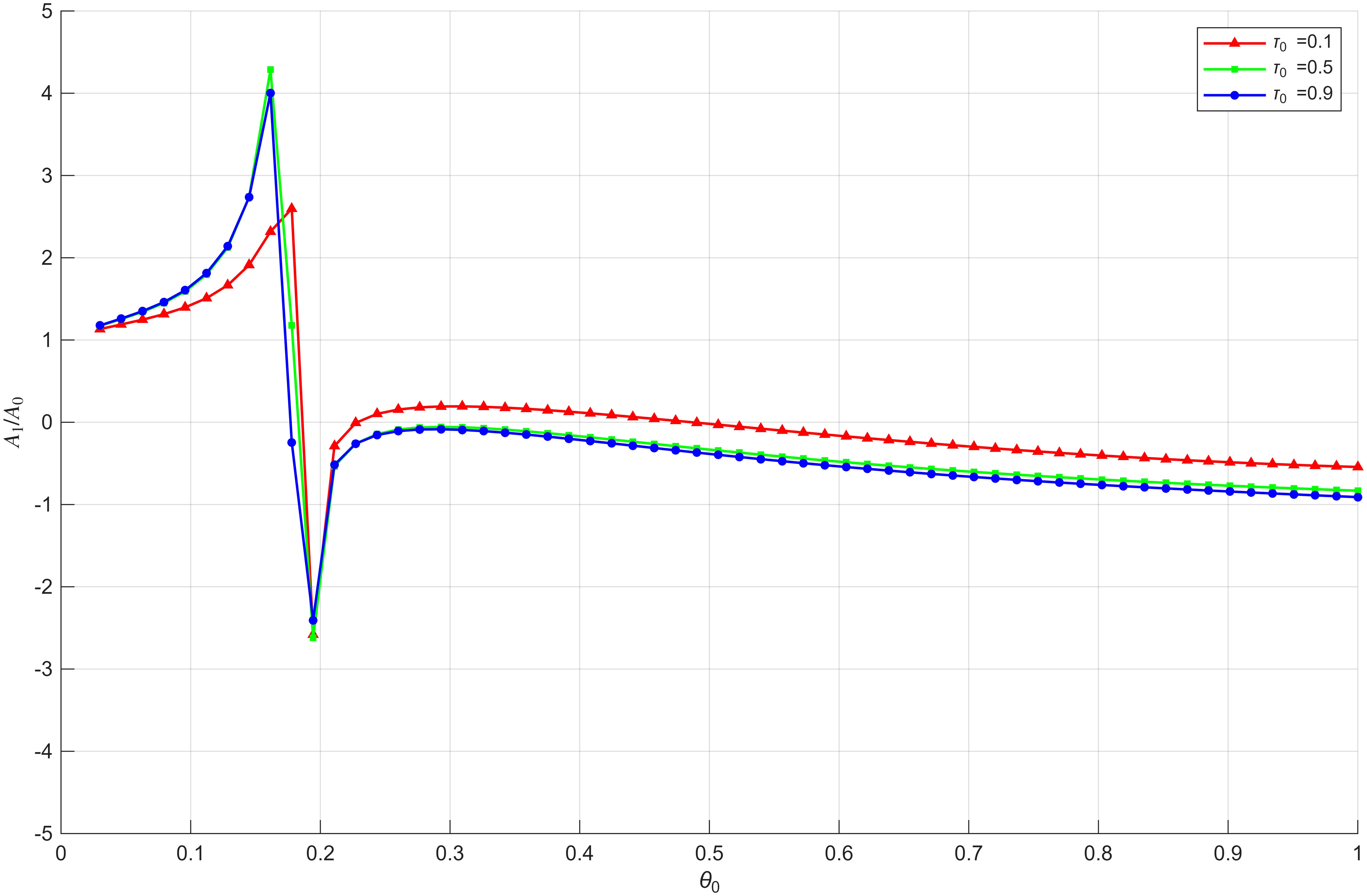}
\caption{}
\label{6a}
\end{subfigure}
\hfill
\begin{subfigure}[b]{0.47\textwidth}
\includegraphics[width=\textwidth]{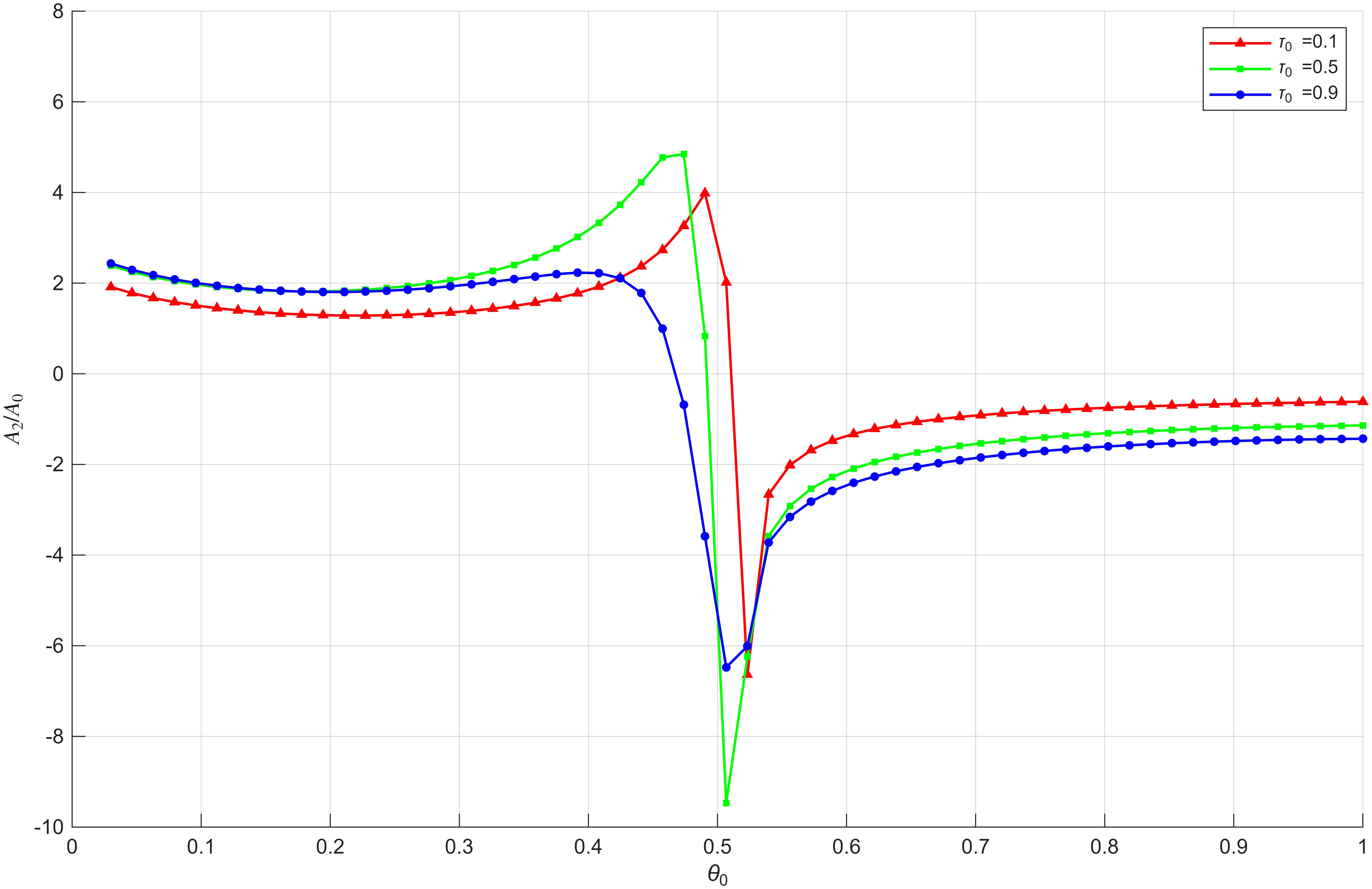}
\caption{}
\label{6b}
\end{subfigure}

\medskip

\begin{subfigure}[b]{0.47\textwidth}
\includegraphics[width=\textwidth]{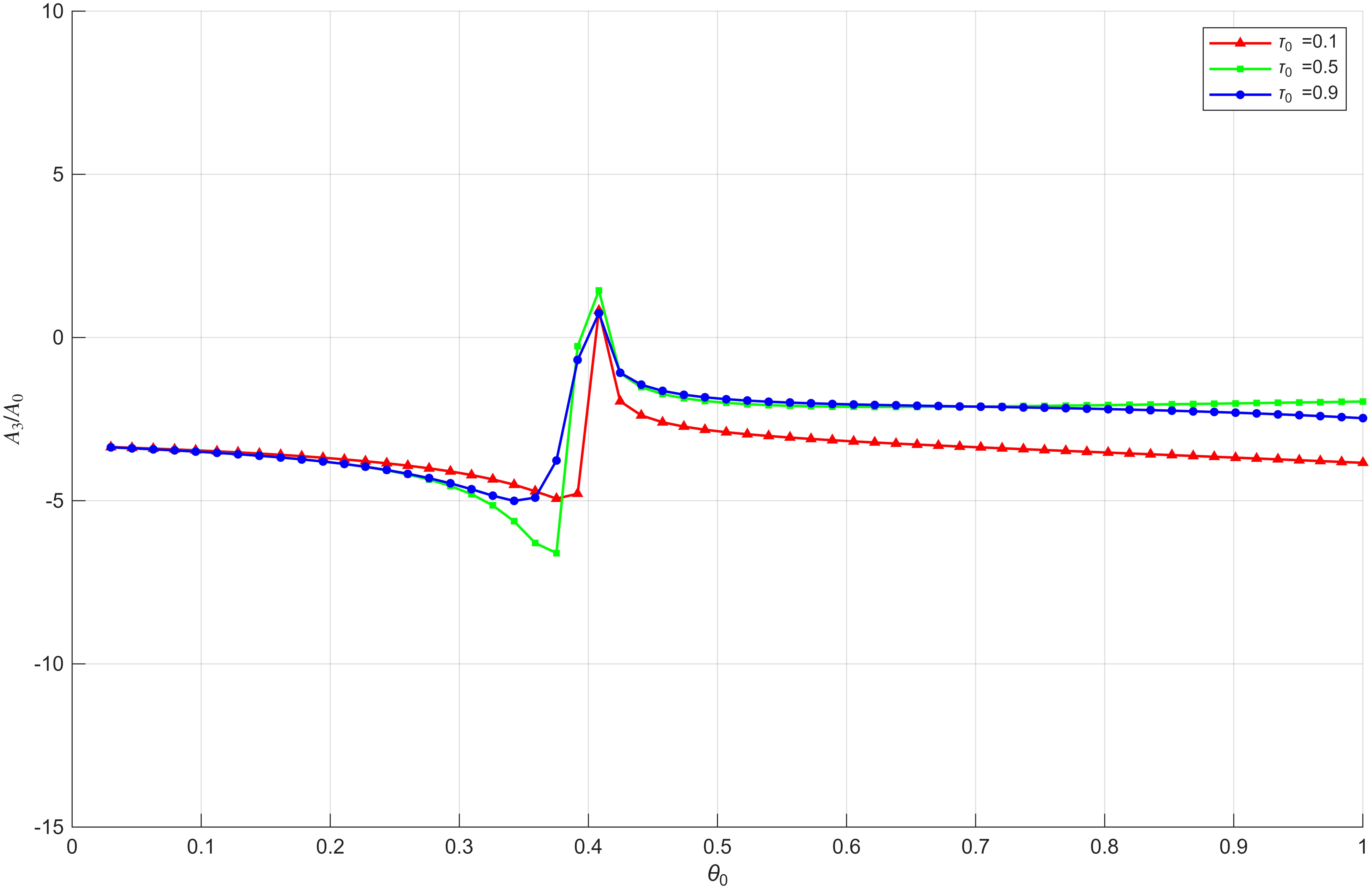}
\caption{}
\label{6c}
\end{subfigure}
\hfill
\begin{subfigure}[b]{0.47\textwidth}
\includegraphics[width=\textwidth]{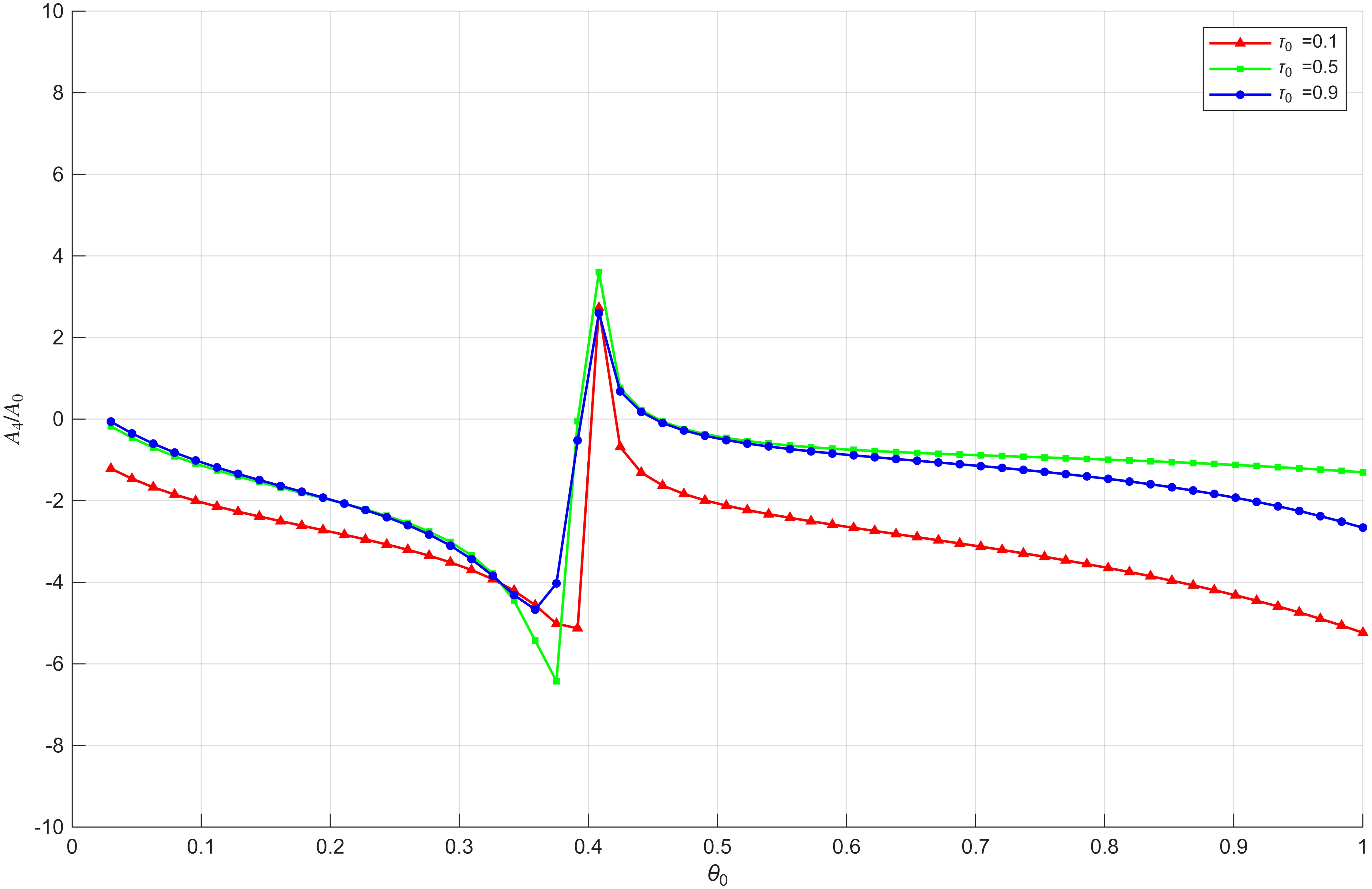}
\caption{}
\label{6d}
\end{subfigure}

\caption{Amplitude ratio versus incidence angle $\theta_0$ for varying relaxation time in the thermo- piezoelectric media}
\label{Figure 6}
\end{figure}

 and continues to fall into the negative region for larger angles. In figure \ref{5c}, all the curves starts from the same point. The curves shows almost same behavior for all the values of $\gamma$. The curves reaches its maximum value between $\theta_0=0.5$ to $\theta_0=0.6$. The curve becomes almost constant after a point for $\gamma=0.1$ and $\gamma=0.5$, but for $\gamma=0.9$, the curve decreases after reaching its maximum value. Figure \ref{5d} shows almost the same effect as \ref{5b}. For $\theta_0<0.4$, the value of refraction coefficient $\frac{A_4}{A_0}$ increases gradually seems almost constant. For $\gamma=0.1$ the curve shows constant behavior till $\theta_0=0.7$ and the it starts decreasing gradually with increasing $\theta_0$. For $\gamma=0.5$, as we are increasing angle of incidence, value of refraction coefficient also increases and for $\gamma=0.9$, the curve reaches its maximum value at $\theta_0=0.4$ and then starts decreasing faster than for other values of $\gamma$. Figure \ref{Figure 6} shows how amplitude ratio vary with the angle of incidence $\theta_0$ for different values of relaxation time when $\Omega^\prime=1$ and $\sigma_{33}=1\times10^{11}$. In figure \ref{6a}, the curves shows the same behavior for different values of relaxation time. For small values of $\tau_0$ the curves almot overlap. The value of reflection coefficient $\frac{A_1}{A_0}$ increases gradually with increasing $\theta_0$ and reaching its maximum value between $\theta_0=0.1$ to $\theta_0=0.2$. After reaching their maximum value, the curves suddenly decreased and reached their minimum value around at $\theta_0=1.9$. The value of reflection coefficient increases suddenly and eventually becomes constant.

In figure \ref{6b}, the value of reflection coefficient $\frac{A_2}{A_0}$ decreasing slowly and after $\theta_0=0.3$, for $\gamma=0.1$ and $\gamma=0.5$, the curve starts increasing and reaches their maximum value between $\theta_0=0.4$ to $\theta_0=0.5$ and then suddenly decreases. After reaching its minimum value between $\theta_0=0.5$ to $\theta_0=0.55$, it increases and eventually approaches a constant value. On the other hand, for $\theta_0=0.9$, with increasing angle of incidence, value of reflection coefficient decreases. This curve reaches its minimum value at $\theta_0=0.5$ and then starts increasing and becomes constant after $\theta_0=0.9$. All the curves shows same behavior for different value of $\tau_0$ in figure \ref{6c}. The refraction coefficient $\frac{A_3}{A_0}$ is negative for small values of $\theta_0$ and decreasing with increasing $\theta_0$. The curves attaining their minimum value between $\theta_0=0.3$ to $\theta_0=0.4$. After attaining their minimum value the value of refraction coefficient $\frac{A_3}{A_0}$ undergoes a rapid increase and reaches their maximum value between $\theta_0=0.4$ to $\theta_0=0.45$. The curves starts decreasing before becoming almost constant. Figure \ref{6d} shows almost the same behavior as \ref{6c}, where for the small values of $\theta_0$ the value of refraction coefficient $\frac{A_4}{A_0}$ is negative. The curves attain their minimum value between $\theta_0=0.35$ to $\theta_0=0.4$ and then the value increases suddenly and reaches its maximum value between $\theta_0=0.4$ to $\theta_0=0.42$. After reaching its maximum value the value of refraction coefficient decreasing gradually.

\subsection{Wave reflection and transmission under displacement, shear, electric, and thermal boundary conditions (Case 2)}
\begin{figure}[htbp]
\centering

\begin{subfigure}[b]{0.47\textwidth}
\includegraphics[width=\textwidth]{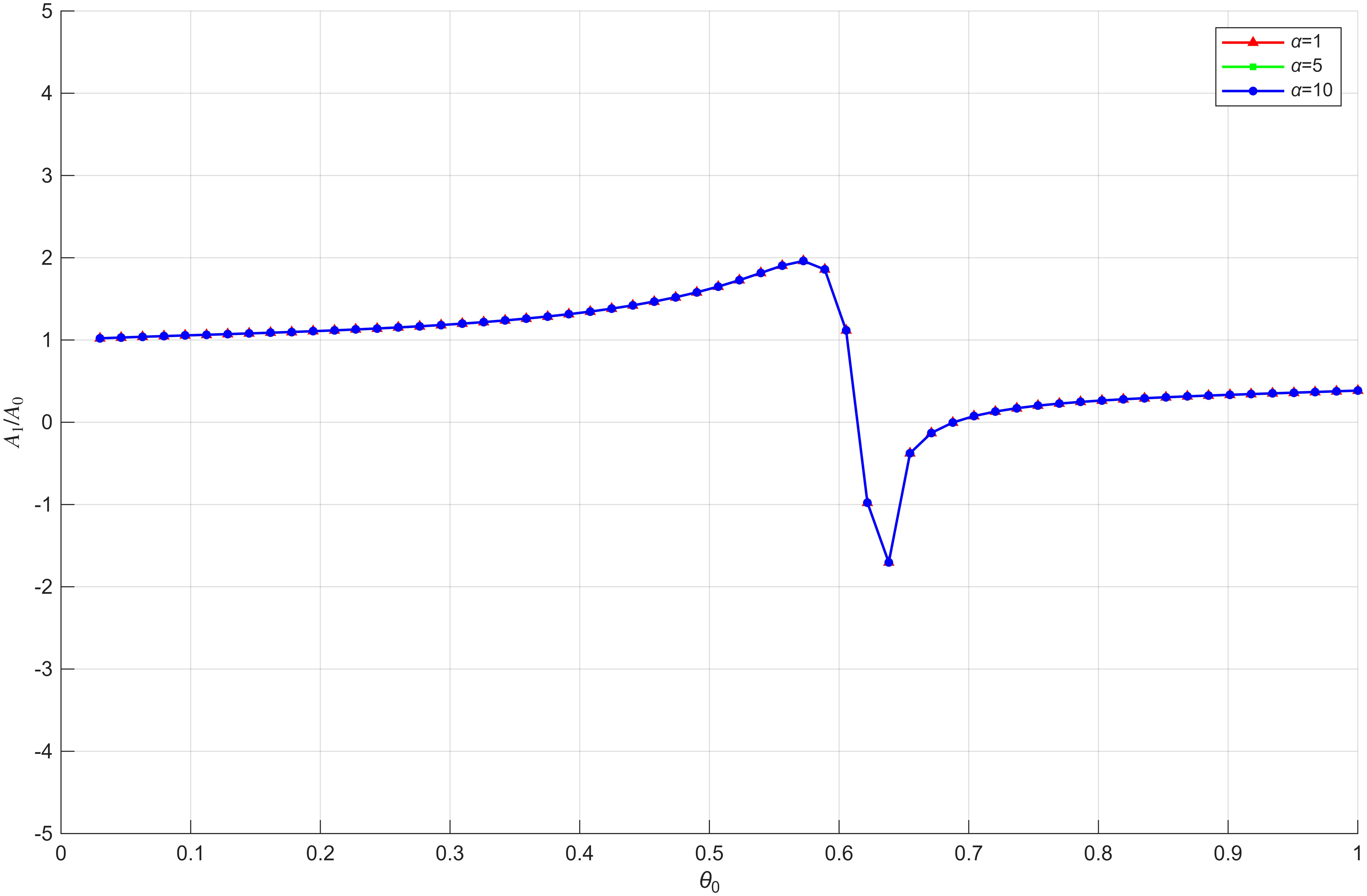}
\caption{}
\label{7a}
\end{subfigure}
\hfill
\begin{subfigure}[b]{0.47\textwidth}
\includegraphics[width=\textwidth]{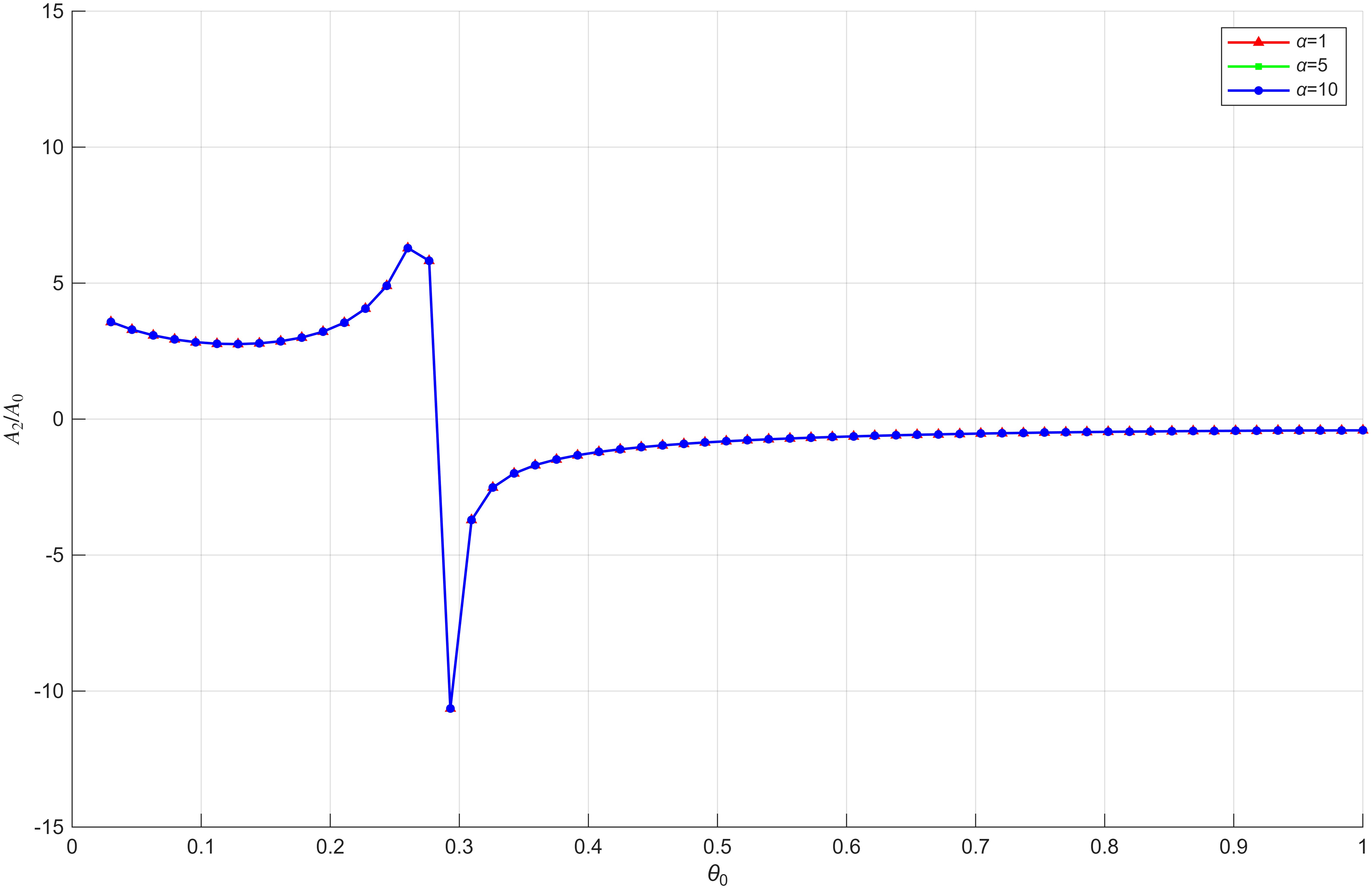}
\caption{}
\label{7b}
\end{subfigure}

\medskip

\begin{subfigure}[b]{0.47\textwidth}
\includegraphics[width=\textwidth]{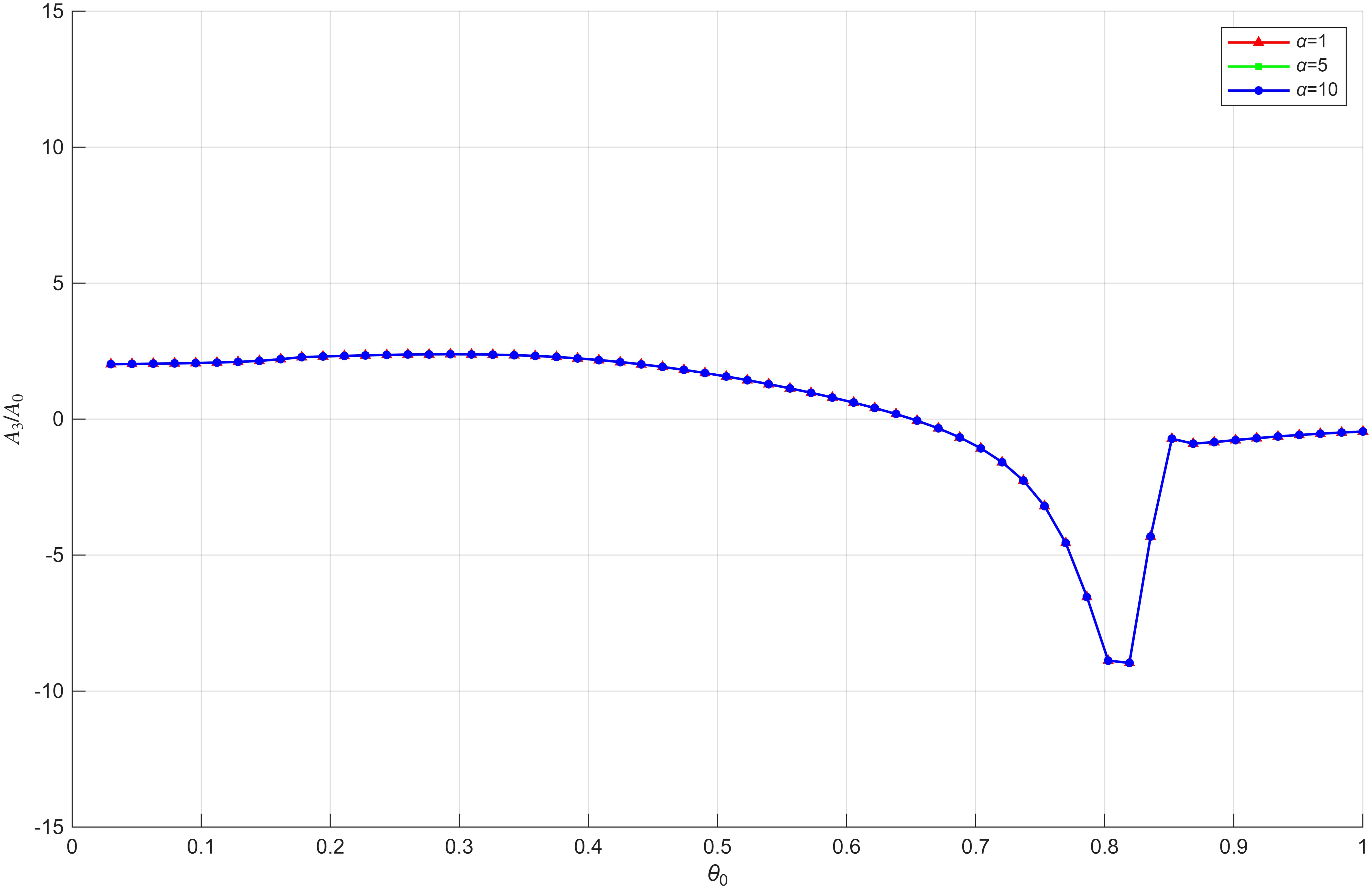}
\caption{}
\label{7c}
\end{subfigure}
\hfill
\begin{subfigure}[b]{0.47\textwidth}
\includegraphics[width=\textwidth]{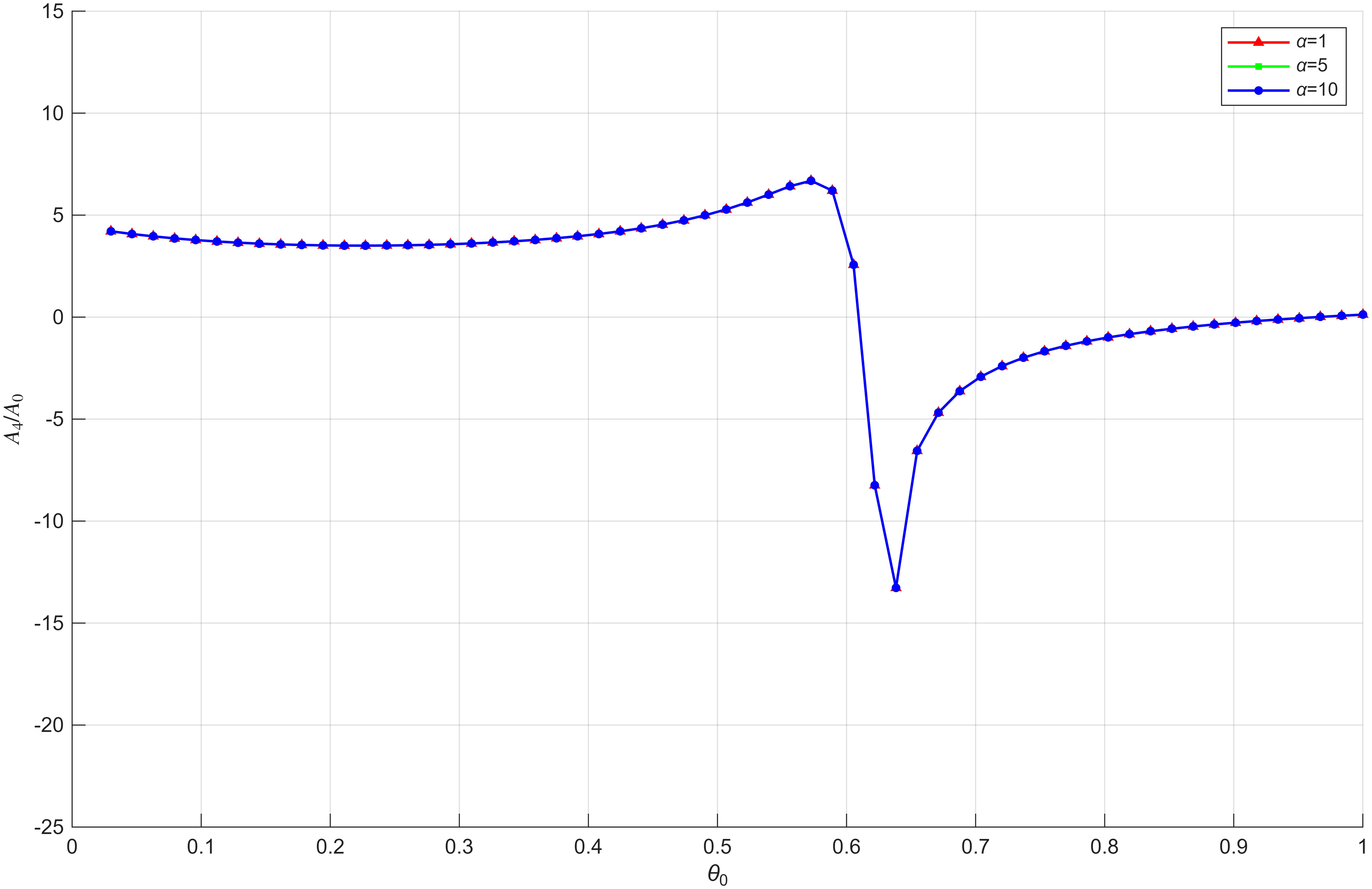}
\caption{}
\label{7d}
\end{subfigure}

\caption{Amplitude ratio versus incidence angle $\theta_0$ for varying gradient parameters in FGPM media }
\label{Figure 7}
\end{figure}
\begin{figure}[htbp]
\centering

\begin{subfigure}[b]{0.47\textwidth}
\includegraphics[width=\textwidth]{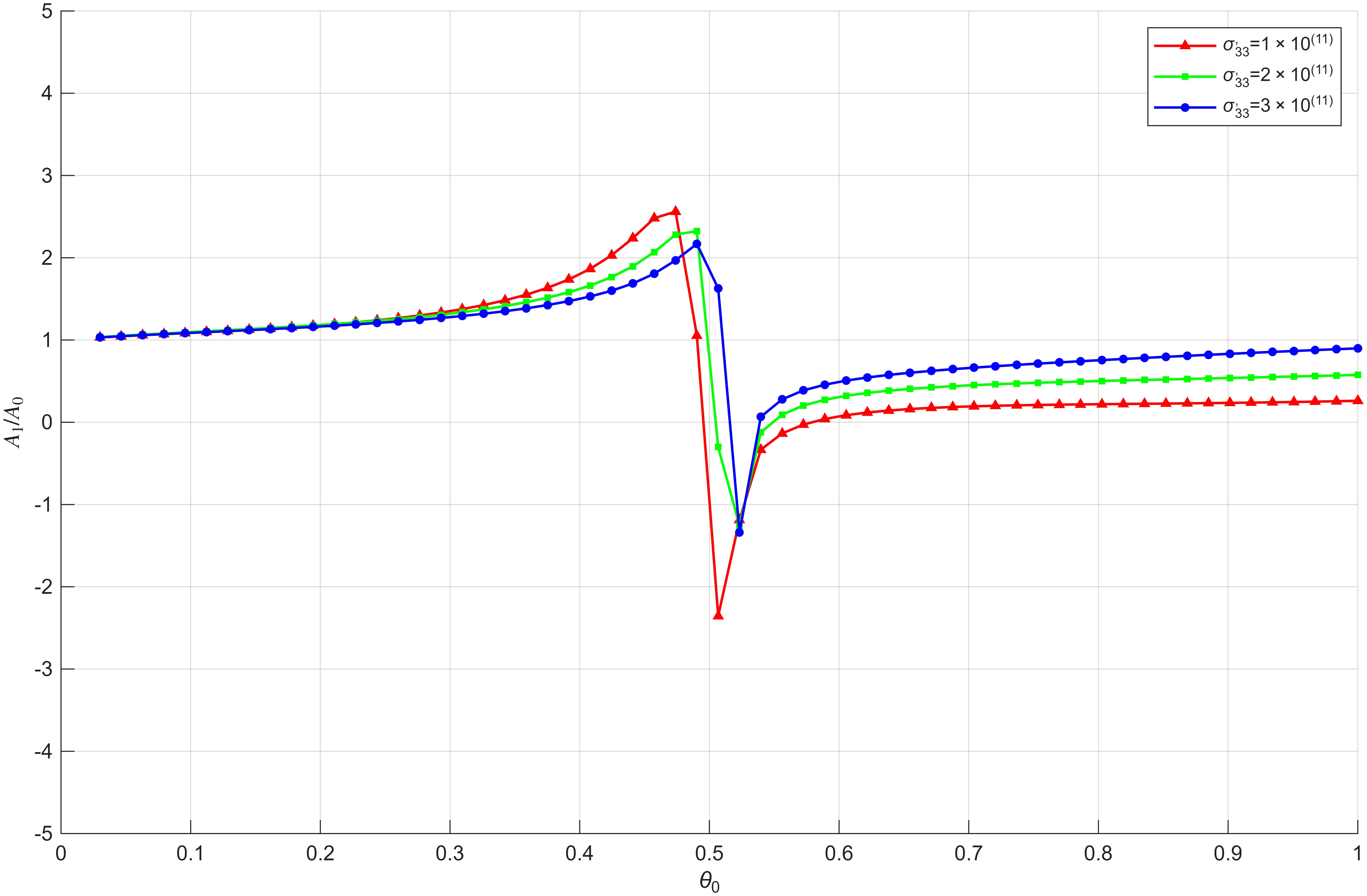}
\caption{}
\label{8a}
\end{subfigure}
\hfill
\begin{subfigure}[b]{0.47\textwidth}
\includegraphics[width=\textwidth]{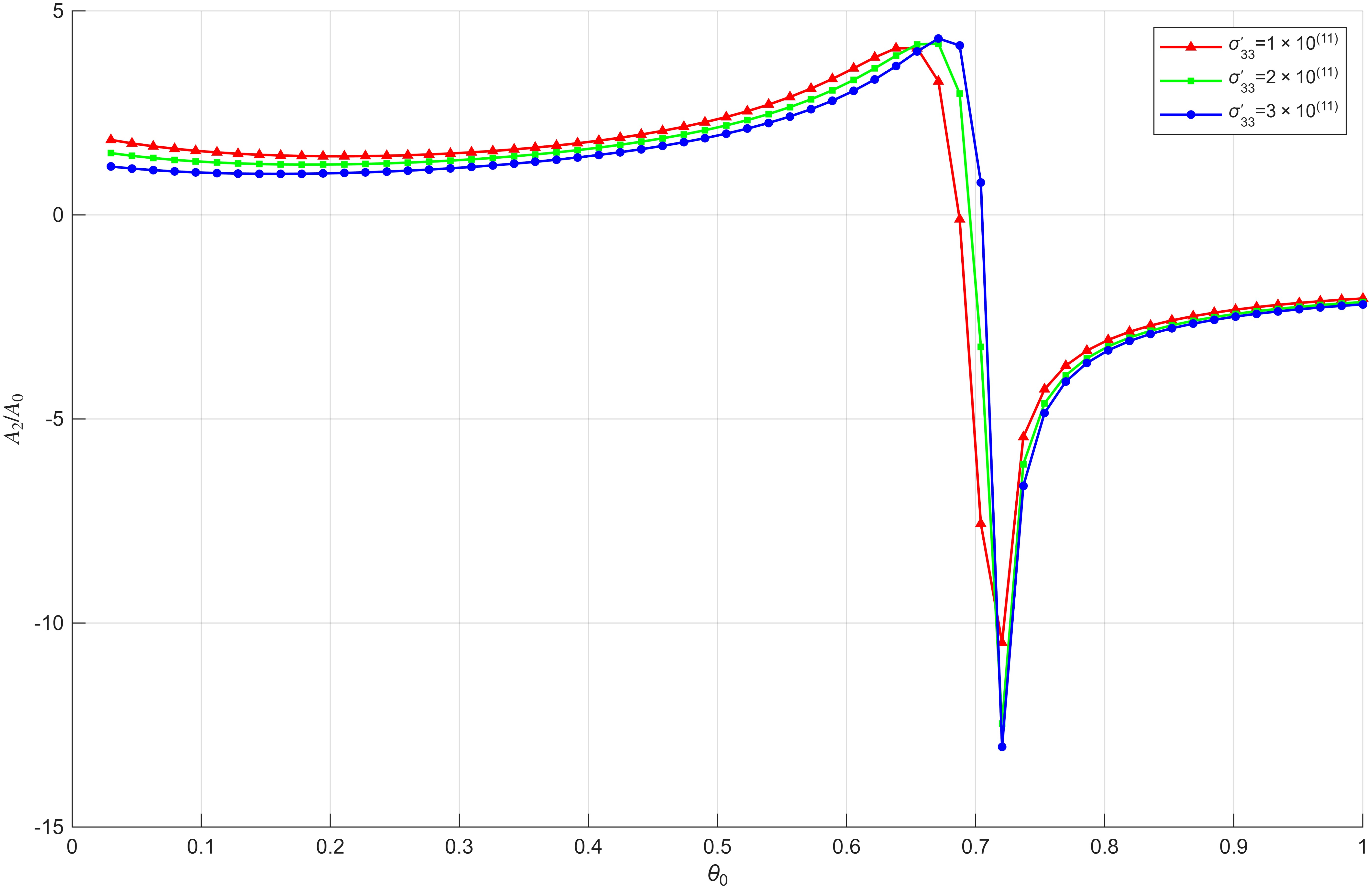}
\caption{}
\label{8b}
\end{subfigure}

\medskip

\begin{subfigure}[b]{0.47\textwidth}
\includegraphics[width=\textwidth]{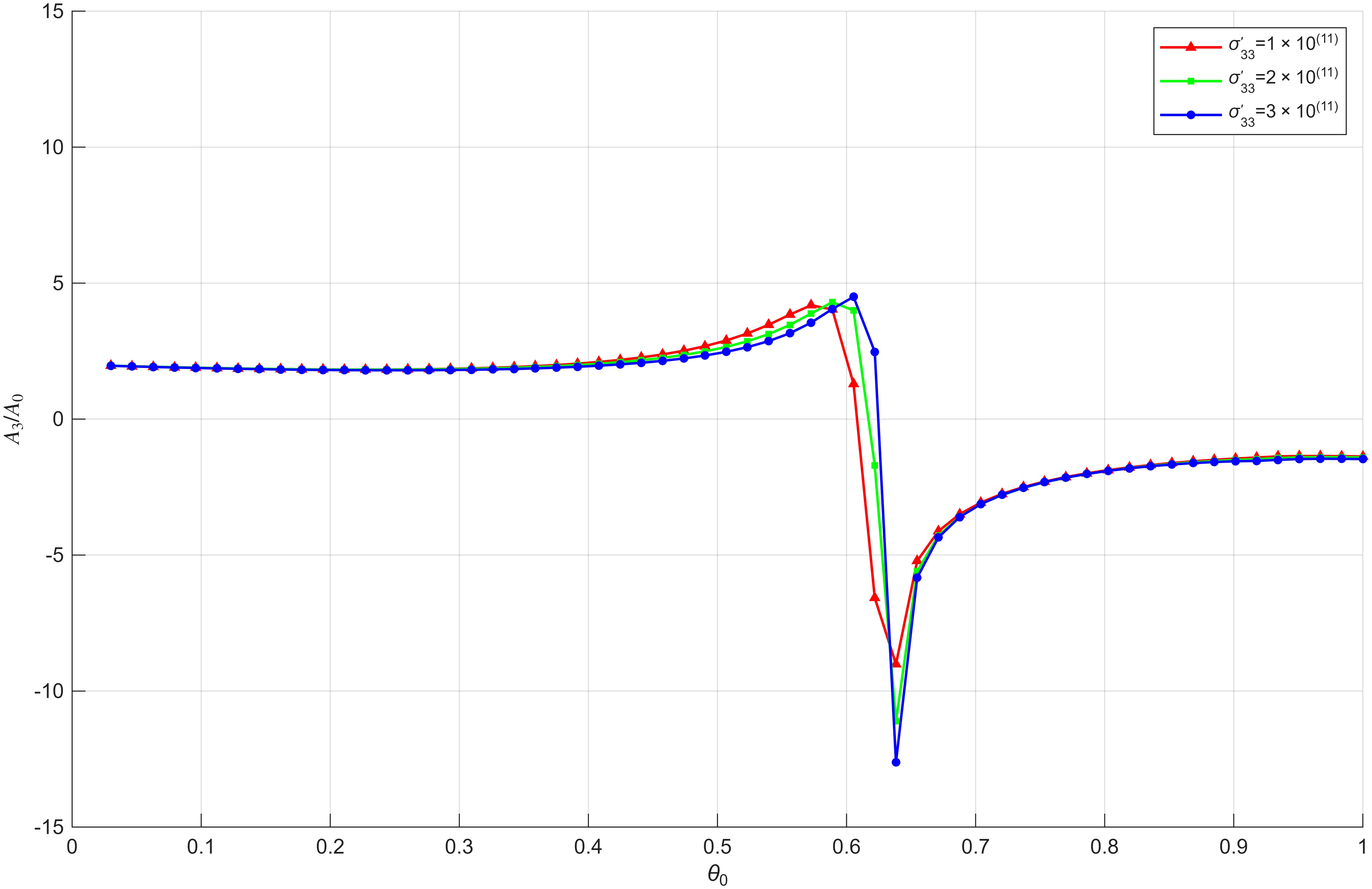}
\caption{}
\label{8c}
\end{subfigure}
\hfill
\begin{subfigure}[b]{0.47\textwidth}
\includegraphics[width=\textwidth]{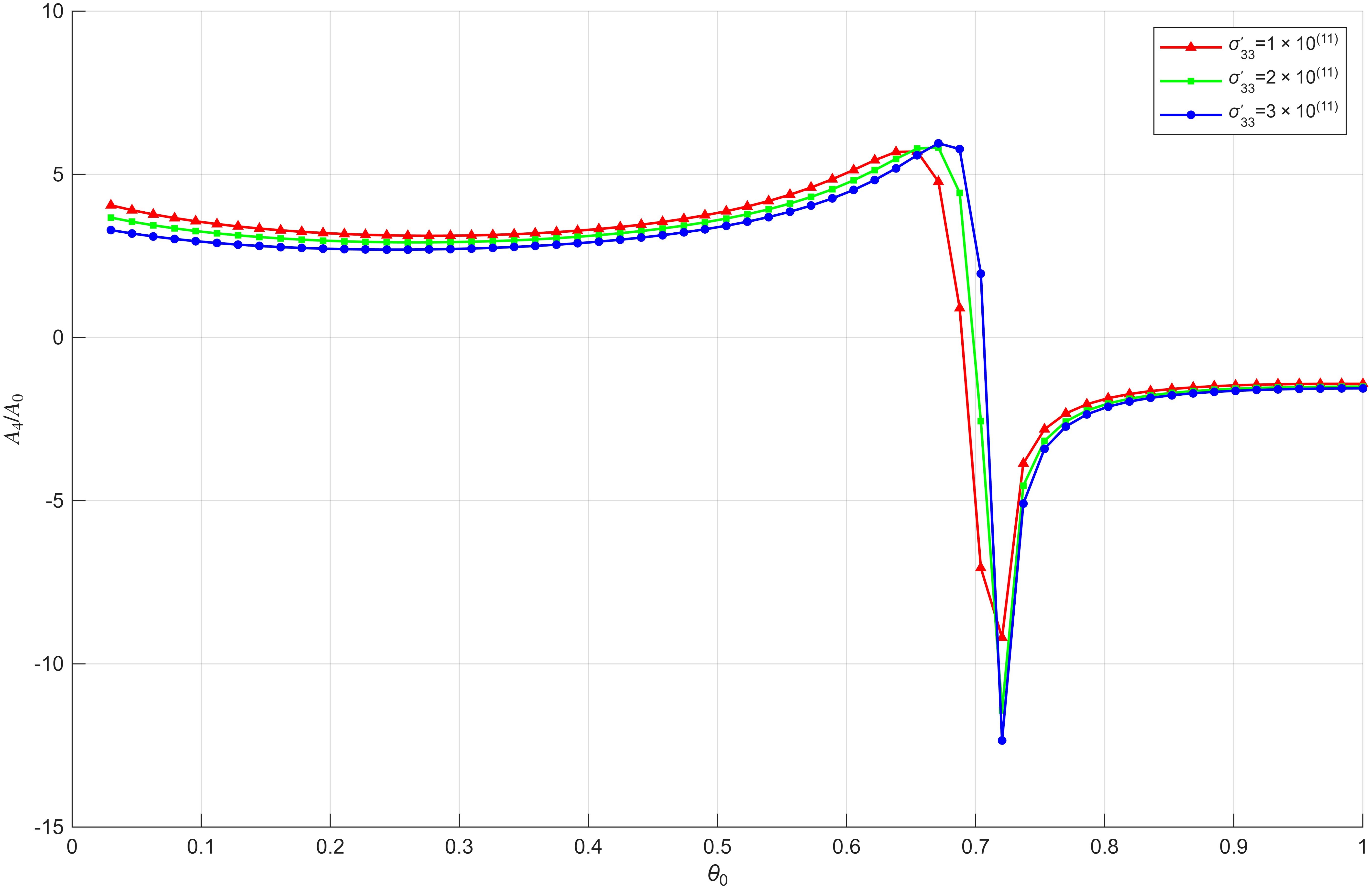}
\caption{}
\label{8d}
\end{subfigure}

\caption{Amplitude ratio versus incidence angle $\theta_0$ for varying initial stress in the piezoelectric media}
\label{Figure 8}
\end{figure}

Figure \ref{Figure 7} depicts the variation of amplitude ratio as a function of the angle of incidence $\theta_0$ under different material gradient when $\sigma_{33}=1\times10^{11}$ and $\Omega^\prime=1$.  The curves for different values of the material gradient $\alpha$ essentially overlap, indicating that $\alpha$ has negligible influence on the behavior of $A_1$ in this configuration. Figure \ref{7a},\ref{7b} and \ref{7d} exhibit almost the same qualitative behavior, showing a smooth rise at smaller values of incidence angle $\theta_0$, followed by a sharp peak and a subsequent deep dip at nearly the same critical point, after which the curves gradually approach a constant value. In figure \ref{7c}, the refraction coefficient $\frac{A_3}{A_0}$ is positive for small value of $\theta_0$, gradually decreasing and reached its minimum value at $\theta_0=0.82$. After attaining its minimum value the curve starts increasing and becomes constant for $\theta_0>0.85$. Figure \ref{Figure 8} depicts how amplitude ratio vary with $\theta_0$ under the influence of initial stress in the thermo-piezoelectric media when $\alpha=1$ and $\Omega^\prime=1$. It is evident that all the graphs exhibit an almost identical pattern. The curves remains positive for small values of angle of incidence $\theta_0$, exhibit a gradual increase attained its maximum values and then showed a sharp and immediate decline and reached its minimum value. After reaching its minimum value, the curve exhibits a slight increase before eventually approaching a constant value.

\begin{figure}[htbp]
\centering

\begin{subfigure}[b]{0.47\textwidth}
\includegraphics[width=\textwidth]{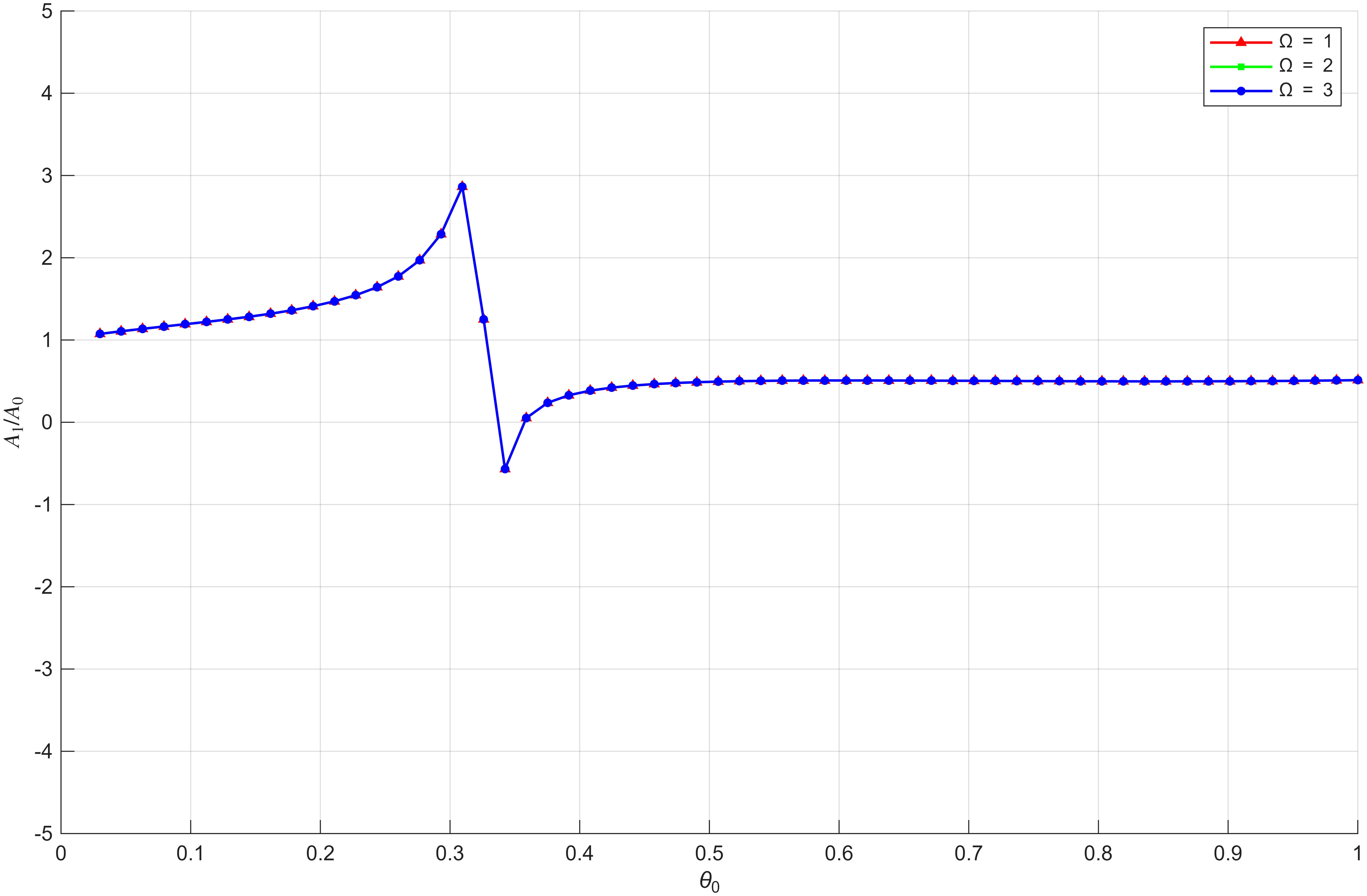}
\caption{}
\label{9a}
\end{subfigure}
\hfill
\begin{subfigure}[b]{0.47\textwidth}
\includegraphics[width=\textwidth]{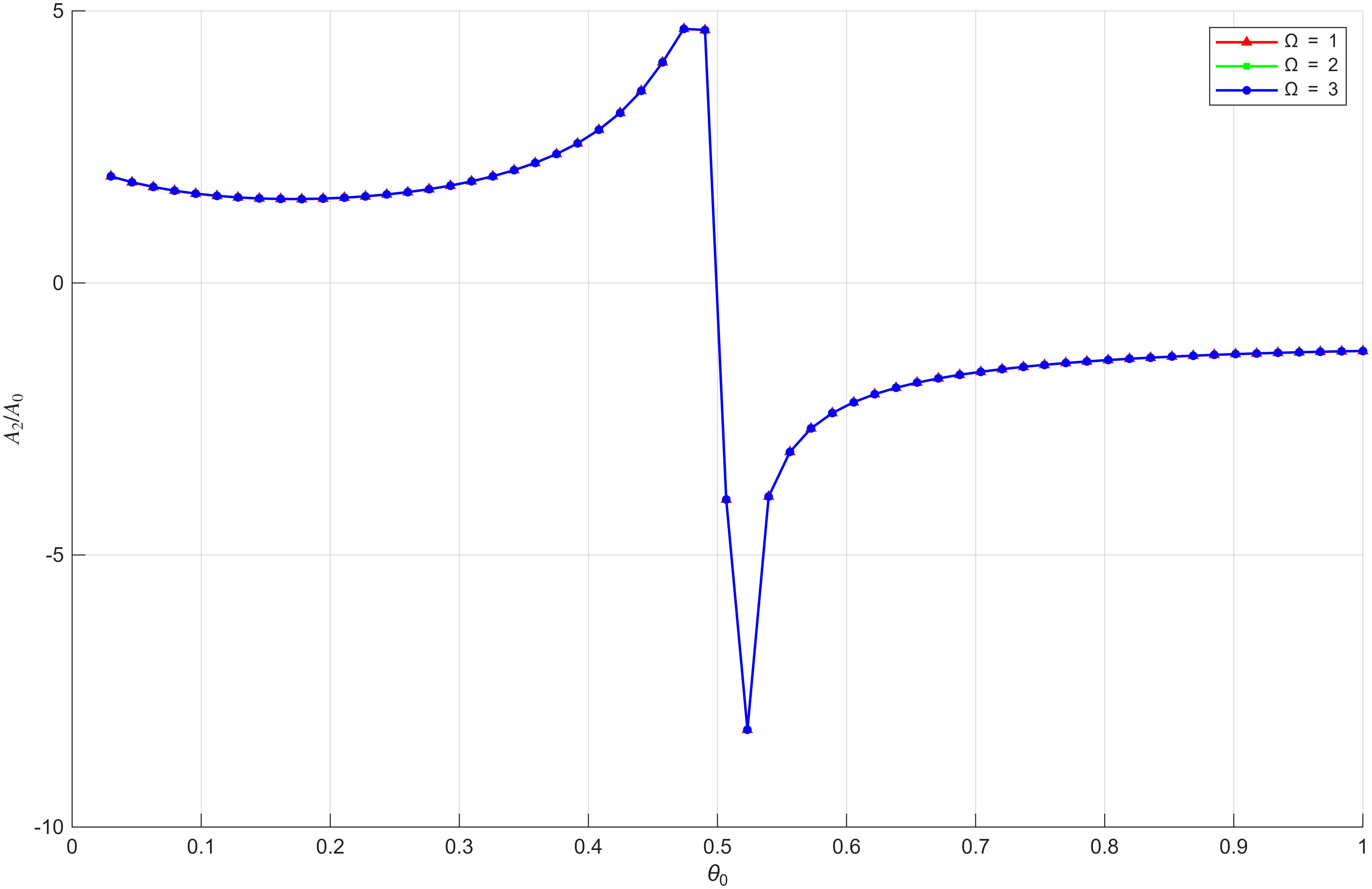}
\caption{}
\label{9b}
\end{subfigure}

\medskip

\begin{subfigure}[b]{0.47\textwidth}
\includegraphics[width=\textwidth]{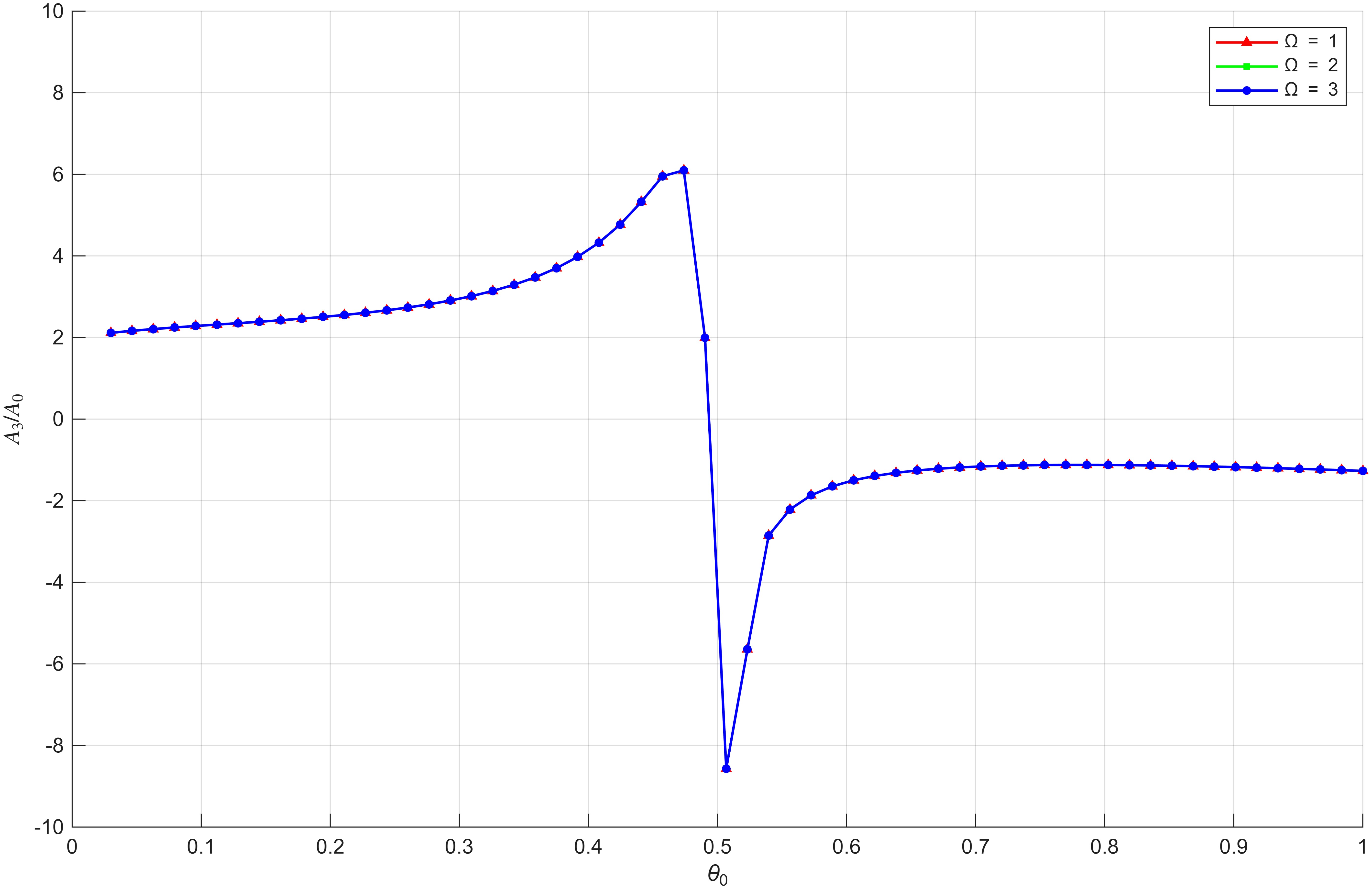}
\caption{}
\label{9c}
\end{subfigure}
\hfill
\begin{subfigure}[b]{0.47\textwidth}
\includegraphics[width=\textwidth]{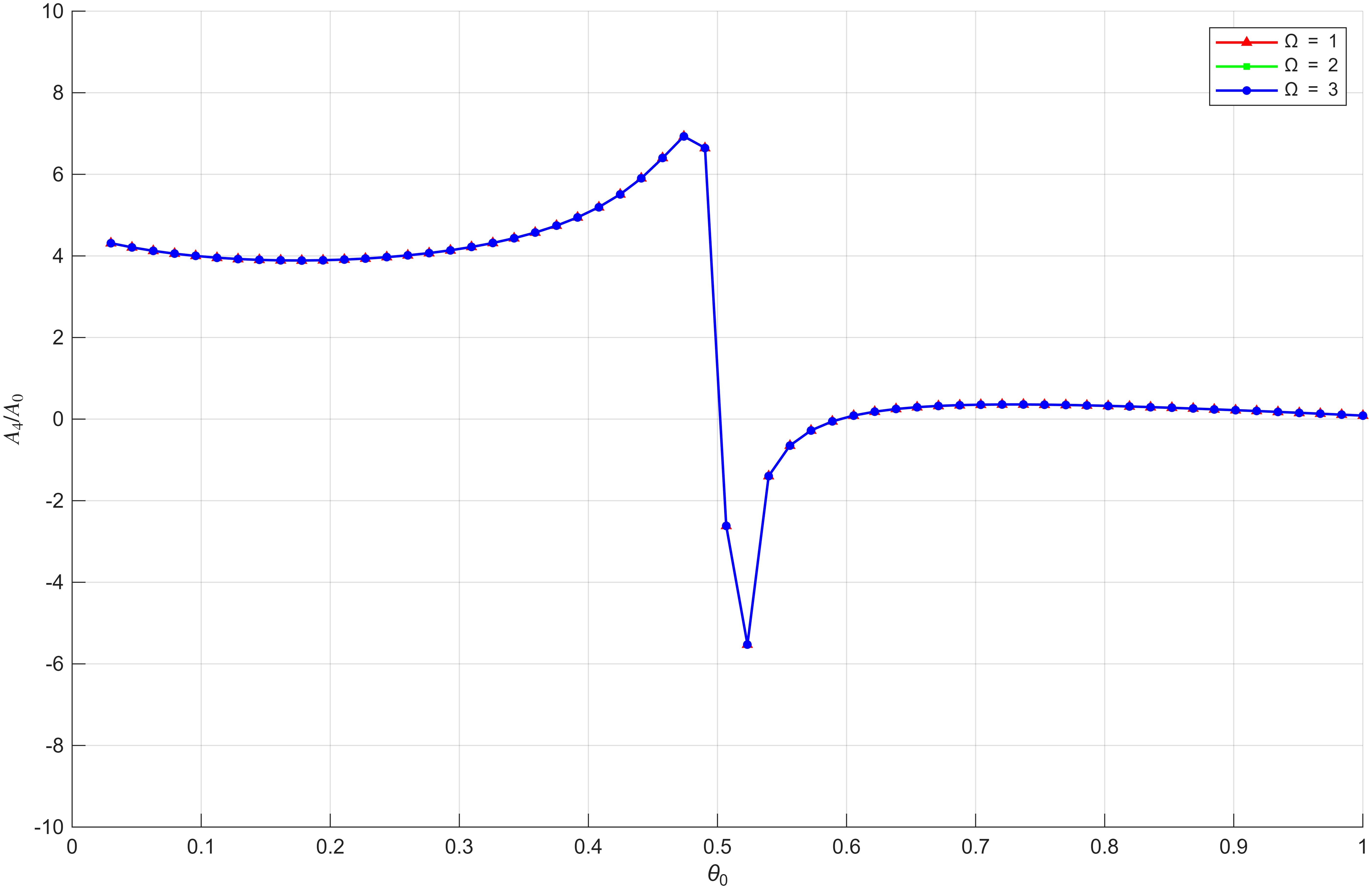}
\caption{}
\label{9d}
\end{subfigure}

\caption{Amplitude ratio versus incidence angle $\theta_0$ for varying rotation parameter in the piezoelectric media}
\label{Figure 9}
\end{figure}

\begin{figure}[htbp]
\centering

\begin{subfigure}[b]{0.47\textwidth}
\includegraphics[width=\textwidth]{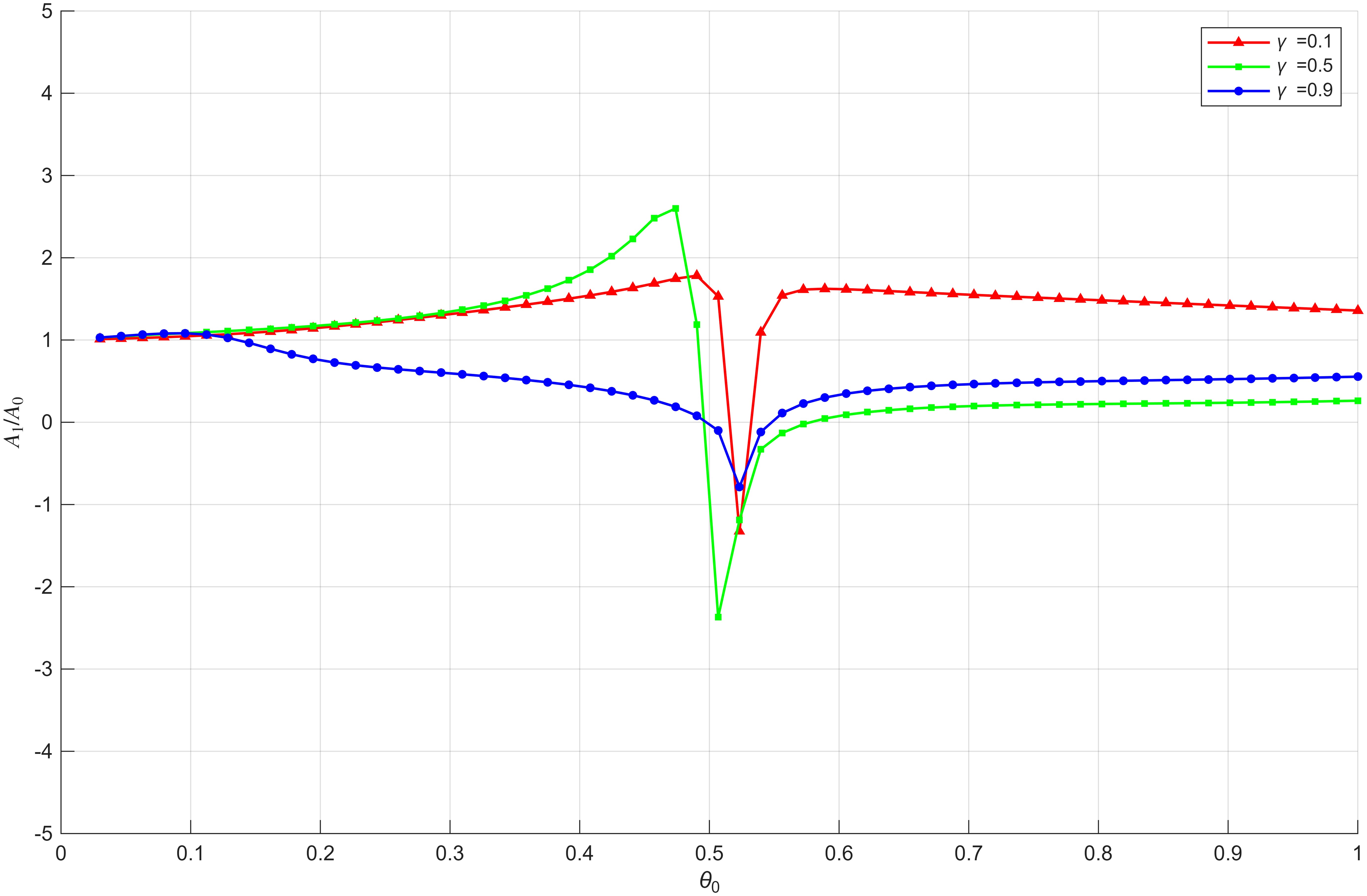}
\caption{}
\label{10a}
\end{subfigure}
\hfill
\begin{subfigure}[b]{0.47\textwidth}
\includegraphics[width=\textwidth]{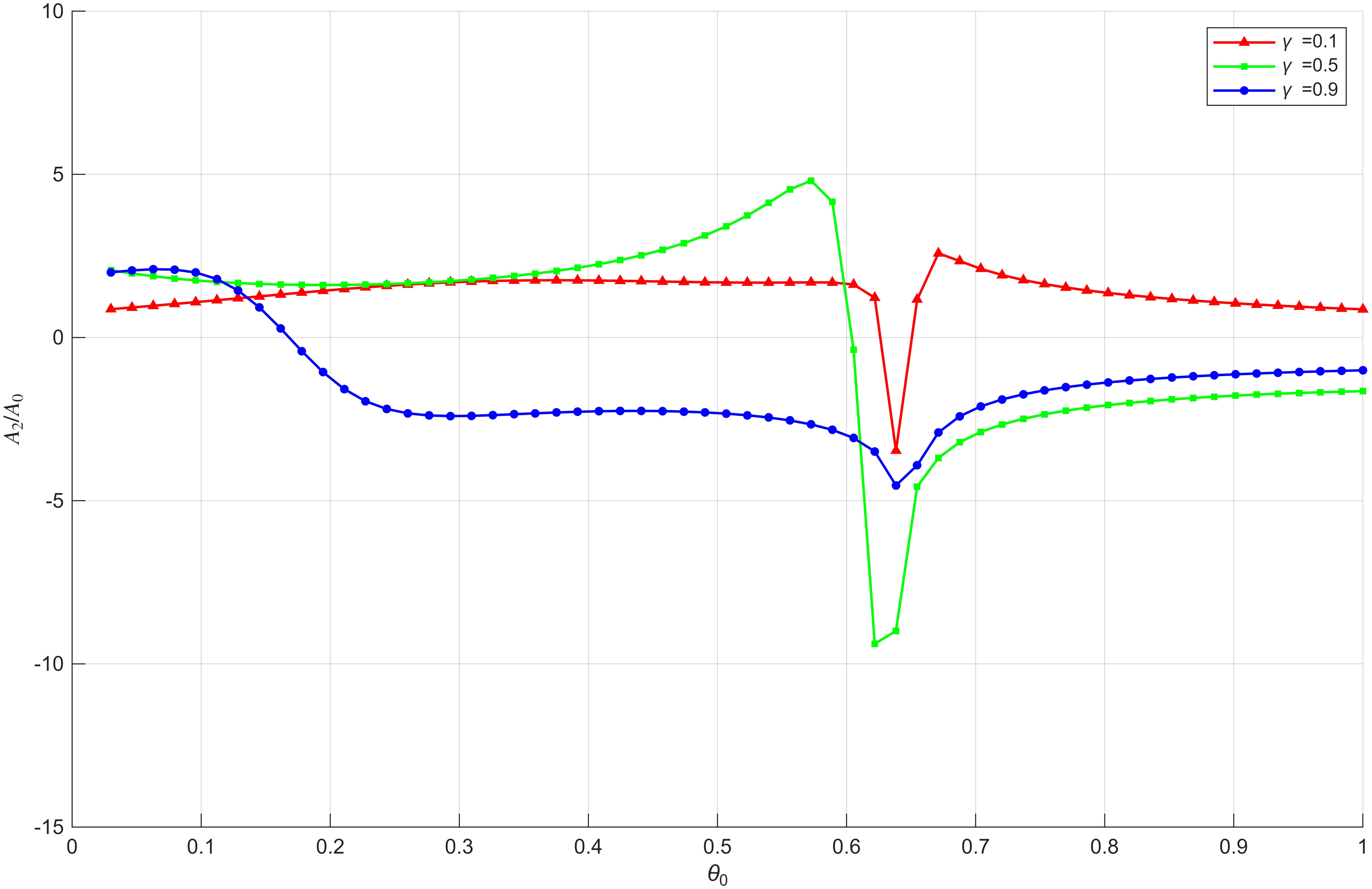}
\caption{}
\label{10b}
\end{subfigure}

\medskip

\begin{subfigure}[b]{0.47\textwidth}
\includegraphics[width=\textwidth]{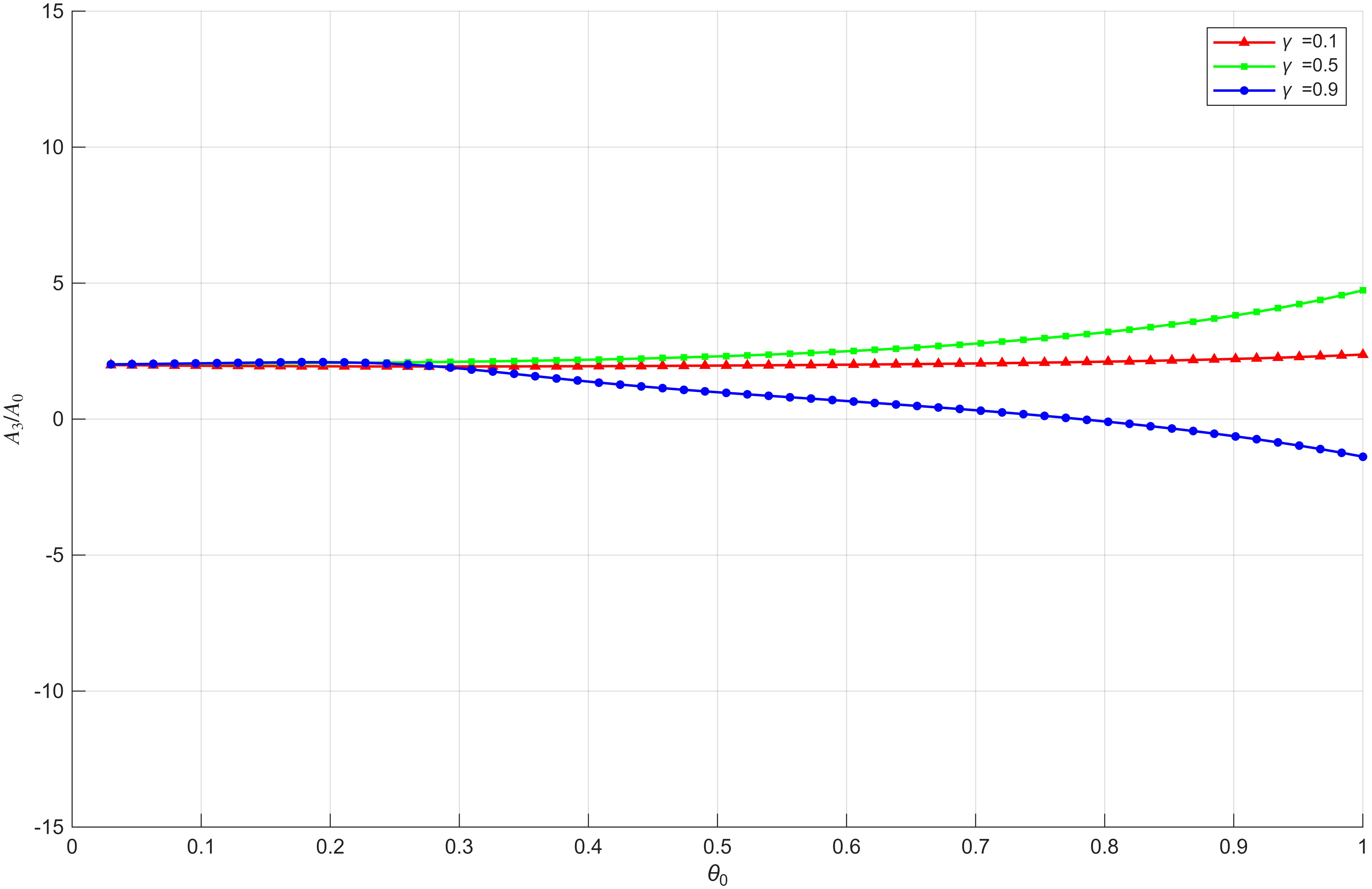}
\caption{}
\label{10c}
\end{subfigure}
\hfill
\begin{subfigure}[b]{0.47\textwidth}
\includegraphics[width=\textwidth]{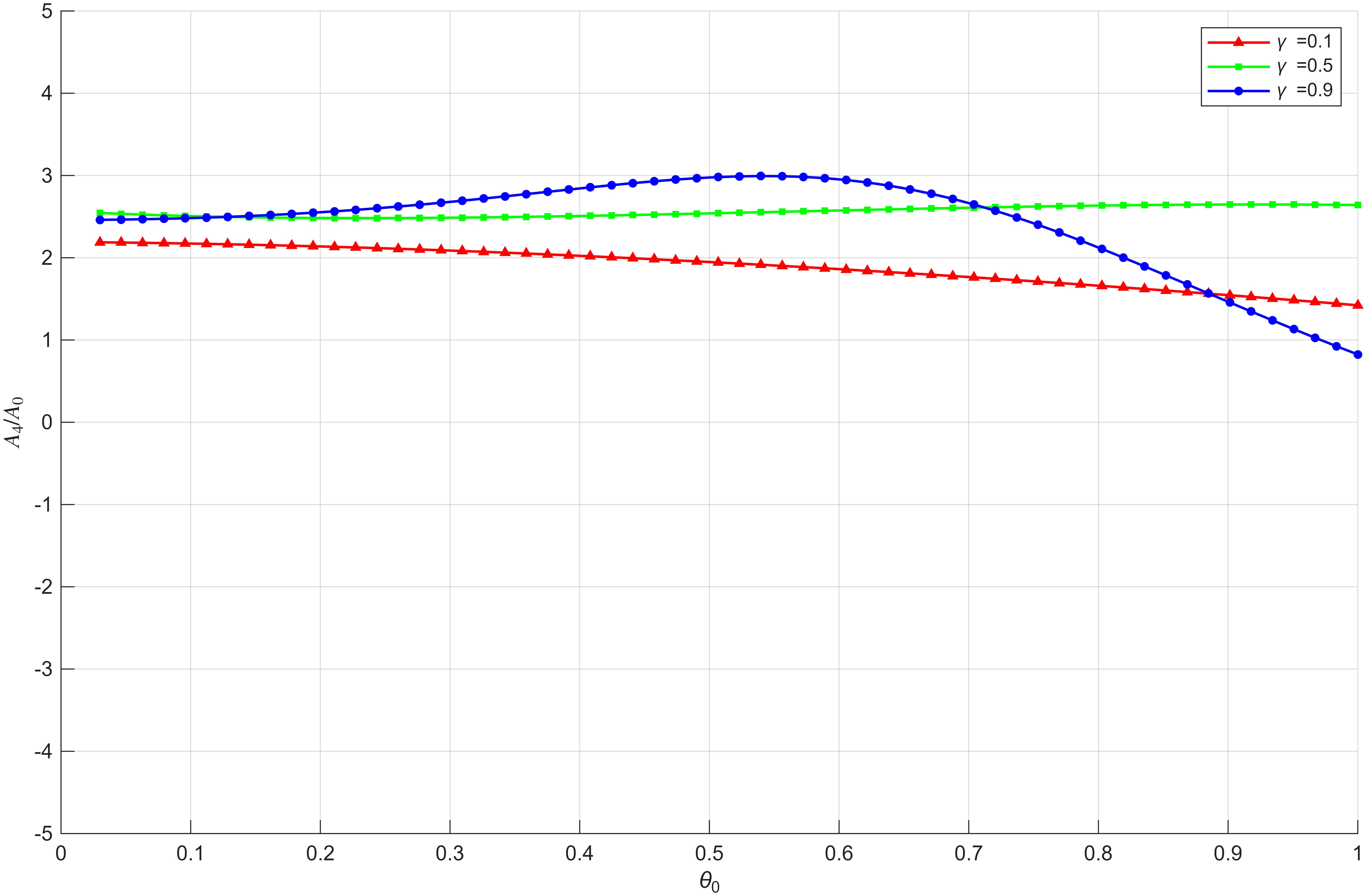}
\caption{}
\label{10d}
\end{subfigure}

\caption{Amplitude ratio versus incidence angle $\theta_0$ for varying order of fractional derivative in the thermo-piezoelectric media}
\label{Figure 10}
\end{figure}

Figure \ref{Figure 9} shows how amplitude ratio vary with respect to the angle of incidence under different values of rotation parameter when $\sigma_{33}=1\times10^{11}$ and $\Omega=1$. All the four graphs from \ref{9a}-\ref{9d} exhibit almost the same quantitative behavior showing a smooth rise at smaller values of
$\theta_0$,followed by a sharp peak and a subsequent deep dip at nearly the same critical point, after which the curves gradually approach a constant value. The most notable observation is that the curves corresponding to different values of $\Omega^\prime$ almost completely overlap in every sub-figure, indicating that the parameter $\Omega^\prime$ has very little influence on the overall pattern of the response. As a result, each figure displays the same characteristic peak–dip–flattening trend, and the overlap of the curves confirms that the system’s behavior remains essentially unchanged for the considered variations in $\Omega^\prime$. Figure \ref{Figure 10} shows how amplitude ratio vary with the angle of incidence for distinct values of order of fractional derivative. In figure \ref{10a}, for $\theta_0<0.45$, the value of $\gamma$ remains positive. For $\gamma=0.1$ and $\gamma=0.5$ the value of reflection coefficient $\frac{A_1}{A_0}$ increases slowly and after reaching $\theta_0=0.48$ it started decreasing and reaching its minimum value at $\theta_0=0.5$ while the curve for $\gamma=0.9$ started decreasing slowly from the beginning and reaching its minimum value at $\theta_0=0.53$. The curve for all the values of $\gamma$ starts increasing and ultimately becomes constant. Figure \ref{10b} exhibits the same overall behavior as figure \ref{10a}. For $\gamma=0.9$, the curve starts decreasing gradually. However, for $\gamma=0.1$ and $\gamma=0.5$, the value initially increases slowly, then drops sharply, reaching its minimum value at $\theta_0=0.63$ after which it rises abruptly and eventually settles to a constant value. In figure \ref{10c}, for $\gamma=0.9$, the amplitude begins with a positive value but decreases steadily as $\theta_0$ increases, ultimately becoming strongly negative. For $\gamma=0.5$, the amplitude ratio $\frac{A_3}{A_0}$ 

\begin{figure}[htbp]
\centering

\begin{subfigure}[b]{0.47\textwidth}
\includegraphics[width=\textwidth]{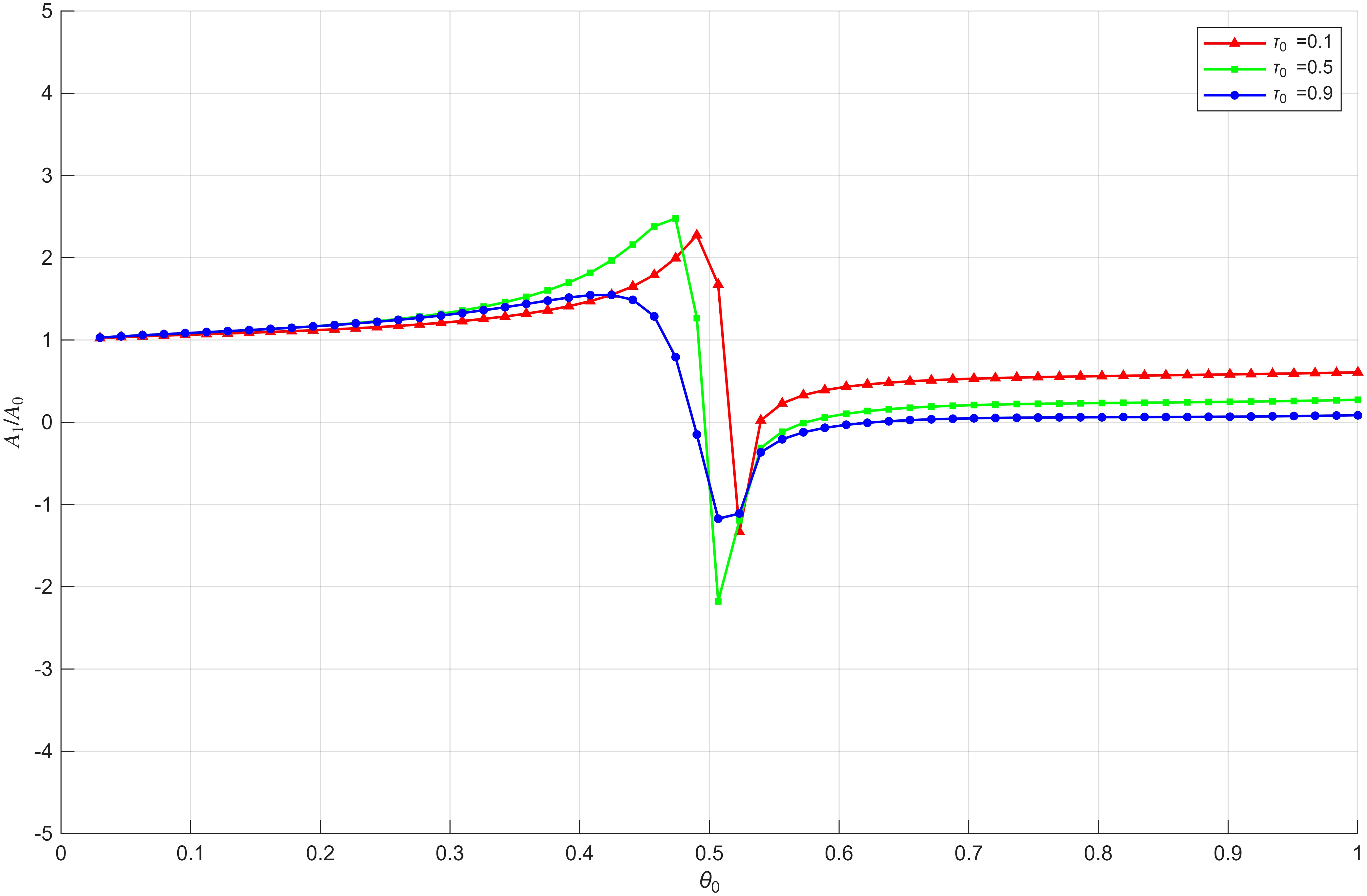}
\caption{}
\label{11a}
\end{subfigure}
\hfill
\begin{subfigure}[b]{0.47\textwidth}
\includegraphics[width=\textwidth]{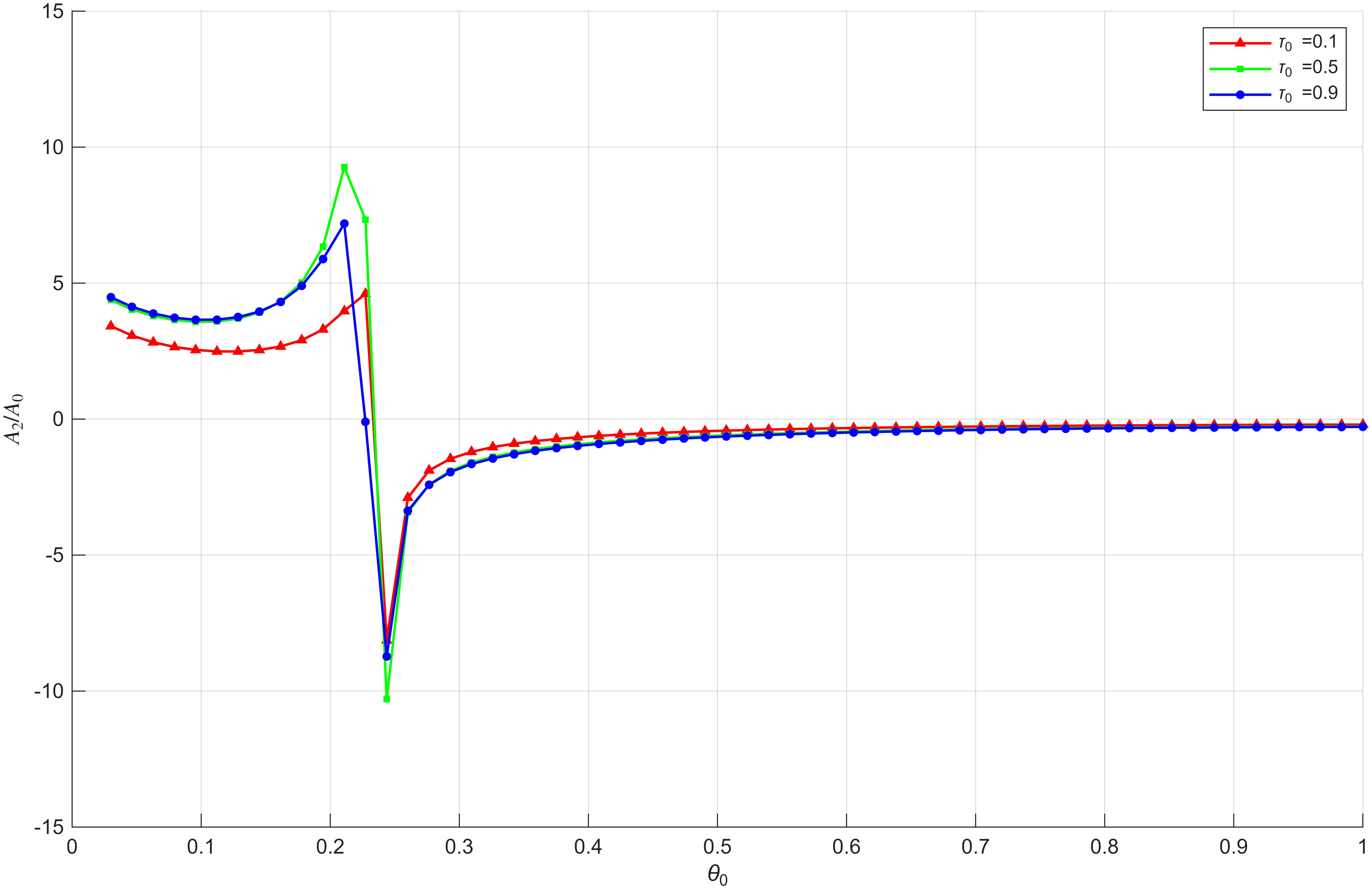}
\caption{}
\label{11b}
\end{subfigure}

\medskip

\begin{subfigure}[b]{0.47\textwidth}
\includegraphics[width=\textwidth]{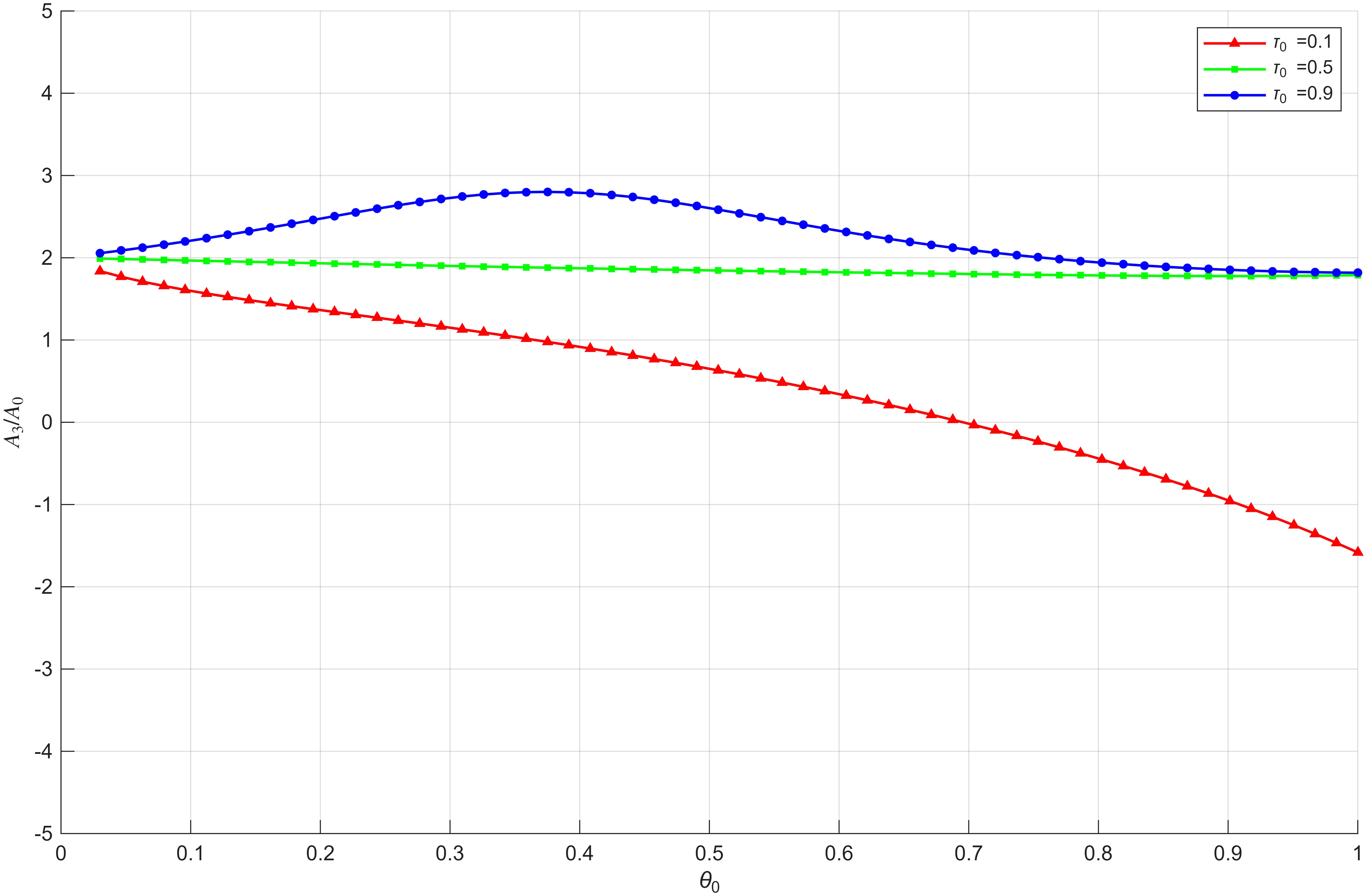}
\caption{}
\label{11c}
\end{subfigure}
\hfill
\begin{subfigure}[b]{0.47\textwidth}
\includegraphics[width=\textwidth]{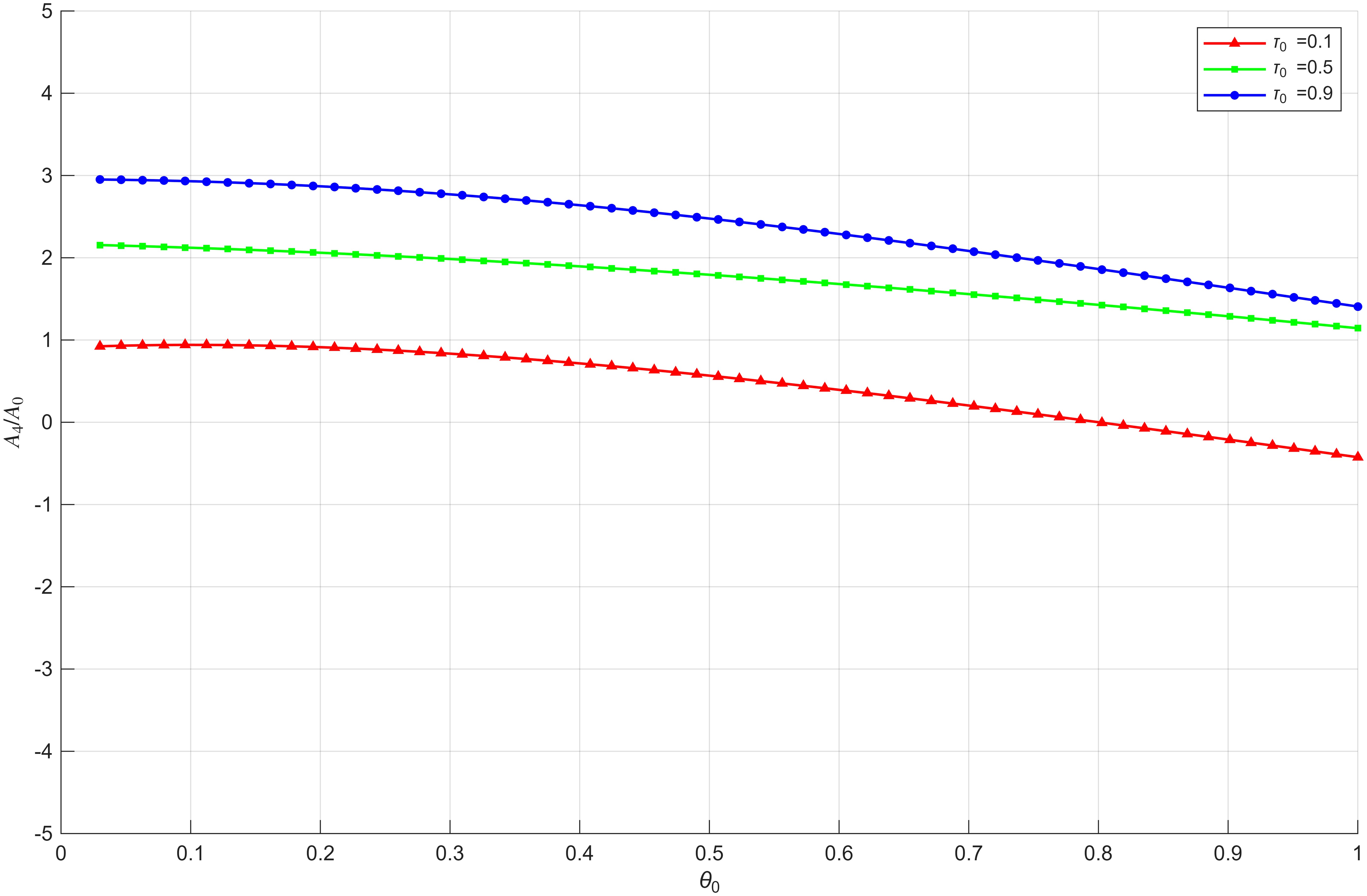}
\caption{}
\label{11d}
\end{subfigure}

\caption{Amplitude ratio versus incidence angle $\theta_0$ for varying relaxation time in the thermo- piezoelectric media}
\label{Figure 11}
\end{figure}

initially remains nearly constant but soon begins to rise gradually and then more sharply with increasing $\theta_0$. In contrast, for $\theta_0=0.1$, the amplitude ratio remains almost unchanged throughout, with only a very slight upward trend. In figure \ref{10d}, for $\gamma=0.1$, the amplitude decreases slowly and almost linearly as $\theta_0$ increases, indicating a mild attenuation effect. The case $\gamma=0.5$, remains nearly constant throughout the entire range of $\theta_0$, showing that this fractional order produces almost no change in the amplitude behavior. For $\gamma=0.9$, the amplitude ratio initially increases and reaches its maximum around $\theta_0=0.5$, after which it decreases sharply, ultimately crossing below zero at higher $\theta_0$. This demonstrates that increasing $\gamma$ enhances the amplitude in the beginning but induces significant attenuation for larger $\theta_0$. Figure  \ref{Figure 11} illustrates how the amplitude ratio vary with angle of incidence for different values of relaxation time $(\tau_0)$. In figure\ref{11a}, we can see that for small incidence angle, all the three curves remain close to unity, indicating that $\tau_0$ have only a mild influence on the reflection coefficient $\frac{A_1}{A_0}$, when the wave approaches the interface nearly normally. As $\theta_0$ increases, each curve exhibits a pronounced peak followed immediately by a sharp negative dip near $\theta_0=0.5$. After the resonance region, the amplitude ratio gradually stabilizes and becomes nearly constant for larger values of $\theta_0$. In figure \ref{11b}, amplitude ratio is positive for small values of $\theta_0$. With the increasing angle of incidence, the curve starts increasing and reaches its maximum value at $\theta_0=0.22$. After reaching its maximum value, the curve suddenly starts decreasing and reaches to it minimum value at $\theta_0=0.24$. After that the curve increases a little bit before becoming constant for higher values of $\theta_0$. The variation of the amplitude ratio $\frac{A_3}{A_0}$ with respect to the angle of incident $\theta_0$ is shown in \ref{11c} for different values of relaxation 
time. For $\tau_0=0.1$, the amplitude ratio decreases almost monotonically with increasing incident angle. Starting from the positive value, the ratio gradually becomes negative for higher angles. For $\tau_0=0.5$, the amplitude ratio remains nearly constant throughout the angle range, fluctuating very slightly around a value close to 1.8. For $\tau_0=0.9$, the amplitude ratio exhibits a distinct non-monotonic variation. The curve rises to a maximum value of about 2.8 around $\theta_0=0.38$, after which it gradually decreases for larger angles. In figure \ref{11d}, for $\tau_0=0.1$, the amplitude ratio initially increases slightly, reaching a peak close to $\theta_0=0.15$. Beyond this angle, the curve decreases steadily and becomes significantly lower for larger values of $\theta_0$. For $\tau_0=0.5$, the amplitude ratio maintains a nearly linear decreasing trend. The value change only moderately. This shows that at $\tau_0=0.5$, the moderate relaxation time, the refracted amplitude becomes less sensitive to angle and exhibits smooth and predictable behavior. For $\tau_0=0.9$, the amplitude ratio starts at the highest value and gradually decreases with the increasing incident angle. Although the reduction is steady, the amplitude remains substantially larger than in the other two cases throughout the entire angle range. This highlights the significant influence of relaxation time on the transmission characteristics at the interface.

\begin{figure}[htbp]
\centering

\begin{subfigure}[b]{0.47\textwidth}
\includegraphics[width=\textwidth]{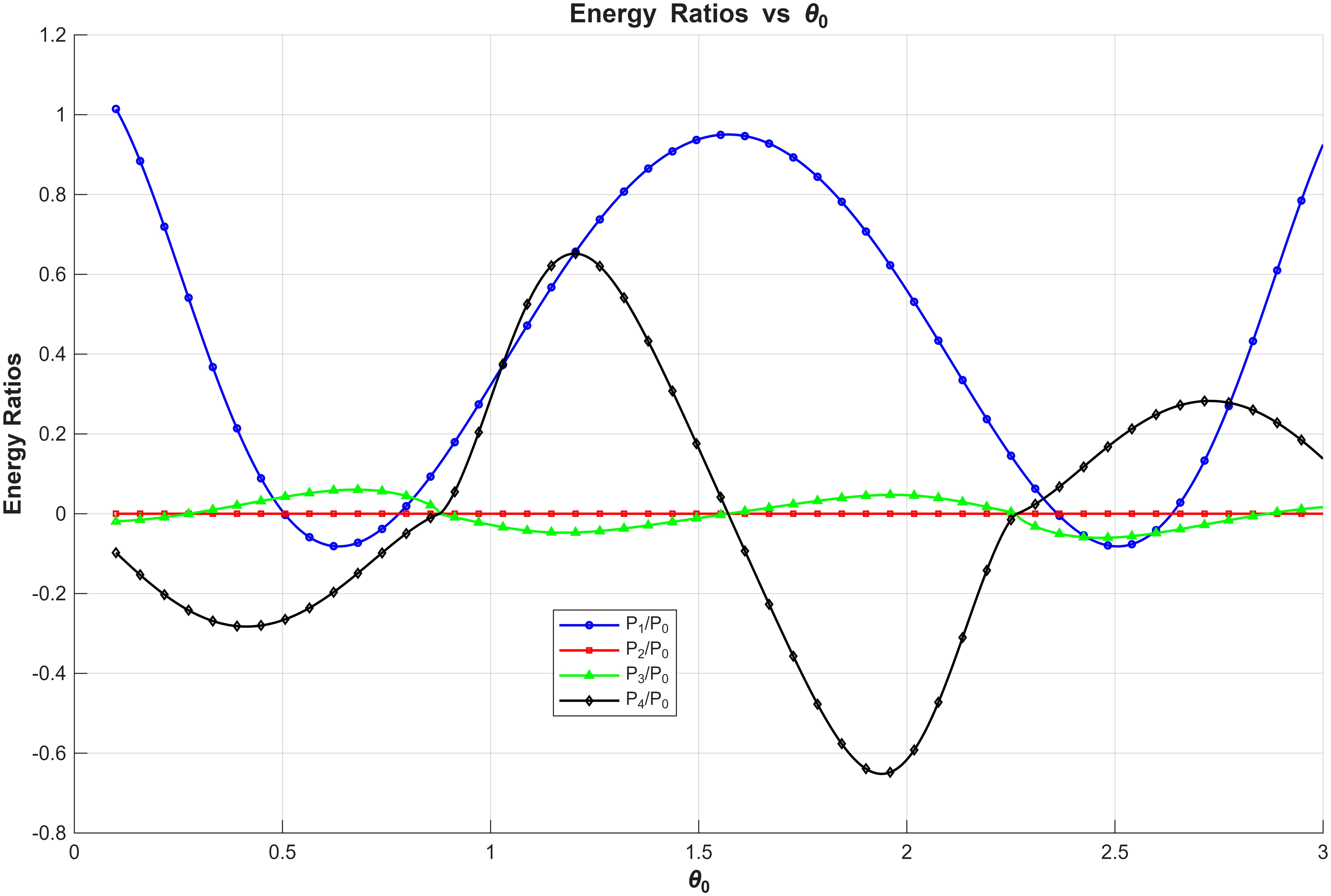}
\caption{}
\label{12a}
\end{subfigure}
\hfill
\begin{subfigure}[b]{0.47\textwidth}
\includegraphics[width=\textwidth]{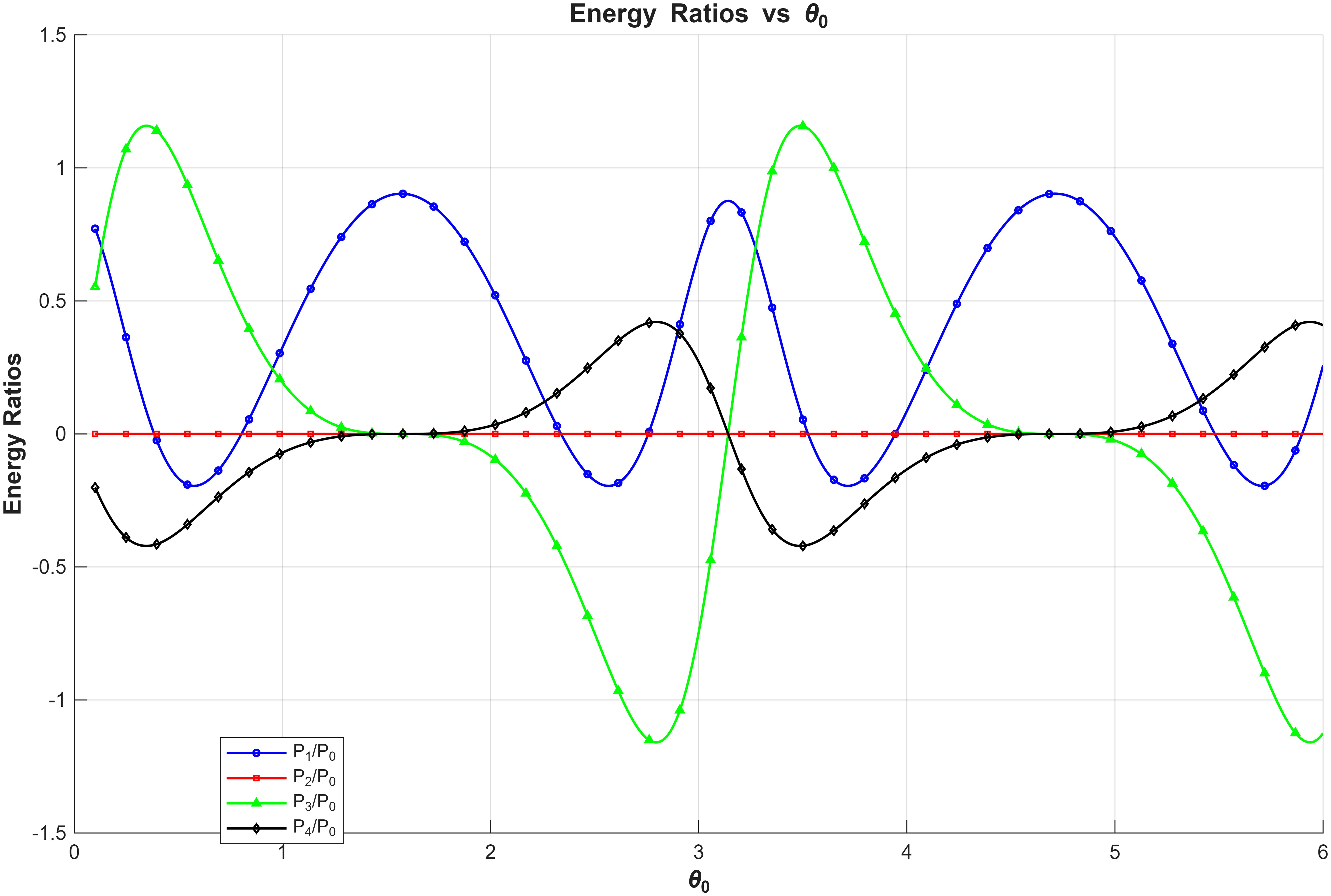}
\caption{}
\label{12b}
\end{subfigure}
\caption{Energy ratios versus incident angle}

\label{Figure 12}
\end{figure}
Figure \ref{12a} and \ref{12b} shows the variation of energy ratio with the incident angle for case 1 and case 2. From the curves in \ref{12a} and \ref{12b}, the law of energy conservation is verified for distinct angles of incidence, as the sum of the energy ratios remains approximately equal to unity.
\section{Conclusion}
\label{07}
The scattering behavior of quasi-longitudinal (qP) waves at the interface between FGPM and thermo-piezoelectric half-spaces is analyzed based on LS theory. Two scenarios were examined: in Case 1, the normal and shear stresses, continuity of electric potential, and thermal insulated boundary conditions were applied; in Case 2, continuity of displacement, shear stress, continuity of electric potential, and thermal insulated boundary conditions were enforced. An analytical approach was adopted to evaluate the amplitude and energy ratios of the scattered waves. The obtained results verify that the evaluated energy ratios are consistent with principle of conservation of energy. The main observations of the present work are outlined below:
 \begin{itemize}
     \item The effects of rotation and material gradation are negligible in both cases considered. This conclusion is corroborated by the complete overlap of the corresponding energy ratio curves for different values of these parameters, indicating that they exert no significant influence on the wave energy partitioning.
     \item In contrast, the effects of initial stress, relaxation time, and the fractional-order derivative are significant, indicating that these parameters strongly affect wave propagation characteristics and energy distribution in the medium.
     \item  The fractional-order parameter $(\gamma)$ influences the amplitude ratios in both cases, reflecting the material’s memory and hereditary effects, which alter the wave propagation behavior and energy distribution across the interface.
     \item The numerical results demonstrate that the total energy ratios of the reflected and transmitted waves remain nearly approximately to unity for all angles of incidence, confirming the law of energy conservation and there by validating the accuracy and consistency of the proposed mathematical model.
 \end{itemize}
 This study significantly enhances the understanding of wave propagation in FGPM and thermo-piezoelectric media in the presence of rotation and initial stress. The insights gained are relevant to real-world applications, such as seismic wave analysis, where accounting for initial stress and rotational effects is essential for accurate interpretation of subsurface structures. The findings have significant implications for applications in geophysics and engineering.\\
\noindent \textbf{Acknowledgement}\\
\noindent The authors thank the Department of Mathematics and Scientific Computing, NIT Hamirpur, for providing research facilities to Hriticka Dhiman during her Ph.D., and gratefully acknowledge for the research fellowship.

 \newpage
\bibliographystyle{unsrt}
\bibliography{References}

@book{balachandran2023introduction,
  title={An introduction to fractional differential equations},
  author={Balachandran, Krishnan},
  year={2023},
  publisher={Springer}
}

@article{sahu2019scattering,
  title={Scattering phenomenon of qP wave at the interface of FGPM and piezoelectric medium},
  author={Sahu, Sanjeev Anand and Chaudhary, Soniya and Paswan, Brijendra},
  journal={Waves in Random and Complex Media},
  volume={29},
  number={3},
  pages={435--455},
  year={2019},
  publisher={Taylor \& Francis}
}

@article{kumar2017effect,
  title={Effect of fractional order on energy ratios at the boundary surface of elastic-piezothermoelastic media},
  author={Kumar, Rajneesh and Sharma, Poonam},
  journal={Coupled systems mechanics},
  volume={6},
  number={2},
  pages={157--174},
  year={2017},
  publisher={Techno-Press}
}

@article{abd2014mathematical,
  title={The mathematical model of reflection and refraction of longitudinal waves in thermo-piezoelectric materials},
  author={Abd-alla, Abo-el-nour N and Hamdan, Abdelmonam M and Giorgio, Ivan and Del Vescovo, Dionisio},
  journal={Archive of Applied Mechanics},
  volume={84},
  number={9},
  pages={1229--1248},
  year={2014},
  publisher={Springer}
}

@article{abd2009reflection,
  title={Reflection and refraction of plane quasi-longitudinal waves at an interface of two piezoelectric media under initial stresses},
  author={Abd-alla, Abo-el-nour N and Alsheikh, Fatimah A},
  journal={Archive of Applied Mechanics},
  volume={79},
  number={9},
  pages={843--857},
  year={2009},
  publisher={Springer}
}

@article{abd2012reflection,
  title={The reflection phenomena of quasi-vertical transverse waves in piezoelectric medium under initial stresses},
  author={Abd-alla, Abo-el-nour N and Al-sheikh, Fatimah A and Al-Hossain, Abdullah Y},
  journal={Meccanica},
  volume={47},
  number={3},
  pages={731--744},
  year={2012},
  publisher={Springer}
}

@article{yuan2012reflection,
  title={Reflection and refraction of plane waves at interface between two piezoelectric media},
  author={Yuan, X and Zhu, ZH2996578},
  journal={Acta Mechanica},
  volume={223},
  number={12},
  pages={2509--2521},
  year={2012},
  publisher={Springer}
}

@article{fang2001surface,
  title={Surface acoustic waves propagating over a rotating piezoelectric half-space},
  author={Fang, Huiyu and Yang, Jiashi and Jiang, Qing},
  journal={IEEE transactions on ultrasonics, ferroelectrics, and frequency control},
  volume={48},
  number={4},
  pages={998--1004},
  year={2001},
  publisher={IEEE}
}

@article{yang2006review,
  title={A review of a few topics in piezoelectricity},
  author={Yang, Jiashi},
  journal={Applied Mechanics Review},
  year={2006}
}

@article{singh2011effect,
  title={Effect of initial stresses on incident qSV-waves in pre-stressed elastic half-spaces},
  author={Singh, SS},
  journal={The ANZIAM Journal},
  volume={52},
  number={4},
  pages={359--371},
  year={2011},
  publisher={Cambridge University Press}
}

@article{weis1985lithium,
  title={Lithium niobate: Summary of physical properties and crystal structure},
  author={Weis, RS and Gaylord, TK},
  journal={Applied Physics A},
  volume={37},
  number={4},
  pages={191--203},
  year={1985},
  publisher={Springer}
}

@article{pang2008reflection,
  title={Reflection and refraction of plane waves at the interface between piezoelectric and piezomagnetic media},
  author={Pang, Yu and Wang, Yue-Sheng and Liu, Jin-Xi and Fang, Dai-Ning},
  journal={International Journal of Engineering Science},
  volume={46},
  number={11},
  pages={1098--1110},
  year={2008},
  publisher={Elsevier}
}

@article{cao2008dispersion,
  title={On dispersion relations of Rayleigh waves in a functionally graded piezoelectric material (FGPM) half-space},
  author={Cao, Xiaoshan and Jin, Feng and Wang, Zikun},
  journal={Acta Mechanica},
  volume={200},
  number={3},
  pages={247--261},
  year={2008},
  publisher={Springer}
}

@article{kumar2021response,
  title={Response of two-temperature on the energy ratios at elastic-piezothermoelastic interface},
  author={Kumar, R and Sharma, P},
  journal={Journal of Solid Mechanics},
  volume={13},
  number={2},
  pages={186--201},
  year={2021}
}

@article{bleustein1968new,
  title={A new surface wave in piezoelectric materials},
  author={Bleustein, Jeffrey L},
  journal={Applied Physics Letters},
  volume={13},
  number={12},
  pages={412--413},
  year={1968}
}

@article{saroj2015love,
  title={Love-type waves in functionally graded piezoelectric material (FGPM) sandwiched between initially stressed layer and elastic substrate},
  author={Saroj, Pradeep K and Sahu, SA and Chaudhary, S and Chattopadhyay, A},
  journal={Waves in Random and Complex Media},
  volume={25},
  number={4},
  pages={608--627},
  year={2015},
  publisher={Taylor \& Francis}
}

@article{arani2011effect,
  title={Effect of material in-homogeneity on electro-thermo-mechanical behaviors of functionally graded piezoelectric rotating shaft},
  author={Arani, A Ghorbanpour and Kolahchi, R and Barzoki, AA Mosallaie},
  journal={Applied Mathematical Modelling},
  volume={35},
  number={6},
  pages={2771--2789},
  year={2011},
  publisher={Elsevier}
}

@article{kumar2013plane,
  title={Plane wave propagation in an anisotropic thermoelastic medium with fractional order derivative and void},
  author={Kumar, Rajneesh and Gupta, Vandana},
  journal={Journal of thermoelasticity},
  volume={1},
  number={1},
  pages={21--34},
  year={2013}
}

@article{kaur2019effect,
  title={Effect of hall current on propagation of plane wave in transversely isotropic thermoelastic medium with two temperature and fractional order heat transfer},
  author={Kaur, Iqbal and Lata, Parveen},
  journal={SN Applied Sciences},
  volume={1},
  number={8},
  pages={900},
  year={2019},
  publisher={Springer}
}

@article{yadav2022reflection,
  title={Reflection of plane waves in a fraction-order generalized magneto-thermoelasticity in a rotating triclinic solid half-space},
  author={Yadav, Anand Kumar},
  journal={Mechanics of Advanced Materials and Structures},
  volume={29},
  number={25},
  pages={4273--4290},
  year={2022},
  publisher={Taylor \& Francis}
}

@article{kaur2022reflection,
  title={Reflection and refraction of plane wave in piezo-thermoelastic diffusive half spaces with three phase lag memory dependent derivative and two-temperature},
  author={Kaur, Iqbal and Lata, Parveen and Singh, Kulvinder},
  journal={Waves in Random and Complex Media},
  volume={32},
  number={5},
  pages={2499--2532},
  year={2022},
  publisher={Taylor \& Francis}
}

@article{kang2021modeling,
  title={Modeling elastic wave propagation through a partially saturated poroviscoelastic interlayer by fractional order derivatives},
  author={Kang, Yonggang and Wei, Peijun and Li, Yueqiu and Zhang, Peng},
  journal={Applied Mathematical Modelling},
  volume={100},
  pages={612--631},
  year={2021},
  publisher={Elsevier}
}

@article{bibi2023propagation,
  title={Propagation and reflection of thermoelastic wave in a rotating nonlocal fractional order porous medium under Hall current influence},
  author={Bibi, Farhat and Ali, Hashmat and Azhar, Ehtsham and Jamal, Muhammad and Ahmed, Iftikhar and Ragab, Adham E},
  journal={Scientific Reports},
  volume={13},
  number={1},
  pages={17703},
  year={2023},
  publisher={Nature Publishing Group UK London}
}

@book{atanackovic2014fractional,
  title={Fractional calculus with applications in mechanics: wave propagation, impact and variational principles},
  author={Atanackovic, Teodor M and Pilipovic, Stevan and Stankovic, Bogoljub and Zorica, Dusan},
  year={2014},
  publisher={John Wiley \& Sons}
}

@article{said2024reflection,
  title={Reflection of waves in a two-temperature magneto-fiber-reinforced solid with memory-dependent derivative using different theories},
  author={Said, Samia M and Abd-Elaziz, El-sayed M and Othman, Mohamed IA},
  journal={Journal of Vibration Engineering \& Technologies},
  volume={12},
  number={7},
  pages={8517--8527},
  year={2024},
  publisher={Springer}
}

@article{barak2023behavior,
  title={Behavior of higher-order MDD on energy ratios at the interface of thermoelastic and piezothermoelastic mediums},
  author={Barak, MS and Ahmad, Hijaz and Kumar, Rajesh and Kumar, Rajneesh and Gupta, Vipin and Awwad, Fuad A and Ismail, Emad AA},
  journal={Scientific Reports},
  volume={13},
  number={1},
  pages={17170},
  year={2023},
  publisher={Nature Publishing Group UK London}
}

@article{li2025effects,
  title={Effects of external magnetic field on the reflection and transmission of thermoelastic coupled waves with consideration of fractional order thermoelasticity},
  author={Li, Ying and Li, Yueqiu and Yue, Tiantian and Cui, Jixian and Wang, Hong},
  journal={Mechanics of Advanced Materials and Structures},
  volume={32},
  number={10},
  pages={2381--2392},
  year={2025},
  publisher={Taylor \& Francis}
}

\appendix
\section{}
\label{appendixa}
$ \chi_0=(\mu_{11}^\prime+\sigma_{11}^\prime)\xi_0^2\sin^3\theta_0+(\mu_{31}^\prime+\sigma_{33}^\prime+2\mu_{44}^\prime)\xi_0^2\cos^2\theta_0\sin\theta_0-\rho^\prime(\xi_0^2c_{I_0}^2\sin\theta_0+{{\Omega}^\prime}^2\sin\theta_0+2{{\Omega}^\prime} i\xi_0c_{I_0}\cos\theta_0)$\\
$M_0=(e_{15}^\prime+e_{31}^\prime)\xi_0^2\cos\theta_0\sin\theta_0$\\
$\nu_0=\beta_{11}i\xi_0\sin\theta_0$\\
$\chi_1=-(\mu_{11}^\prime+\sigma_{11}^\prime)\xi_1^2\sin^3\theta_1+(\mu_{31}^\prime+\sigma_{33}^\prime+2\mu_{44}^\prime)\xi_1^2\cos^2\theta_1\sin\theta_1-\rho^\prime(\xi_1^2c_{I_1}^2\sin\theta_1+{{\Omega}^\prime}^2\sin\theta_1+2{{\Omega}^\prime} i\xi_1c_{I_1}\cos\theta_1) $\\
$M_1=-(e_{31}^\prime+e_{15}^\prime)\xi_1^2\cos\theta_1\sin\theta_1$\\
$\nu_1=\beta_{11}i\xi_1\sin\theta_1$\\
We can see ,\\ 
$\chi_0=-\chi_1, M_0=M_1, \nu_0=-\nu_1$\\
$ \chi_2=(\mu_{44}^\prime+\sigma_{11}^\prime)\xi_2^2\sin^3\theta_2-(\mu_{31}^\prime+\mu_{44}^\prime)\xi_2^2\cos^2\theta_2\sin\theta_2+(\mu_{33}^\prime+\sigma_{33}^\prime)\xi_2^2\cos^2\theta_2\sin\theta_2-\rho^\prime(\xi_2^2\sin\theta_2c_{T_2}^2+{{\Omega}^\prime}^2\cos\theta_2+2\Omega i\xi_2c_{T_2}\cos\theta_2)$
$M_2=(e_{15}^\prime \xi_2^2\sin^2\theta_2+e_{33}^\prime \xi_2^2\cos^2\theta_2)$\\
$\nu_2=-\beta_{33}i\xi_2\cos\theta_2$\\
$\chi_3=\mu_{11}^0\xi_3^2\sin^3\theta_3+(\mu_{13}^0+\mu_{44}^0)\xi_3^2\cos^2\theta_3\sin\theta_3-\mu_{44}^0\alpha 2i\xi_3\cos\theta_3\sin\theta_3+\mu_{44}^0\xi_3^2\sin\theta_3\cos^2\theta_3-\rho^0(\xi_3^2c_{I_3}\sin\theta_3+\Omega^2\sin\theta_3+2\Omega i\xi_3c_{I_3}\cos\theta_3)$\\
$M_3=(e_{15}^0+e_{31}^0)\xi_3^2\cos\theta_3\sin\theta_3-e_{15}^0\alpha i\xi_3\sin\theta_3$\\
$\nu_3=0 $\\
$\chi_4=-(\mu_{13}^0+\mu_{44}^0)\xi_4^2\sin\theta_4\cos^2\theta_4+\mu_{44}^0\xi_4^2\sin^3\theta_4+\mu_{13}^0\alpha i \xi_4\cos\theta_4\sin\theta_4+\mu_{33}^0 i \xi_4\cos\theta_4\sin\theta_4-\mu_{33}^0\alpha i \xi_4\sin^2\theta_4-\rho^0(\xi_4^2c_{T_4}^2\sin\theta_4+\Omega^2\sin\theta_4+2\Omega i\xi_4c_{T_4}\cos\theta_4)$\\
$M_4=e_{33}^0\alpha i\xi_4\cos\theta_4-e_{33}^0\xi_4^2\cos^2\theta_4-e_{15}^0\xi_4^2\sin^2\theta_4$\\
$\nu_4=0$\\

\section{}
\label{B}
$L_0= (2e_{15}^\prime+e_{31}^\prime)\xi_0^2\sin^2\theta_0\cos\theta_0+e_{33}^\prime \xi_0^2\cos^3\theta_0$\\
$G_0=-\epsilon_{11}^\prime \xi_0^2\sin^2\theta_0-\epsilon_{33}^\prime \xi_0^2\cos^2\theta_0$\\
$S_0=-p_3 i\xi_0\cos\theta_0$\\
$L_1=(2e_{15}^\prime+e_{31}^\prime)\xi_1^2\sin^2\theta_1\cos\theta_1-e_{33}^\prime \xi_1^2\cos^3\theta_1$\\
$G_1=\epsilon_{11}^\prime \xi_1^2\sin^2\theta_1+\epsilon_{33}^\prime \xi_1^2\cos^2\theta_1$\\
$S_1=-p_3 i\xi_1\cos\theta_1$\\
$L_2=e_{15}^\prime \xi_2^2\sin^3\theta_2-(e_{15}^\prime+e_{31}^\prime)\xi_2^2\cos^2\theta_2\sin\theta_2+e_{33}^\prime \xi_2^2\sin\theta_2\cos^2\theta_2$\\
$G_2=-\epsilon_{11}^\prime \xi_2^2\sin^2\theta_2-\epsilon_{33}^\prime \xi_2^2\cos^2\theta_2$\\
$S_2=p_3 i\xi_2\cos\theta_2$\\
$L_3=(e_{31}^0+2e_{15}^0)\xi_3^2\sin^2\theta_3\cos\theta_3+e_{33}^0\cos^3\theta_3-e_{31}^0\alpha i\xi_3\sin^2\theta_3-e_{33}^0\alpha i\xi_3\cos^2\theta_3$\\
$G_3=-\epsilon_{11}^0\xi_3^2\sin^2\theta_3-\epsilon_{33}^0\xi_3^2\cos^2\theta_3+\epsilon_{33}^0\alpha i\xi_3\cos\theta_3$\\
$S_3=0$\\
$L_4=-(e_{31}^0+e_{15}^0)\xi_4^2\cos^2\theta_4+e_{15}^0\sin^3\theta_4+e_{33}^0\xi_4^2\sin\theta_4\cos^2\theta_4+e_{31}^0\alpha i\xi_4\sin\theta_4\cos\theta_4-e_{33}^0\alpha i\xi_4\cos\theta_4\sin\theta_4$\\
$G_4=-\epsilon_{11}^0\xi_4^2\sin^2\theta_4-\epsilon_{33}^0\xi_4^2\cos^2\theta_4+\epsilon_{33}^0\alpha i\xi_4\cos\theta_4$\\
$S_4=0$\\
\section{}
\label{C}
$E_0=T_0\beta_{11}\xi_0^2c_{I_0}\sin^2\theta_0+T_0\beta_{33}\xi_0^2c_{I_0}\cos^2\theta_0-T_0\beta_{11}\tau_0\sin^2\theta_0(-c_{I_0})^{\gamma+1}(i\xi_0)^{\gamma+2}+T_0\tau_0\beta_{33}\cos^2\theta_0(i\xi_0)^{\gamma+2}(-c_{I_0})^{\gamma+1}$\\
$D_0=-T_0p_3\xi_0^2c_{I_0}\cos\theta_0-\tau_0T_0p_3\cos\theta_0(i\xi_0)^{\gamma+2}(-c_{I_0})^{\gamma+1}$\\
$F_0=K_{11}\xi_0^2\sin^2\theta_0+K_{33}\xi_0^2\cos^2\theta_0-\rho^\prime C_ei\xi_0c_{I_0}-\tau_0\rho^\prime C_e(-i\xi_0c_{I_0})^{\gamma+1}$\\
$E_1=T_0\beta_{11}\xi_1^2c_{I_1}\sin^2\theta_1+T_0\beta_{33}\xi_1^2c_{I_1}\cos^2\theta_1-T_0\tau_0\beta_{11}\sin^2\theta_1(i\xi_1)^{\gamma+1}(-c_{I_1})^{\gamma+1}-T_0\tau_0\beta_{33}\cos^2\theta_1(-i\xi_1)^{\gamma+2}(-c_{I_1})^{\gamma+1}$\\
$D_1=T_0p_3i\xi_1c_{I_1}\cos\theta_1(-i\xi_1)^{\gamma+2}c_{I_1}^{\gamma+1}$\\
$F_1=K_{11}\xi_1^2\sin^2\theta_1+K_{33}\xi_1^2\cos^2\theta_1-\rho^\prime C_ei\xi_1c_{I_1}-\tau_0\rho^\prime C_e(-i\xi_1c_{I_1})^{\gamma+1}$\\
$E_2=T_0\beta_{11}\xi_2^2c_{T_2}\cos\theta_2\sin\theta_2-T_0\beta_{33}\xi_2^2c_{T_2}\cos\theta_2\sin\theta_2-T_0\tau_0\beta_{11}\cos\theta_2\sin\theta_2(i\xi_2)^{\gamma+2}(-c_{T_2})^{\gamma+1}\\-T_0\tau_0\beta_{33}\cos^2\theta_2(-i\xi_2)^{\gamma+2}(c_{T_2})^{\gamma+1}$\\
$D_2=T_0p_3\xi_2^2c_{T_2}\cos\theta_2+T_0\tau_0p_3\cos\theta_2(-i\xi_2)^{\gamma+2}(c_{T_2})^{\gamma+1}$\\
$F_2=K_{11}\xi_2^2\sin^2\theta_2+K_{33}\xi_2^2\cos\theta_2-\rho^\prime C_ei\xi_2c_{T_2}-\tau_0\rho^\prime C_e(-i\xi_2c_{T_2})^{\gamma+1}$\\
$E_3= 0$,
$D_3=-0$,
$F_3=0$,
$E_4=0$,
$D_4=0$,
$F_4=0$

\section{}
\label{D}
$a_{11}=1, a_{12}=\frac{J_{11}}{J_1}, a_{13}=-\frac{J_{12}}{J_1}, a_{14}=-\frac{J_{13}}{J_1}$,\\
$a_{21}= 1, a_{22}=-\frac{J_{21}}{J_2}, a_{23}= -\frac{J_{22}}{J_2},a_{24}=\frac{J_{23}}{J_2}$,\\
$a_{31}=1, a_{32}=-\frac{N_2}{N_0}, a_{33}=0, a_{34}=0$,\\
$a_{41} = 1, a_{42} = \frac{R_2}{R_0}, a_{43}= 0, a_{44}=0$,\\
$b_{11}=1,b_{12}=\frac{I_{11}}{I_1},b_{13}=-\frac{I_{12}}{I_1}, b_{14}=-\frac{I_{13}}{I_1},$\\
$b_{21}= 1, b_{22}=-\frac{J_{21}}{J_2}, b_{23}= -\frac{J_{22}}{J_2},b_{24}=\frac{J_{23}}{J_2}$,\\
$b_{31}=1, b_{32}=-\frac{N_2}{N_0}, b_{33}=0, b_{34}=0$,\\
$b_{41} = 1, b_{42} = \frac{R_2}{R_0}, b_{43}= 0, b_{44}=0$,\\
$m_1=-1,m_2=1,m_3=1,m_4=-1$,\\
$s_1=-1,s_2=1,s_3=1,s_4=-1$,\\
$J_{11}=\zeta_1[(\mu_{31}^\prime-\mu_{33}^\prime)\cos\theta_2\sin\theta_2-N_2e_{33}^\prime\cos\theta_2-R_2\beta_{33}]$,\\
$J_{12}=\zeta_2[\mu_{31}^0\sin^2\theta_3+\mu_{33}^0\cos^2\theta_3]$,\\
$J_{13}=\zeta_3[(\mu_{31}^0-\mu_{33}^0)\sin\theta_4\cos\theta_4]$,\\
$J_1=\mu_{31}^\prime\sin^2\theta_0+\mu_{33}^\prime\cos^2\theta_0+N_0e_{33}^\prime\cos\theta_0-R_0\beta_{33}$,\\
$J_{21}=\zeta_1[\mu_{44}^\prime\sin2\theta_2+N_2e_{15}^\prime\sin\theta_2]$,\\
$J_{22}=\zeta_2[\mu_{44}^0\sin2\theta_3]$,\\
$J_{23}=\zeta_3[\mu_{44}^0\sin2\theta_4]$,\\
$J_2=\mu_{44}^\prime\sin2\theta_0+N_0e_{15}^\prime\sin\theta_0$,\\
$I_{11}=\sin\theta_2, I_{12}=\cos\theta_3,I_{13}=\sin\theta_4,I_1=\cos\theta_0$\\
$N_0=\frac{(E_0\nu_0-F_0\chi_0)}{(F_0M_0-D_0\nu_0)}$,\\
$N_2=\frac{(E_2\nu_2-F_2\chi_2)}{(F_2M_2-D_2\nu_2)}$,\\
$R_0=\frac{(M_0E_0-D_0\chi_0)}{(D_0\nu_0-F_0M_0)}$,\\
$R_2=\frac{(M_2E_2-D_2\chi_2)}{(D_2\nu_2-F_2M_2)}$,\\

\section{}
\label{E}

$\mathbf{X} = 
\begin{vmatrix}
a_{11} & a_{12} & a_{13} & a_{14} \\
a_{21} & a_{22} & a_{23} & a_{24} \\
a_{31} & a_{32} & a_{33} & a_{34} \\
a_{41} & a_{42} & a_{43} & a_{44}
\end{vmatrix}$,
$\mathbf{X_1} = 
\begin{vmatrix}
m_1 & a_{12} & a_{13} & a_{14} \\
m_2 & a_{22} & a_{23} & a_{24} \\
m_3 & a_{32} & a_{33} & a_{34} \\
m_4 & a_{42} & a_{43} & a_{44}
\end{vmatrix}$,
$\mathbf{X_2} = 
\begin{vmatrix}
a_{11} & m_1 & a_{13} & a_{14} \\
a_{21} & m_2 & a_{23} & a_{24} \\
a_{31} & m_3 & a_{33} & a_{34} \\
a_{41} & m_4 & a_{43} & a_{44}
\end{vmatrix}$\\
$\mathbf{X_3} = 
\begin{vmatrix}
a_{11} & a_{12} & m_1 & a_{14} \\
a_{21} & a_{22} & m_2 & a_{24} \\
a_{31} & a_{32} & m_3 & a_{34} \\
a_{41} & a_{42} & m_4 & a_{44}
\end{vmatrix}$,
$\mathbf{X_4} = 
\begin{vmatrix}
a_{11} & a_{12} & a_{13} & m_1 \\
a_{21} & a_{22} & a_{23} & m_2 \\
a_{31} & a_{32} & a_{33} & m_3 \\
a_{41} & a_{42} & a_{43} & m_4
\end{vmatrix}$\\
\section{}
\label{F}

$\mathbf{Y} = 
\begin{vmatrix}
b_{11} & b_{12} & b_{13} & b_{14} \\
b_{21} & b_{22} & b_{23} & b_{24} \\
b_{31} & b_{32} & b_{33} & b_{34} \\
b_{41} & b_{42} & b_{43} & b_{44}
\end{vmatrix}$,
$\mathbf{Y_1} = 
\begin{vmatrix}
s_1 & b_{12} & b_{13} & b_{14} \\
s_2 & b_{22} & b_{23} & b_{24} \\
s_3 & b_{32} & b_{33} & b_{34} \\
s_4 & b_{42} & b_{43} & b_{44}
\end{vmatrix}$,
$\mathbf{Y_2} = 
\begin{vmatrix}
b_{11} & s_1 & b_{13} & b_{14} \\
b_{21} & s_2 & b_{23} & b_{24} \\
b_{31} & s_3 & b_{33} & b_{34} \\
b_{41} & s_4 & b_{43} & b_{44}
\end{vmatrix}$\\
$\mathbf{Y_3} = 
\begin{vmatrix}
b_{11} & b_{12} & s_1 & b_{14} \\
b_{21} & b_{22} & s_2 & b_{24} \\
b_{31} & b_{32} & s_3 & b_{34} \\
a_{41} & b_{42} & s_4 & b_{44}
\end{vmatrix}$,
$\mathbf{Y_4} = 
\begin{vmatrix}
b_{11} & b_{12} & b_{13} & s_1 \\
b_{21} & b_{22} & b_{23} & s_2 \\
b_{31} & b_{32} & b_{33} & s_3 \\
b_{41} & b_{42} & b_{43} & s_4
\end{vmatrix}$\\
\end{document}